\documentclass[universe,article,accept,oneauthor,pdftex,12pt,a4paper]{mdpi}

\usepackage{graphicx}
\usepackage{amsmath}
\usepackage{amssymb}
\usepackage{amsfonts}
\usepackage{bm}
\usepackage{color}
\usepackage{soul}
\usepackage{upgreek}
\graphicspath{{fig/}}

\lastpage{x}
\doinum{10.3390/------}
\pubvolume{xx}
\pubyear{2015}
\externaleditor{Academic Editor: Kazuharu Bamba}
\history{Received: 28 June 2015  / Accepted: 10 October 2015 / Published: xx}

\Title{A Cosmological Model Describing the Early
Inflation, the Intermediate Decelerating Expansion, and the Late Accelerating
Expansion of the Universe by a Quadratic Equation of State}

\Author{Pierre-Henri Chavanis}

\address[1]{%
Laboratoire de Physique Th\'eorique, Universit\'e Paul Sabatier,
118 route de Narbonne, 31062 Toulouse, France; E-Mail: chavanis@irsamc.ups-tlse.fr.
\vspace{-12pt}}

\abstract{We develop a cosmological model based on a quadratic equation of state
\mbox{$p/c^2=-(\alpha+1){\rho^2}/{\rho_P}+\alpha\rho-(\alpha+1)\rho_
{\Lambda}$}, where
$\rho_P$ is the Planck density and $\rho_{\Lambda}$ the cosmological density,
``unifying'' vacuum energy and dark energy in the spirit of a~generalized
Chaplygin gas model. For   $\rho\rightarrow
\rho_P$, it reduces to $p=-\rho_P c^2$ leading to a phase of
early accelerating expansion (early
inflation) with a constant density equal to the Planck density
$\rho_P=5.16 \times 10^{99}\, {\rm g}/{\rm m}^3$ (vacuum energy). For
$\rho_{\Lambda}\ll\rho\ll \rho_P$, we recover  the standard linear equation
of
state
$p=\alpha \rho c^2$ describing radiation
($\alpha=1/3$) or pressureless matter
($\alpha=0$) and leading to an intermediate phase of decelerating expansion. For
$\rho\rightarrow \rho_{\Lambda}$, we get
$p=-\rho_{\Lambda} c^2$ leading to a phase of late accelerating
expansion (late inflation) with a constant density equal to the
cosmological density $\rho_{\Lambda}=7.02\times 10^{-24}\, {\rm g}/{\rm
m}^3$ (dark energy). The pressure is successively negative
(vacuum energy), positive (radiation and matter),
and negative again (dark
energy). We show a nice ``symmetry'' between the early universe (vacuum energy
$+$ $\alpha$-fluid) and the late universe
($\alpha$-fluid $+$ dark energy). In~our model, they are described by two
polytropic equations of state with index $n=+1$ and $n=-1$
respectively. Furthermore, the Planck density $\rho_P$ in the early
universe plays a~role similar to the cosmological density
$\rho_{\Lambda}$ in the late universe. They represent fundamental
upper and lower density bounds differing by $122$ orders of
magnitude. The cosmological constant ``problem'' may be a false problem.
We study the evolution of the scale factor, density, and pressure.
Interestingly, our quadratic equation of state leads to a fully analytical
model describing the evolution of the universe from the early inflation (Planck
era) to
the late accelerating expansion (de Sitter era). These two phases are bridged
by a decelerating algebraic expansion ($\alpha$-era).
Our model does not present any
singularity at $t=0$ and exists eternally in the past (although it may
be incorrect to extrapolate the solution to the infinite past).
On the other hand, it admits a scalar field interpretation based
on an~inflaton, quintessence, or tachyonic field. Our model generalizes the
standard $\Lambda$CDM model by incorporating naturally a phase of early
inflation that avoids the primordial singularity. Furthermore, it describes the
early inflation, the intermediate decelerating expansion, and the late
accelerating expansion of the universe simultaneously in terms of a single
equation of state. We determine the corresponding scalar field potential that
unifies the inflaton and quintessence potentials.
}


\begin{document}

\vspace{-12pt}
\section{Introduction}

The evolution of the universe may be divided into four main periods \cite{bt}.
In the vacuum energy era (Planck era), the
universe undergoes a phase of early
inflation  that brings it from the Planck size $l_P=1.62\times 10^{-35}\, {\rm
m}$
to an almost ``macroscopic'' size $a\sim 10^{-6}~{\rm m}$ in a tiniest fraction
of a second \cite{guth1,guth2,guth3,linde}.  The universe then enters in the
radiation era
and, when the temperature cools down below approximately $10^3~{\rm K}$, in the
matter era \cite{weinberg}. Finally, in the dark energy era (de Sitter era), the
universe undergoes a phase of late inflation \cite{cst}. The early inflation  is
necessary  to solve notorious difficulties such as the singularity problem, the
flatness problem, and the horizon problem
\cite{guth1,guth2,guth3,linde}. The late inflation is necessary to account for
the
observed accelerating expansion of the universe
\cite{novae1,novae2,novae3,novae4}. At present, the
universe is composed of approximately $5\%$ baryonic matter, $20\%$
dark matter, and $75\%$ dark energy \cite{bt}.  Despite the success of the
standard $\Lambda$CDM model, the nature
of vacuum energy, dark matter, and dark energy remains very mysterious and leads
to many
speculations.

The phase of inflation in the early universe is usually described by some
hypothetical scalar field $\phi$, called inflaton, with its
origin in the quantum fluctuations of the
vacuum \cite{guth1,guth2,guth3,linde}. This leads to an equation of state
$p=-\rho c^2$,
implying a constant energy density,  called the vacuum energy. This energy
density is usually identified with the Planck density  $\rho_P=c^5/G^2\hbar=5.16
\times
10^{99}\,
{\rm g}/{\rm
m}^3$. As a result of the vacuum energy, the universe expands exponentially
rapidly on a timescale of the order of the Planck
time $t_P=1/(G\rho_P)^{1/2}=(\hbar
G/c^5)^{1/2}=5.39\times
10^{-44}~{\rm s}$.

The phase of acceleration in the late universe is usually ascribed to
the cosmological constant $\Lambda$ which is equivalent to a constant
energy density $\rho_{\Lambda}=\Lambda/(8\pi G)=7.02\times 10^{-24}\, {\rm
g}/{\rm m}^3$ called the dark energy \cite{cst}. This acceleration can be
modeled
by an equation of state $p=-\rho c^2$ implying a constant energy
density identified with the cosmological density $\rho_{\Lambda}$. As
a result of the dark energy, the universe expands exponentially
rapidly on a timescale of the order of the cosmological time
$t_{\Lambda}=1/(G\rho_{\Lambda})^{1/2}=(8\pi/\Lambda)^{1/2}=1.46\times
10^{18}\, {\rm s}$ (de Sitter solution). This
leads to a phase of late inflation. Instead of introducing a cosmological
constant, some authors have proposed to explain the
acceleration of the universe in terms of a dark
energy with a time-varying density
associated with a canonical scalar field called
quintessence
\cite{quintessence1,quintessence2,quintessence3,quintessence4,
quintessence5,quintessence6,quintessence7,quintessence8,quintessence9,
quintessence10,quintessence11,quintessence12,quintessence13}, or in terms of a
tachyonic field \cite{sen,tachyon1,tachyon2,tachyon3}.

Between the phase of early inflation and the phase of late accelerating
expansion, the universe is in the radiation era, then in the matter era
\cite{weinberg}.  These
phases are described by a linear equation of state
$p=\alpha\rho c^2$ with $\alpha=1/3$ for the radiation and $\alpha=0$ for
the
pressureless matter (including baryonic matter and dark matter). The scale
factor increases algebraically as $t^{2/[3(1+\alpha)]}$ and the density
decreases algebraically as $t^{-2}$. For $\alpha>-1/3$, the universe is
decelerating.

In recent works \cite{chavanis1,chavanis2,chavanis3,chavanisAIP}, we have
proposed to
describe the transition
between a phase of algebraic expansion ($p=\alpha\rho c^2$) and a phase of
exponential expansion ($p=-\rho c^2$) by a generalized polytropic equation of
state of the form
 \begin{equation}
\label{intro1}
p=(\alpha \rho+k\rho^{1+1/n}) c^2.
\end{equation}
This is the sum of a standard linear equation of state $p=\alpha\rho c^2$ and
a
polytropic equation of state $p=k\rho^{\gamma}c^2$ with $\gamma=1+1/n$.
Polytropic equations of state play an important role in astrophysics
\cite{chandra,st}, statistical physics \cite{tsallis} and mathematical biology
\cite{murray}, and they may also be relevant in cosmology. We~have studied the
equation of state (\ref{intro1}) for any values of $\alpha$, $k$ and
$n$ and
found the following structure. Positive indices $n>0$ describe the early
universe where the polytropic component dominates the linear component because
the density is high. Negative indices $n<0$
describe the late universe where the
polytropic component dominates the linear component because the density is low.
On the other hand, a positive polytropic pressure ($k>0$) leads to past or
future singularities (or peculiarities) while a negative polytropic pressure
($k<0$) leads to a phase of exponential expansion (inflation) in
the past or in the future [Note 1: the polytropic equation of
state (\ref{intro1}) with $k<0$ and $n<0$
is equivalent to the generalized Chaplygin gas model
$p/c^2=B\rho-A/\rho^a$ with $a\ge -1$ that has been proposed to describe the
late accelerating expansion of the universe (the original
Chaplygin gas model corresponds to  $B=0$ and $a=1$)
\cite{chaplygin1,chaplygin2,chaplygin3,chaplygin4,chaplygin5,chaplygin6,
chaplygin7,chaplygin8,chaplygin9}. Therefore,
the polytropic equation of state (\ref{intro1}) with $k<0$ and
$n>0$ introduced in \cite{chavanis1} can be seen as an extension of the
generalized Chaplygin gas model to describe the early accelerating
expansion of the universe
(inflation).]. In
the early universe ($n>0$, $k<0$), the
generalized polytropic equation of state (\ref{intro1}) leads to a
maximum bound
for the density that it is natural to identify with the Planck density
$\rho_P=5.16 \times 10^{99}\, {\rm g}/{\rm m}^3$ (vacuum energy). In the
late
universe ($n<0$,
$k<0$), it leads to a minimum bound for the density that it is natural to
identify with the cosmological density $\rho_{\Lambda}=7.02\times 10^{-24}\,
{\rm g}/{\rm m}^3$ (dark energy). These bounds differ by $122$ orders of
magnitude. Taking
$n=+1$, $k=-4/(3\rho_P)$ and $\alpha=1/3$ we obtain an equation of state
$p/c^2=-4{\rho^2}/3{\rho_P}+\rho/3$ that describes the transition between
the
vacuum energy era ($\rho_P$) and the radiation era. On the other hand, taking
$n=-1$,
$k=-\rho_{\Lambda}$ and $\alpha=0$ we obtain an equation of state
$p/c^2=-\rho_{\Lambda}$  that describes the transition between the matter era
and the dark energy era ($\rho_{\Lambda}$). More~generally, the equation of
state
$p/c^2=-(\alpha+1){\rho^2}/{\rho_P}+\alpha\rho$ describes the transition
between
the vacuum energy era and an $\alpha$-era in the early universe, and the
equation of state $p/c^2=\alpha\rho-(\alpha+1)\rho_{\Lambda}$ describes
the
transition
between an $\alpha$-era and the dark energy era in the late
universe.

In this paper, we propose to describe the vacuum energy, the $\alpha$-fluid, and
the dark energy in a ``unified'' manner by a single, quadratic, equation of
state of the form [Note 2: This idea was sketched in
\cite{chavanis2,chavanisAIP} and
is here systematically developed.]:
\begin{equation}
\label{intro2}
p=-(\alpha+1)\frac{\rho^2}{\rho_P} c^2+\alpha\rho
c^2-(\alpha+1)\rho_{\Lambda}
c^2
\end{equation}
involving the Planck density and the cosmological density [Note 3: It is
oftentimes
argued that the dark energy
(cosmological constant) corresponds to the vacuum
energy. This leads to the so-called \emph{cosmological constant problem}
\cite{weinbergcosmo,paddycosmo} because the cosmological density
$\rho_{\Lambda}=7.02\times 10^{-24}\, {\rm g}/{\rm m}^3$ and the Planck
density $\rho_P=5.16 \times 10^{99}\, {\rm g}/{\rm m}^3$ differ by about
$122$ orders of magnitude. We think it is a mistake to
identify the dark energy with the vacuum energy. In this
paper, we regard the vacuum energy and the dark energy as two
distinct entities. We call vacuum energy the energy associated
with the Planck density and dark energy the energy associated with the
cosmological density. The vacuum energy is responsible for the
inflation of the early universe and the dark energy for the inflation
(acceleration) of the late universe. In this viewpoint, the
vacuum energy is due to
quantum mechanics and the dark energy is an effect of general
relativity. The cosmological constant $\Lambda$ is interpreted as a
fundamental constant of nature applying to the cosmophysics in the
same way the Planck constant $\hbar$ applies to the
microphysics.]. In the early universe ($\rho\gg\rho_{\Lambda}$), we recover
the
equation of state
$p/c^2=-(\alpha+1){\rho^2}/{\rho_P}+\alpha\rho$ unifying the vacuum
energy
($\rho_P$) and
the $\alpha$-fluid. In the late universe ($\rho\ll\rho_{P}$), we recover
the equation of state
$p/c^2=\alpha\rho-(\alpha+1)\rho_{\Lambda}$
unifying the $\alpha$-fluid and the dark energy ($\rho_{\Lambda}$). The
$\alpha$-fluid may
represent  radiation ($\alpha=1/3$)
or pressureless dark matter ($\alpha=0$).
Actually, some works \cite{muller} indicate that dark matter may be described by
an isothermal equation of state $p=\alpha\rho c^2$  with a small value of
$\alpha\ll 1$. It is therefore useful to leave $\alpha$ unspecified and
treat the general case of arbitrary $\alpha$. However, to
simplify the discussion, we shall assume $0\le \alpha\le 1$. Moreover, for
illustrations, we will select the values $\alpha=1/3$ (radiation) and
$\alpha=0$ (pressureless matter) considered in our previous
studies \cite{chavanis1,chavanis2,chavanis3,chavanisAIP}.

The quadratic equation of state (\ref{intro2}) leads to a fully
analytical
cosmological model
describing the evolution of the universe from the initial inflation (Planck era)
to the late accelerating expansion (de Sitter era). These two phases are bridged
by an algebraic decelerating expansion ($\alpha$-era). The~pressure is
successively negative
(vacuum energy), positive ($\alpha$-era),
and negative again (dark
energy). Our~model does not
present any
singularity at $t=0$---the phase of early  inflation avoids the primordial Big
Bang singularity---and exists eternally in the past (although it may
be incorrect to extrapolate the solution to the infinite past).
On the other hand, our model admits a scalar field interpretation
based on an~inflaton, quintessence, or  tachyonic
field. This correspondence is interesting because the early
inflation and the late
acceleration of the universe are usually described in terms of a scalar field.
There~exist general techniques \cite{cst,bamba}  to
represent
a fluid model in terms of a (canonical or tachyonic) scalar field. Although~the
hydrodynamic and field representations are equivalent at the background level,
they may totally differ at the level of
the perturbations when the fluid has a negative pressure. As a result, it~is
important to give the two representations and study the evolution of the
perturbations in each case. It~is known that the scalar field representation
is more complete than the hydrodynamic
representation, and provides a more realistic
model [Note 4: The perfect fluid approach is more
rough than the scalar field approach because for given perfect fluid variables
$\rho$ and $P$ one cannot restore the scalar field variables $\phi$,
$\nabla\phi$ and $V(\phi)$  in the inhomogeneous case (while the converse is
always
possible).]. In this paper, we only
consider the evolution of the
background where the hydrodynamic and field  representations are equivalent. The~study of the perturbations will be considered in future works.

The paper is organized as follows. In Section \ref{sec_basic}, we recall the basic
equations of cosmology that are needed in our study. In Section
\ref{sec_early}, we describe the transition between the vacuum energy era
($\rho_P$) and
the $\alpha$-era in the early universe.  In Section \ref{sec_late}, we describe
the
transition between the $\alpha$-era and the dark energy era ($\rho_{\Lambda}$)
in the late
universe. In Section \ref{sec_g}, we introduce the general model where vacuum
energy $+$ $\alpha$-era $+$ dark energy are described by the quadratic equation
of state (\ref{intro2}). This model reveals a  nice ``symmetry'' between
the
early universe (vacuum energy $+$ $\alpha$-era) and the late universe
($\alpha$-era $+$ dark energy). These two phases are described by two polytropic
equations of state with index $n=+1$ and $n=-1$ respectively. The mathematical
formulae in the early and in the late universe are strikingly symmetric.
Furthermore, the cosmological density $\rho_{\Lambda}$ in the late universe
plays a role similar to the Planck density $\rho_P$ in the early universe. They
represent fundamental lower and upper density bounds differing by $122$ orders
of magnitude. Interestingly, these densities $\rho_P$ and $\rho_{\Lambda}$
(together with $\alpha$) appear as the coefficients of the  equation of state
(\ref{intro2}). Therefore, this equation of state provides a~``unification'' of
vacuum energy and dark energy. We propose to interpret vacuum energy, radiation
and dark energy as a ``generalized radiation'' described by the equation of
state (\ref{intro2}) with $\alpha=1/3$ and treat baryonic matter and
dark matter
as independent species. This leads to Equation (\ref{smx2}) that
generalizes Equation (\ref{smx1}) of the
standard $\Lambda$CDM
model. This equation avoids the primordial singularity and describes the
whole evolution of the universe from the early inflation to the late
accelerating expansion. In Appendix \ref{sec_sft}, we develop a
scalar field theory and derive the potential $V(\phi)$ associated with the
quadratic equation of state (\ref{intro2}). This potential  (see
Equations
(\ref{sft6})--(\ref{unif1})) unifies the inflaton
potential in the early universe (see Equation (\ref{early15b})) and the
quintessence
potential in the late
universe (see Equation (\ref{hel2})).

\section{Basic Equations of Cosmology}
\label{sec_basic}

In a space with uniform curvature, the line element of the expanding universe is
given by the
Friedmann-Lema\^itre-Roberston-Walker (FLRW) metric
\begin{eqnarray}
\label{b1}
ds^2=c^2 dt^2-a(t)^2\left\lbrace \frac{dr^2}{1-kr^2}+r^2\,  (d\theta^2+
\sin^2\theta\,  d\phi^2)\right \rbrace,
\end{eqnarray}
where $a(t)$ represents the radius of curvature of the $3$-dimensional
space, or the scale factor. By an abuse of language, we
shall sometimes call it the ``radius of the universe''. On the other
hand, $k$ determines the curvature of space. The universe may be
closed ($k>0$), flat ($k=0$), or open ($k<0$).

If the universe is isotropic and homogeneous at all points in conformity with
the line element of Equation~(\ref{b1}), and contains a uniform perfect fluid of
energy
density $\epsilon(t)=\rho(t) c^2$ and isotropic pressure $p(t)$, the
energy-momentum tensor $T^{i}_{j}$ is
\begin{eqnarray}
\label{ei2}
T^0_0=\rho c^2,\qquad T^1_1=T^2_2=T^3_3=-p.
\end{eqnarray}
The Einstein equations
\begin{eqnarray}
\label{ei3}
R^i_j-\frac{1}{2}g^i_j R-\Lambda g^i_j=-\frac{8\pi G}{c^2}T^i_j
\end{eqnarray}
relate the geometrical structure of the spacetime ($g_{ij}$) to the
content of the universe ($T_{ij}$). For the sake of generality, we have
accounted for a possibly non-zero cosmological constant $\Lambda$. Given Equations
(\ref{b1}) and (\ref{ei2}), these
equations reduce to
\begin{eqnarray}
\label{ei4}
8\pi G\rho+\Lambda=3 \frac{{\dot a}^2+kc^2}{a^2},\qquad \frac{8\pi
G}{c^2}p-\Lambda=-\frac{2 a \ddot a+{\dot a}^2+kc^2}{a^2},
\end{eqnarray}
where dots denote differentiation with respect to time. These are the well-known
cosmological equations describing a non-static universe first derived by
Friedmann \cite{weinberg}.

The Friedmann equations are usually written in the form
\begin{equation}
\label{b2}
\frac{d\rho}{dt}+3\frac{\dot a}{a}\left (\rho+\frac{p}{c^2}\right )=0,
\end{equation}
\begin{equation}
\label{b3}
\frac{\ddot a}{a}=-\frac{4\pi G}{3} \left (\rho+\frac{3p}{c^2}\right
)+\frac{\Lambda}{3},
\end{equation}
\begin{equation}
\label{b4}
H^2=\left (\frac{\dot a}{a}\right )^2=\frac{8\pi
G}{3}\rho-\frac{kc^2}{a^2}+\frac{\Lambda}{3},
\end{equation}
where we have introduced the Hubble parameter $H=\dot a/a$.  Among
these three equations, only two are independent. The
first equation, which can be viewed as an ``equation of continuity'',
can be directly derived from the conservation of the energy momentum
tensor $\partial_i T^{ij}=0$ which results from the Bianchi
identities. For a given barotropic equation of state $p=p(\rho)$, it
determines the relation between the density and the scale
factor. Then, the temporal evolution of the scale factor is given by
Equation (\ref{b4}).

In this paper, we consider a flat universe ($k=0$) in agreement
with the
observations of the cosmic microwave background (CMB)
\cite{planck2013,planck2015}. On the other
hand, we set $\Lambda=0$ because the contribution of the dark energy
will be taken into account in the equation of state. The Friedmann equations
then reduce
to
\begin{equation}
\label{b7}
\frac{d\rho}{dt}+3\frac{\dot a}{a}\left (\rho+\frac{p}{c^2}\right )=0,
\end{equation}
\begin{equation}
\label{b8}
\frac{\ddot a}{a}=-\frac{4\pi G}{3} \left (\rho+\frac{3p}{c^2}\right ),
\end{equation}
\begin{equation}
\label{b9}
H^2=\left (\frac{\dot a}{a}\right )^2=\frac{8\pi G}{3}\rho.
\end{equation}
The deceleration parameter is defined by
\begin{equation}
\label{b10}
q(t)=-\frac{{\ddot a}a}{{\dot a}^2}.
\end{equation}
The universe is decelerating when $q>0$ and accelerating when $q<0$.
Introducing the equation of state parameter $w=p/\rho c^2$, and using the
Friedmann equations (\ref{b8}) and (\ref{b9}), we obtain for \mbox{a flat
universe}
\begin{equation}
\label{b11}
q(t)=\frac{1+3w(t)}{2}.
\end{equation}
We see from Equations (\ref{b8}) and (\ref{b11}) that the universe is
decelerating if
$w>-1/3$
(strong energy condition) and accelerating if $w<-1/3$ [Note 5: According to
general relativity, the source for the
gravitational potential is $\rho+3p/c^2$. Indeed, the spatial part ${\bf g}$ of
the geodesic acceleration satisfies the exact equation $\nabla\cdot {\bf
g}=-4\pi G (\rho c^2+3p)$ showing that the source of geodesic acceleration is
$\rho+3p/c^2$ not $\rho$ \cite{paddycosmo}. Therefore, in general
relativity, gravitation becomes ``repulsive'' when $p<-\rho c^2/3$.]. On the
other hand,
according to Equation (\ref{b7}), the density decreases with the scale factor if
$w>-1$ (null dominant energy condition) and increases with the scale factor if
$w<-1$. The latter case corresponds to a ``phantom''~universe
\cite{ghosts1,ghosts2,ghosts3,ghosts4,ghosts5,ghosts6,ghosts7,ghosts8,
ghosts9,ghosts10,ghosts11,ghosts12,ghosts13,ghosts14,ghosts15,ghosts16,ghosts17,
ghosts18}.

\section{The Early Universe}
\label{sec_early}

In Reference \cite{chavanis1}, we have proposed to describe the  transition between
the vacuum energy era and the $\alpha$-era in the early universe by a single
equation of state of the form of Equation (\ref{intro1}) with $k<0$ and $n>0$.
It can be
written as
\begin{equation}
\label{newearly1a}
p/c^2=\alpha\rho-|k|{\rho}^{1+1/|n_e|}.
\end{equation}
Assuming $w>-1$, the equation of continuity (\ref{b7}) can be
integrated into
\begin{equation}
\label{newearly2a}
\rho=\frac{\rho_P}{\lbrack 1+(a/a_1)^{3(1+\alpha)/|n_e|}\rbrack^{|n_e|}},
\end{equation}
where $\rho_P=\lbrack(\alpha+1)/|k|\rbrack^{|n_e|}$ and $a_1$ is a constant of
integration. We see that $\rho_P$ corresponds to an~upper bound (maximum value)
for the density reached for $a\rightarrow 0$. Since this solution describes the
early universe, it is natural to identify $\rho_P$ with the Planck density
$\rho_P=c^5/G^2\hbar=5.16\times 10^{99}\, {\rm g/m^3}$. As~a~result, the
equation of
state (\ref{newearly1a}) can be rewritten as
 \begin{equation}
\label{newearly1}
p/c^2=\alpha\rho-(\alpha+1)\rho \left (\frac{\rho}{\rho_P}\right
)^{1/|n_e|}.
\end{equation}
For the sake of simplicity, and for definiteness, we shall select the index
$n_e=1$. The general case $n_e>0$ has been treated in \cite{chavanis1} and leads
to qualitatively similar results. Therefore, we propose to describe the
transition between the vacuum energy era and the $\alpha$-era in the early
universe by a single equation of state of the form
\begin{equation}
\label{early1}
p=\alpha\rho c^2-(\alpha+1)\frac{\rho^2}{\rho_P}c^2,
\end{equation}
where $\rho_P$  is the Planck density. This equation of state corresponds to a
generalized polytropic equation of state (\ref{intro1}) with $n=+1$ and
$k=-(\alpha+1)/\rho_P$. For $\rho\ll\rho_P$, we recover the linear
equation of
state $p\sim \alpha\rho c^2$. For $\rho\rightarrow \rho_P$, we get
$p\rightarrow
 -\rho_P c^2$ corresponding to the vacuum energy. The relation of
Equation~(\ref{newearly2a}) between the density and the scale factor becomes
\begin{equation}
\label{early2}
\rho=\frac{\rho_P}{1+(a/a_1)^{3(1+\alpha)}}.
\end{equation}
The characteristic scale $a_1$ marks the transition between the vacuum energy
era and the $\alpha$-era. The~equation of state (\ref{early1})
interpolates
smoothly between the vacuum energy era ($p=-\rho c^2$, $\rho=\rho_P$) and
the
$\alpha$-era ($p=\alpha\rho c^2$,
$\rho_{\alpha}=\rho_P/(a/a_1)^{3(1+\alpha)}$). It provides therefore a
``unified'' description of the vacuum energy (Planck) era and
$\alpha$-era in the early universe. This amounts to summing the {\it inverse} of
the densities of these two phases. Indeed, Equation~(\ref{early2}) can be rewritten
as
\begin{equation}
\label{early8vc}
\frac{1}{\rho}=\frac{1}{\rho_{P}}+\frac{1}{\rho_{\alpha}}.
\end{equation}
At $a=a_1$ we have $\rho_{\alpha}=\rho_P$ so that $\rho_1=\rho_P/2$.
Writing $\rho_{\alpha}=\rho_{\alpha,0}/(a/a_0)^{3(1+\alpha)}$, where
$a_0$ is
the present value of the scale factor and $\rho_{\alpha,0}$ is the present
density of the $\alpha$-fluid, and using the asymptotic expression
$\rho_{\alpha}\sim\rho_P/(a/a_1)^{3(1+\alpha)}$, we find that the
transition
scale factor
$a_1$ is determined by the relation $\rho_P
a_1^{3(1+\alpha)}=\rho_{\alpha,0}a_0^{3(1+\alpha)}$.

The equation of state parameter $w=p/\rho c^2$ and the deceleration parameter
$q$ are given by
\begin{equation}
\label{early3a}
w=-(\alpha+1)\frac{\rho}{\rho_P}+\alpha, \qquad
q=\frac{1+3\alpha}{2}-\frac{3}{2}(\alpha+1)\frac{\rho}{\rho_P}.
\end{equation}
The velocity of sound $c_s^2=p'(\rho)$ is given by
\begin{equation}
\label{vs1}
\frac{c_s^2}{c^2}=-2(\alpha+1)\frac{\rho}{\rho_P}+\alpha.
\end{equation}
As the universe expands from $a=0$ to $a=+\infty$, the density decreases from
$\rho_P$ to $0$, the equation of state parameter $w$ increases from $-1$ to
$\alpha$, the deceleration parameter $q$ increases from $-1$ to
$(1+3\alpha)/2$,
and the ratio $(c_s/c)^2$ increases from $-\alpha-2$ to $\alpha$ (see
{Figures} \ref{taLINLIN} and \ref{phasetransition}
 of \cite{chavanis1}).

\subsection{The Vacuum Energy Era: Early Inflation}

When $a\ll a_1$, the density tends to a maximum value
\begin{equation}
\label{early5}
\rho=\rho_{max}=\rho_P
\end{equation}
and the pressure tends to $p=-\rho_P c^2$. The Planck density
$\rho_P=5.16\times 10^{99}{\rm g}/{\rm m}^3$ (vacuum energy) represents
a
fundamental upper bound for the density. A constant value of the
density $\rho\simeq \rho_P$ gives rise to a phase of early
inflation. From the Friedmann equation (\ref{b9}), we find that the
Hubble parameter is constant, $H=(8\pi/3)^{1/2}~t_P^{-1}$, where we have
introduced
the Planck time $t_P=1/(G\rho_P)^{1/2}=(\hbar
G/c^5)^{1/2}=5.39\times
10^{-44}~{\rm s}$. Numerically, $H=5.37\times 10^{43}\, {\rm
s}^{-1}$. Therefore, for $a\ll a_1$, the scale factor increases exponentially
rapidly
with time as
\begin{equation}
\label{early6}
a(t)\sim l_P e^{({8\pi}/{3})^{1/2}t/t_P}.
\end{equation}
The timescale of the exponential growth is the Planck time $t_P=5.39\times
10^{-44}~{\rm s}$. We have defined the ``original'' time $t=0$ such that $a(0)$
is equal to the Planck length $l_P=c t_P=(G\hbar/c^3)^{1/2}=1.62\times
10^{-35}\,
{\rm m}$. Mathematically speaking, the universe exists at any time in the past
($a\rightarrow 0$ and $\rho\rightarrow \rho_P$ for $t\rightarrow -\infty$), so
there is no primordial singularity (Big Bang). However,  when $a\rightarrow 0$,
we cannot ignore the quantum fluctuations associated with the spacetime. In that
case, we cannot use the classical Einstein equations anymore and a theory of
quantum gravity is required. It is not known whether quantum gravity will
remove, or not, the primordial singularity. Therefore, we cannot extrapolate the
solution in Equation~(\ref{early6}) to the infinite past. However, this solution
may provide
a {\it semi-classical} description of the phase of early inflation when
$a\ge l_P$.

\subsection{The $\alpha$-Era}
\label{sec_rad}

When $a\gg a_1$, we recover the equation
\begin{equation}
\label{isa1}
\rho\sim\frac{\rho_P}{(a/a_1)^{3(1+\alpha)}}
\end{equation}
corresponding to the pure linear equation of state $p=\alpha\rho c^2$. When
$a\gg a_1$, the Friedmann equation~(\ref{b9})~yields
\begin{equation}
\label{isa2}
\frac{a}{a_1}\sim \left\lbrack \frac{3}{2}(\alpha+1)\left
(\frac{8\pi}{3}\right
)^{1/2}\frac{t}{t_P}\right\rbrack^{2/[3(\alpha+1)]}.
\end{equation}
We then have
\begin{equation}
\label{isa3}
\frac{\rho}{\rho_P}\sim \frac{1}{\left\lbrack \frac{3}{2}(\alpha+1)\left
(\frac{8\pi}{3}\right )^{1/2}\frac{t}{t_P}\right\rbrack^{2}}.
\end{equation}
During the $\alpha$-era, the scale factor increases algebraically as
$t^{2/[3(\alpha+1)]}$ and the density decreases algebraically as $t^{-2}$.

\subsection{The General Solution}
\label{sec_gen}

For the equation of state (\ref{early1}), the density is related
to the scale factor by Equation~(\ref{early2}). It is possible to solve the Friedmann
equation (\ref{b9}) with the density-radius relation of  Equation~(\ref{early2})
analytically. Introducing $R=a/a_1$, we obtain
\begin{eqnarray}
\label{new1}
\int \sqrt{1+R^{3(1+\alpha)}}\, \frac{dR}{R}=\left (\frac{8\pi}{3}\right
)^{1/2}t/t_P
\end{eqnarray}
which can be integrated into \cite{chavanis1}:
\begin{eqnarray}
\label{early8}
\sqrt{(a/a_1)^{3(\alpha+1)}+1}-\ln \left
(\frac{1+\sqrt{(a/a_1)^{3(\alpha+1)}+1}}{(a/a_1)^{3(\alpha+1)/2}}\right )
=\frac{3}{2}(\alpha+1)\left (\frac{8\pi}{3}\right )^{1/2} \frac{t}{t_P}+C,
\end{eqnarray}
where $C$ is a constant of integration determined such that $a=l_P$ at
$t=0$. Setting $\epsilon=l_P/a_1$, we get
\begin{eqnarray}
\label{gp3}
C(\epsilon)=\sqrt{\epsilon^{3(\alpha+1)}+1}-\ln \left
(\frac{1+\sqrt{\epsilon^{3(\alpha+1)}+1}}{\epsilon^{3(\alpha+1)/2}}\right
).
\end{eqnarray}
For $t\rightarrow +\infty$, Equation~(\ref{early8}) reduces to
Equation~(\ref{isa2}). For
$t\rightarrow -\infty$, we have the exact \mbox{asymptotic result}
\begin{equation}
\label{gp3b}
a(t)\sim a_1 e^{({8\pi}/{3})^{1/2}t/t_P+D}
\end{equation}
with $D=2(C+\ln 2-1)/[3(1+\alpha)]$. In general $\epsilon\ll 1$ (for the
radiation $\alpha=1/3$, we find \mbox{$\epsilon=6.20\times 10^{-30}$
\cite{chavanis1})}, so that $C$
and $D$ can be approximated by $C\simeq 1-\ln 2
+(3/2)(\alpha+1)\ln\epsilon$ and $D\simeq \ln\epsilon$. With this
approximation,
Equation~(\ref{gp3b}) returns Equation~(\ref{early6}). The time $t_1$ marking the end of
the inflation is obtained by substituting  $a=a_1$ in Equation~(\ref{early8}).
According to Equation~(\ref{early3a}),
the universe is accelerating when $a<a_c$ (\emph{i.e.},  $\rho>\rho_c$) and
decelerating when $a>a_c$ (\emph{i.e.},  $\rho<\rho_c$) where
$a_c/a_1=[2/(1+3\alpha)]^{1/[
3(\alpha+1)]}$
and $\rho_c/\rho_P=(1+3\alpha)/[3(\alpha+1)]$. The time $t_c$ at which the
universe starts
decelerating is obtained by substituting $a=a_c$ in Equation~(\ref{early8}).
This corresponds to the time at which the curve $a(t)$ presents an
inflexion point.
For $\alpha=1/3$ (radiation) this inflexion point $a_c$ coincides
with $a_1$ so it also marks the end of the inflation ($t_c=t_1$). For
$\alpha\neq 1/3$ the two points differ.

\subsection{The Pressure}

The pressure is given by Equation~(\ref{early1}). Using Equation~(\ref{early2}), we get
\begin{equation}
\label{press1}
p=\frac{\alpha (a/a_1)^{3(\alpha+1)}-1}{\left\lbrack
(a/a_1)^{3(\alpha+1)}+1\right\rbrack^2}\rho_P c^2.
\end{equation}
The pressure starts from $p=-\rho_P c^2=-4.64\times 10^{116}  \, {\rm
g}/{\rm
m}\,
{\rm s}^2$ at $t=-\infty$, remains approximately constant during the inflation,
increases at the end of the inflation,
becomes  positive,  reaches a maximum value $p_e$, and decreases algebraically
during the $\alpha$-era. At $t=t_i=0$, $p_i\simeq -\rho_P c^2$. The point at
which the pressure vanishes ($w=0$) corresponds to
$\rho_w/\rho_P=\alpha/(\alpha+1)$ and
$a_w/a_1=(1/\alpha)^{1/[3(\alpha+1)]}$. On~the other hand, the pressure
reaches
its maximum ($c_s=0$) when $\rho_e/\rho_P=\alpha/[2(\alpha+1)]$ and
$a_e/a_1=[(\alpha+2)/\alpha]^{1/[3(\alpha+1)]}$ [Note 6: The
velocity of sound is imaginary ($c_s^2<0$) when $a<a_e$ and real  ($c_s^2>0$)
when $a>a_e$. It is always less than the speed of light.]. The maximum pressure
is
$p_{e}/(\rho_P c^2)=\alpha^2/[4(\alpha+1)]$. At the transition point
$t=t_1$, we
have $p_1/(\rho_P
c^2)=-(1-\alpha)/4$. At the deceleration point $t=t_c$, we have $p_c/(\rho_P
c^2)=-(1+3\alpha)/[9(1+\alpha)]$.

\subsection{The Evolution of the Early Universe}

\begin{figure}[H]
\begin{center}
\includegraphics[clip,scale=0.43]{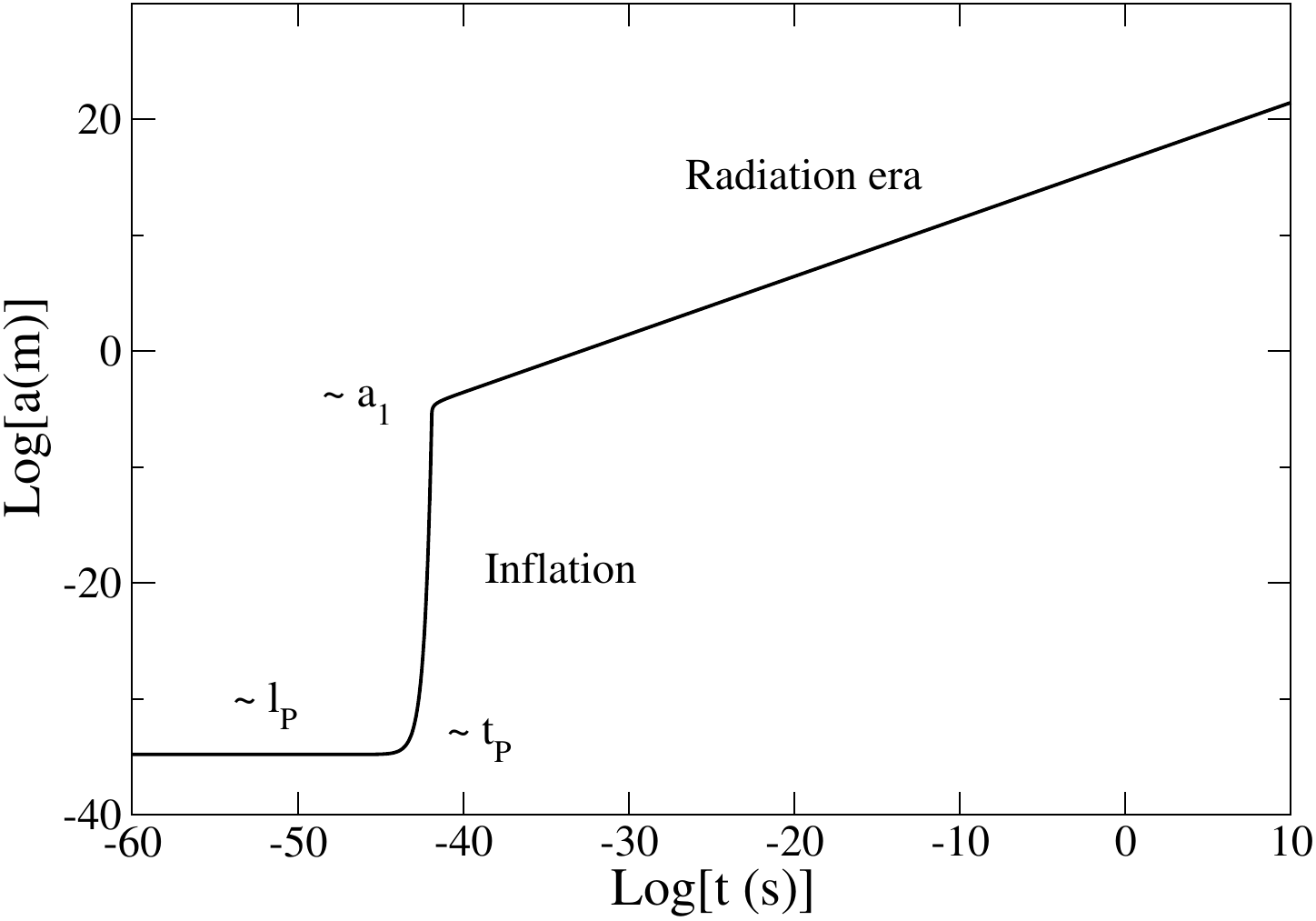}
\caption{Evolution of the scale factor $a$ with the time $t$ in logarithmic
scales. This~figure clearly shows the phase of inflation connecting the vacuum
energy era to the $\alpha$-era (representing here the radiation era). During the
inflation, the scale factor increases
exponentially rapidly on a timescale of the order of the Planck time $t_P$ (for
$\alpha=1/3$
it increases by $29$ orders of magnitude in less than $10^{-42}$ s).  In the
$\alpha$-era, the scale factor increases algebraically as
$t^{2/[3(\alpha+1)]}$.}
\label{taLOGLOG}
\end{center}
\end{figure}

\begin{figure}[H]
\begin{center}
\includegraphics[clip,scale=0.43]{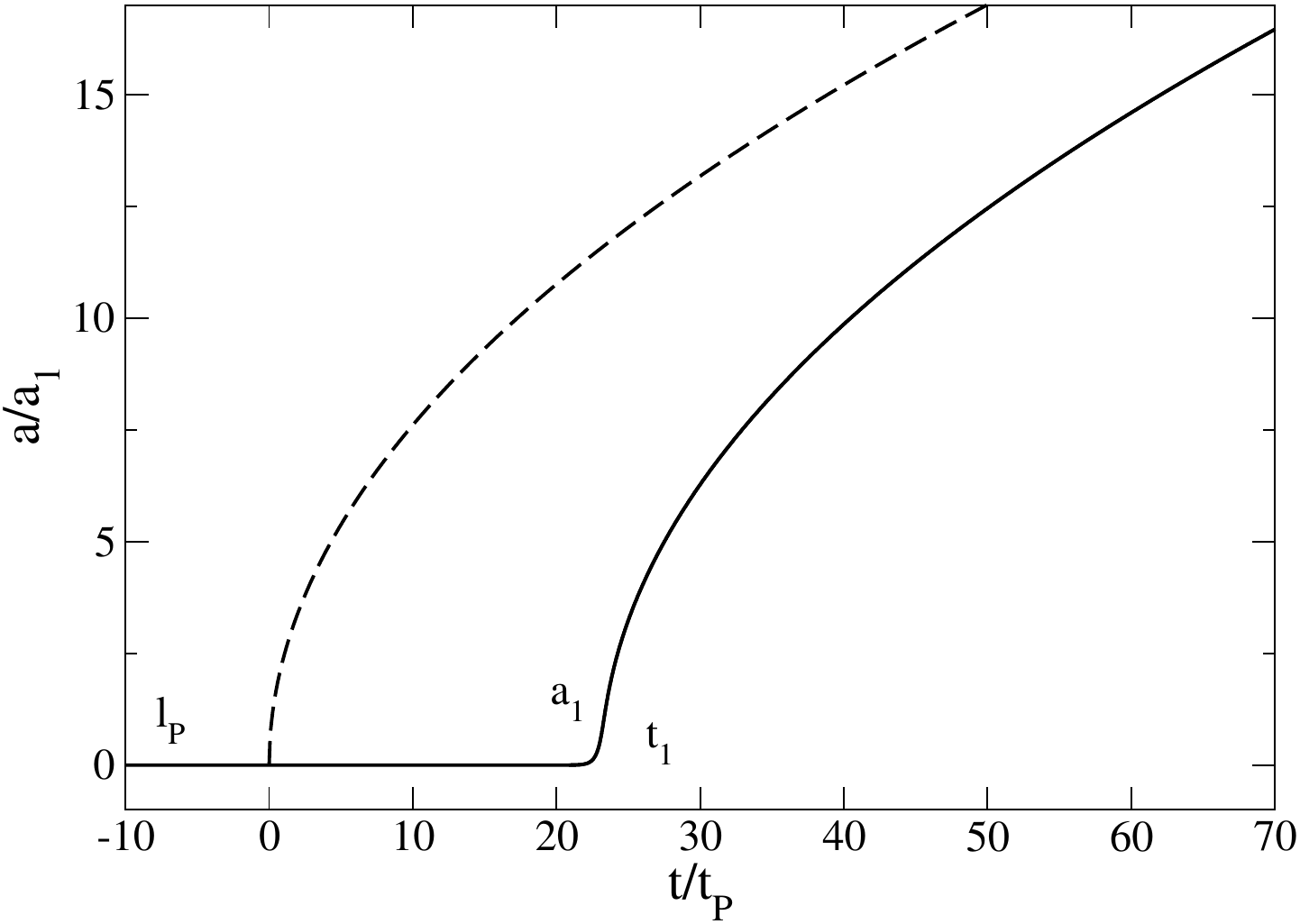}
\caption{Evolution of the scale factor $a$ with the time $t$ in linear
scales. The dashed line corresponds to a pure linear equation of state
$p=\alpha\rho c^2$
 leading to a finite time singularity at $t=0$ where $a=0$
(Big Bang). When quantum mechanics is taken into account (as in our
semi-classical model), the initial singularity is smoothed-out
and the scale factor at $t=0$ is equal to the Planck length
$l_P=1.62\times 10^{-35}\, {\rm m}$. This is similar to a second order
phase transition where the Planck constant plays the role of finite
size effects (see Section \ref{sec_analogy} for a~development of this
analogy).}
\label{taLINLIN}
\end{center}
\end{figure}

\begin{figure}[H]
\begin{center}
\includegraphics[clip,scale=0.45]{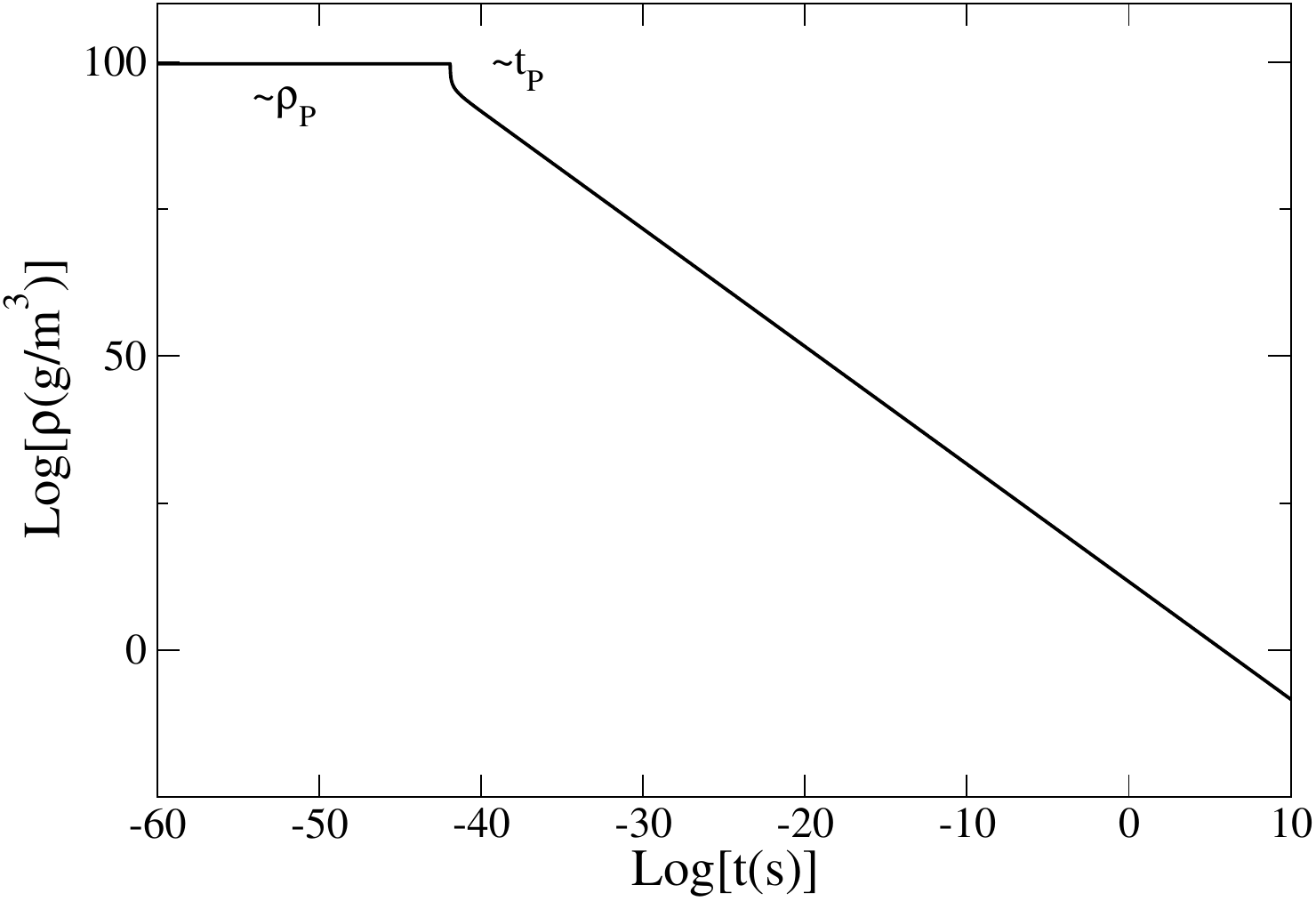}
\caption{Evolution of the density $\rho$ with the time $t$ in logarithmic
scales. During the inflation, the density remains approximately constant with
the Planck value $\rho_{max}=\rho_P$ which represents an upper bound. In the
$\alpha$-era, $\rho$ decreases algebraically as $t^{-2}$.}
\label{trhoLOGLOG}
\end{center}
\end{figure}

\begin{figure}[H]
\begin{center}
\includegraphics[clip,scale=0.45]{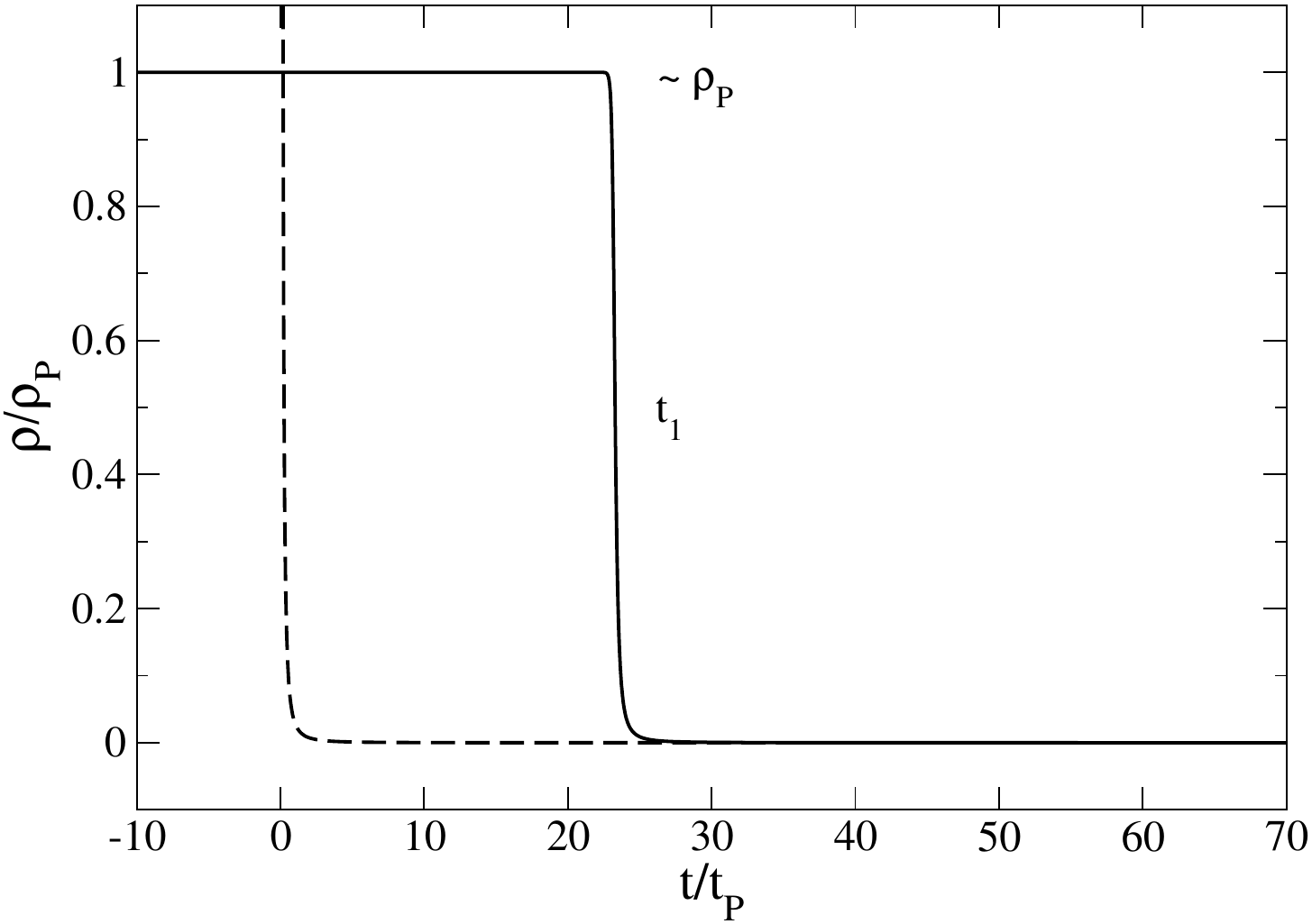}
\caption{Evolution of the density $\rho$ with the time $t$ in linear scales. The
dashed line corresponds to a pure linear equation of state $p=\alpha\rho c^2$
 leading to a finite time singularity at $t=0$. Quantum mechanics limits the
rise of the density to the Planck value $\rho_P=5.16\times 10^{99}\, {\rm
g}/{\rm
m}^3$.}
\label{trhoLINLIN}
\end{center}
\end{figure}

\begin{figure}[H]
\begin{center}
\includegraphics[clip,scale=0.45]{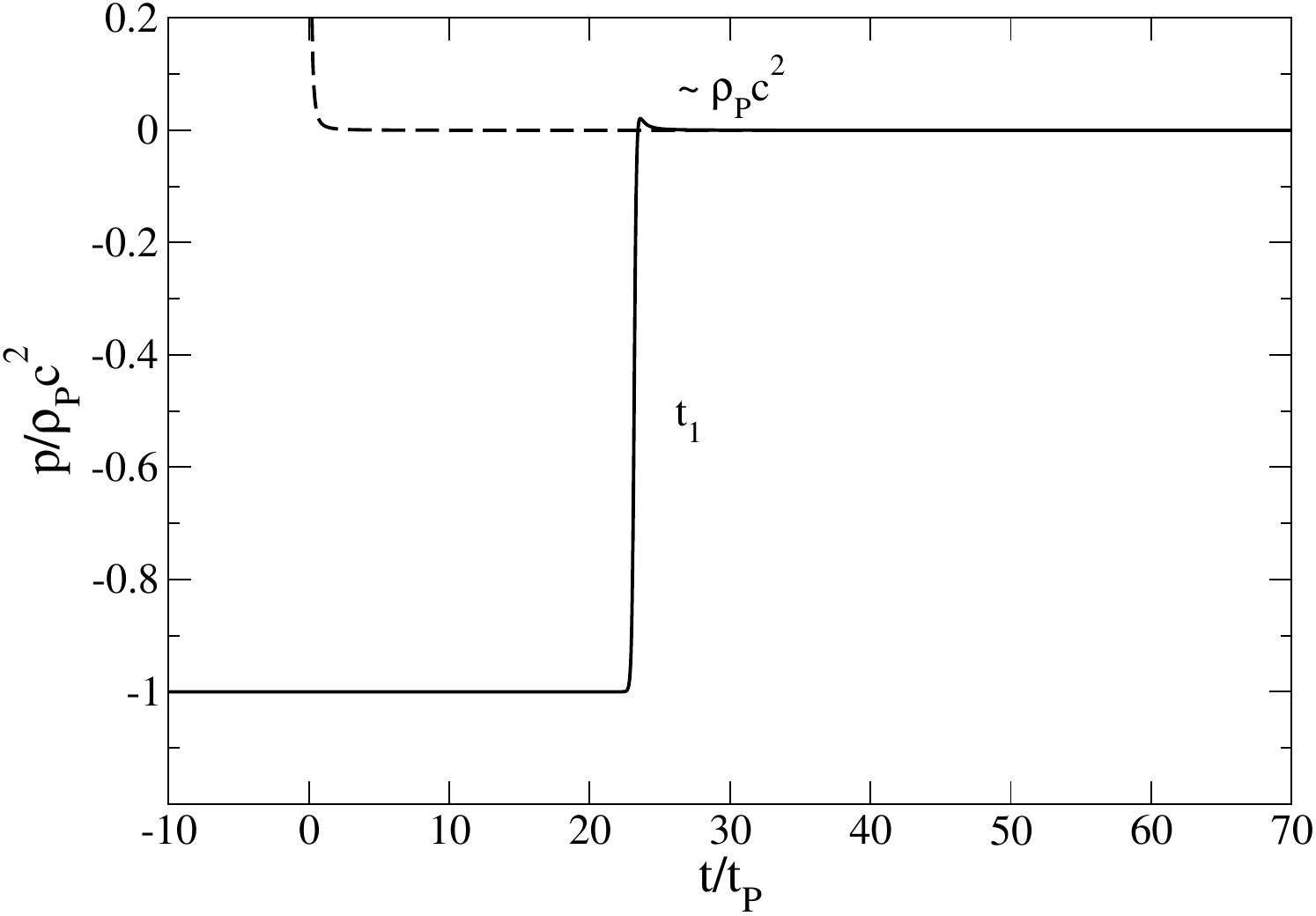}
\caption{Evolution of the pressure $p$ with the time $t$ in linear scales. The
dashed line corresponds to a pure linear equation of state $p=\alpha\rho c^2$
 leading to a finite time singularity at $t=0$. Quantum mechanics limits the
rise of the pressure to  a~maximum value \mbox{$p_{e}/(\rho_P
c^2)=\alpha^2/[4(\alpha+1)]$}. In the vacuum energy era, the pressure
decreases
exponentially rapidly as we go backward in time, becomes negative, and tends to
\mbox{$-\rho_P c^2=-4.64\times
10^{116}  \, {\rm g}/{\rm m}\, {\rm s}^2$ for $t\rightarrow -\infty$}.}
\label{tpressLINLIN}
\end{center}
\end{figure}

In our model, the universe ``starts'' at $t=-\infty$ with a
vanishing
radius $a=0$, a finite density $\rho=\rho_P=5.16\times 10^{99}\, {\rm
g/m^3}$, and a finite pressure $p=-\rho_P c^2=-4.64\times 10^{116} \, {\rm
g}/{\rm m}\, {\rm s}^2$. The
universe exists at any time in the past and does not present any
singularity. For $t<0$, the radius of the universe is less than the
Planck length $l_P=1.62\times 10^{-35}\, {\rm m}$. In the Planck era,
quantum gravity should be taken into account so our semi-classical
approach is probably not valid in the infinite past. At $t=t_i=0$, the
radius of the universe is equal to the Planck length $a_i=l_P=1.62\times
10^{-35}\, {\rm m}$. The corresponding density and
pressure are $\rho_i\simeq \rho_P=5.16\times 10^{99}\, {\rm g}/{\rm
m}^3$ and
$p_i\simeq -\rho_P c^2=-4.64\times 10^{116} \, {\rm g}/{\rm m}\,
{\rm
s}^2$. We~note that quantum mechanics regularizes the finite time
singularity present in the standard Big Bang theory. This~is similar
to finite size effects in second order phase transitions (see
Section \ref{sec_analogy}). The Big Bang theory is recovered for
$\hbar=0$.  The universe first undergoes a phase of inflation
during which its radius increases exponentially rapidly with time
while its density and pressure remain approximately constant. The~inflation
``starts'' at $t_i=0$ and ends at $t_1$. The timescale of the inflation
corresponds to the Planck time $t_P=5.39\times 10^{-44}~{\rm s}$ (for the
radiation
$\alpha=1/3$, we find $t_1=23.3\, t_P=1.25\times 10^{-42}\, {\rm
s}$~\cite{chavanis1}). During~this very short lapse of time, the scale factor
grows
from $a_i=l_P=1.62\times 10^{-35}\, {\rm m}$ to $a_1$ (for $\alpha=1/3$, we
find
$a_1=2.61\times 10^{-6}\, {\rm m}$). By contrast, the density and the pressure
do
not change significatively: they go from $\rho_i\simeq \rho_P$ and $p_i\simeq
-\rho_P c^2$ to $\rho_1=\rho_P/2$ and $p_1=-[(1-\alpha)/4]\rho_P c^2$
(for
$\alpha=1/3$, we find $\rho_1=0.5\, \rho_P=2.58\times 10^{99}\, {\rm
g/m^3}$ and
$p_1=-0.167\rho_P c^2=-7.73\times 10^{115} \, {\rm g}/{\rm m}\, {\rm
s}^2$). The
pressure passes from negative values to positive values at $t_w$ (for
$\alpha=1/3$, we find $t_w=23.4t_P=1.26\times 10^{-42}\, {\rm s}$,
$\rho_w=0.25\rho_P=1.29\times 10^{99}\, {\rm g}/{\rm m}^3$,
$a_w=1.32\,
a_1=3.43\times
10^{-6}\, {\rm m}$). After the inflation, the universe enters in the
$\alpha$-era. The radius increases algebraically as
$a\propto t^{2/[3(\alpha+1)]}$ while the density decreases algebraically
as $\rho\propto t^{-2}$. The pressure achieves its maximum value $p_e$ at $t_e$
(for $\alpha=1/3$, we find $p_e=2.08\times 10^{-2}\rho_P c^2=9.66\times
10^{114}
\, {\rm
g}/{\rm m}\, {\rm s}^2$, $t_{e}=23.6\, t_P=1.27\times 10^{-42}\, {\rm
s}$,
$\rho_{e}=0.125\, \rho_P=6.44\times 10^{98}\, {\rm g}/{\rm m}^3$,
$a_{e}=1.63\,
a_1=4.24\times 10^{-6}\, {\rm m}$). During the inflation, the
universe is accelerating and during the $\alpha$-era it is
decelerating (if $\alpha>-1/3$). The transition (marked by an inflexion point)
takes
place at a time $t_c$ (for $\alpha=1/3$ it coincides
with the end of the inflation $t_1$). The evolution of
the scale factor and density as a function of time are
represented in Figures
\ref{taLOGLOG}--\ref{tpressLINLIN} in  logarithmic and linear scales (the
figures
correspond to the radiation $\alpha=1/3$).

{\it Remark:} If we consider the transition between the
inflation era and the radiation era ($\alpha=1/3$), we find that the
temperature is
given by a generalized Stefan-Boltzmann law (see {Equation~(84a)} in \cite{chavanis1}).
The~evolution of
the temperature is discussed in detail in References
\cite{chavanis1,chavanisAIP,jgrav}. In
our model, the
temperature is initially very low, increases exponentially rapidly during the
inflation up to a fraction ($\sim 0.523$) of the Planck temperature
$T_P=1.42\times 10^{32}\, {\rm K}$ which
is of the order of the Grand Unified Theories (GUT) scale, then decreases
algebraically during the radiation era.  On the other hand, our model generates
a value of the entropy as large as $S/k_B=5.04\times 10^{87}$
\cite{chavanis1}. This is very different from the standard
inflationary scenario \cite{guth1,guth2,guth3,linde}. In that scenario, the
universe is
radiation dominated up to $t_i=10^{-35}\, {\rm s}$ and expands exponentially
rapidly  by a factor $10^{30}$ in the interval $t_i<t<t_f$ with $t_f=10^{-33}\,
{\rm s}$. For $t>t_f$, the evolution is again radiation dominated. At $t=t_i$,
the temperature is about $10^{27}\, {\rm K}$ (this corresponds to the epoch at
which most GUTs have a significant influence on the evolution of the universe).
During~the exponential inflation, the temperature drops drastically and one must
advocate a phase of re-heating  by various high energy processes (not very well
understood) to restore the initial temperature.

\vspace{-24pt}
\subsection{Analogy with Phase Transitions}
\label{sec_analogy}

The standard Big Bang theory is a classical theory in which quantum effects are
neglected. In that case, it exhibits a finite time singularity: the radius of
the universe is equal to zero at $t=0$ while its density is infinite. For $t<0$,
the solution is not defined and we may take $a=0$. For $t>0$ the radius of the
universe increases as $a\propto t^{2/[3(\alpha+1)]}$.  This is similar to a
second order phase transition if we view the time $t$ as the control parameter
({e.g.,} the temperature $T$) and the scale factor $a$ as the order parameter
({e.g.,} the magnetization $M$). For $\alpha=1/3$ (radiation), it is amusing to
note that the exponent in
$a\propto t^{1/2}$ is the same as in mean field theories of second order phase
transitions ({\emph{i.e.},}
$M\sim (T_c-T)^{1/2}$) but this is essentially a coincidence.

\begin{figure}[H]
\begin{center}
\includegraphics[clip,scale=0.45]{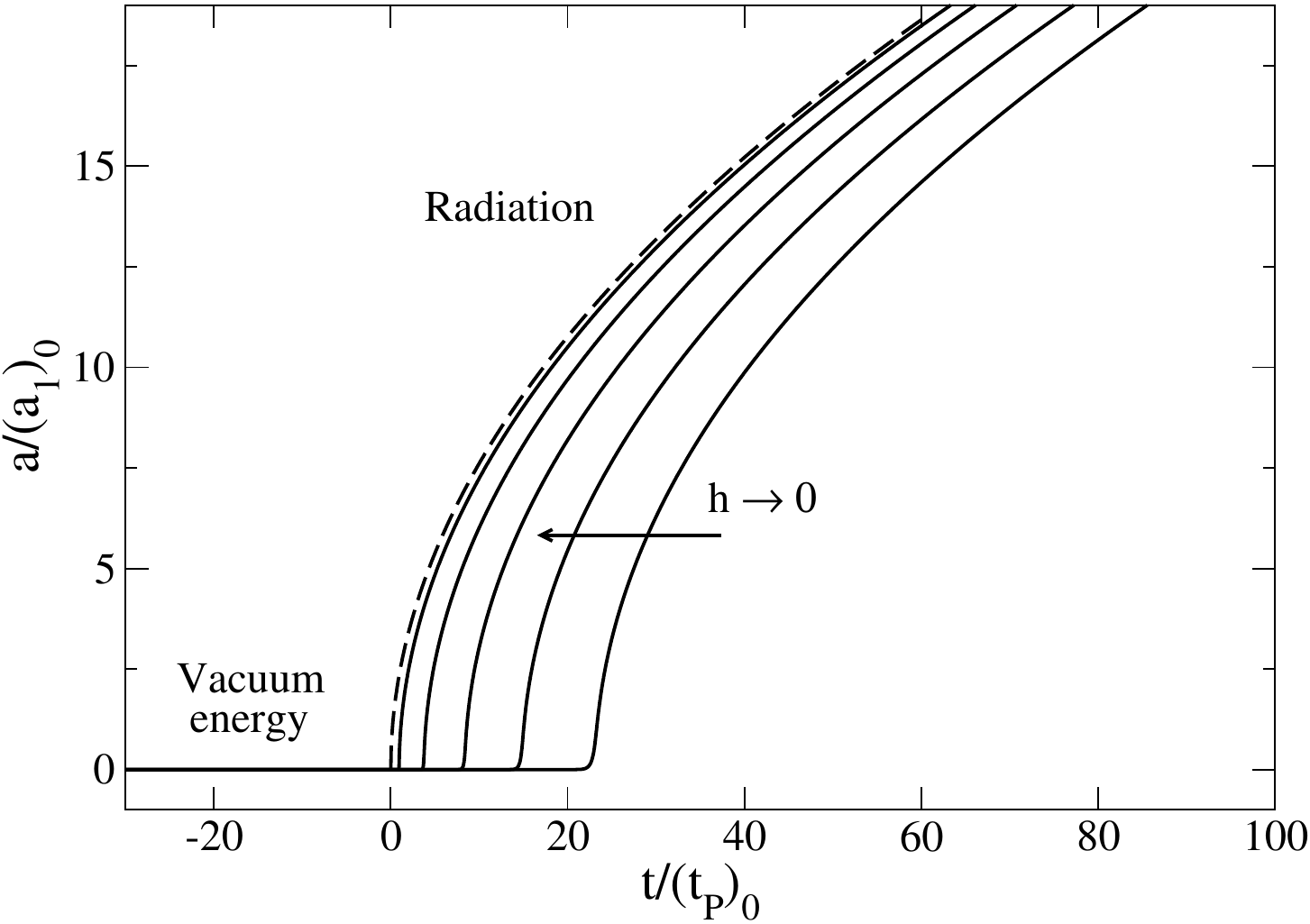}
\caption{Effect of quantum mechanics (finite value of the Planck constant) on
the regularization of the singular Big Bang solution ($\hbar=0$, dashed line) in
our semi-classical model (see \cite{chavanis1} for details about the
construction of this curve). The singularity at $t=0$ is replaced by an
inflationary expansion from the vacuum energy era to the $\alpha$-era. We can
draw an analogy with second order phase transitions where the Planck constant
plays the role of finite size effects.}
\label{phasetransition}
\end{center}
\end{figure}

When quantum mechanics is taken into account, as in our
semi-classical model, the singularity at $t=0$ disappears and the curves $a(t)$
and $\rho(t)$ are regularized. In particular, we find that $a=l_P>0$ at $t=0$,
instead of $a=0$, due to the finite value of $\hbar$. This is similar to the
regularization due to finite size effects ({e.g.,} the system size $L$ or the
number of particles $N$) in ordinary phase transitions. In~this sense, the
classical limit $\hbar\rightarrow 0$ is similar to the thermodynamic limit
($L\rightarrow +\infty$ or $N\rightarrow +\infty$) in ordinary phase
transitions. The convergence of our semi-classical solution towards the
classical
Big Bang solution when
$\hbar\rightarrow 0$ is shown in Figure \ref{phasetransition} for  $\alpha=1/3$
(radiation).

\subsection{Scalar field theory}
\label{sec_sfe}

The phase of inflation in the very early universe is usually described by a
scalar field called~inflaton~\cite{linde}. A canonical scalar
field minimally coupled to
gravity evolves according to the Klein-Gordon equation
\begin{equation}
\label{early13b}
\ddot \phi+3H\dot\phi+\frac{dV}{d\phi}=0,
\end{equation}
where $V(\phi)$ is the potential of the scalar field. The scalar field tends to
run down the potential towards lower energies. The density and the pressure of
the universe are related to the scalar field by
\begin{equation}
\label{early14}
\rho c^2=\frac{1}{2}\dot\phi^2+V(\phi),\qquad
p=\frac{1}{2}\dot\phi^2-V(\phi).
\end{equation}

Using standard techniques \cite{cst,bamba}, we find that the
inflaton potential corresponding to
the equation of state (\ref{early1}) is (see {Section 8.1} of
\cite{chavanis2}):
\begin{equation}
\label{early15b}
V(\psi)=\frac{1}{2}\rho_P c^2
\frac{(1-\alpha)\cosh^2\psi+\alpha+1}{\cosh^4\psi}
\qquad (\psi\ge 0),
\end{equation}
where we have defined
\begin{equation}
\label{early15bv}
\psi=\left (\frac{8\pi G}{3c^2}\right
)^{1/2}\frac{3\sqrt{\alpha+1}}{2}\phi.
\end{equation}
For $\psi\rightarrow 0$,
\begin{equation}
\frac{V(\psi)}{\rho_P c^2}\simeq
1-\frac{3+\alpha}{2}\psi^2+\frac{9+5\alpha}{6}\psi^4+...
\label{inflation19}
\end{equation}
which is consistent with the symmetry breaking scalar field potential used to
describe inflation. \mbox{For~$\psi\rightarrow +\infty$},
\begin{equation}
\frac{V(\psi)}{\rho_P c^2}\sim 2(1-\alpha)e^{-2\psi}\qquad (\alpha\neq
1),
\label{inflation19b}
\end{equation}
\begin{equation}
\frac{V(\psi)}{\rho_P c^2}\sim 16 e^{-4\psi}\qquad (\alpha=1).
\label{inflation19bb}
\end{equation}

\begin{figure}[H]
\centering
\includegraphics[clip,scale=0.45]{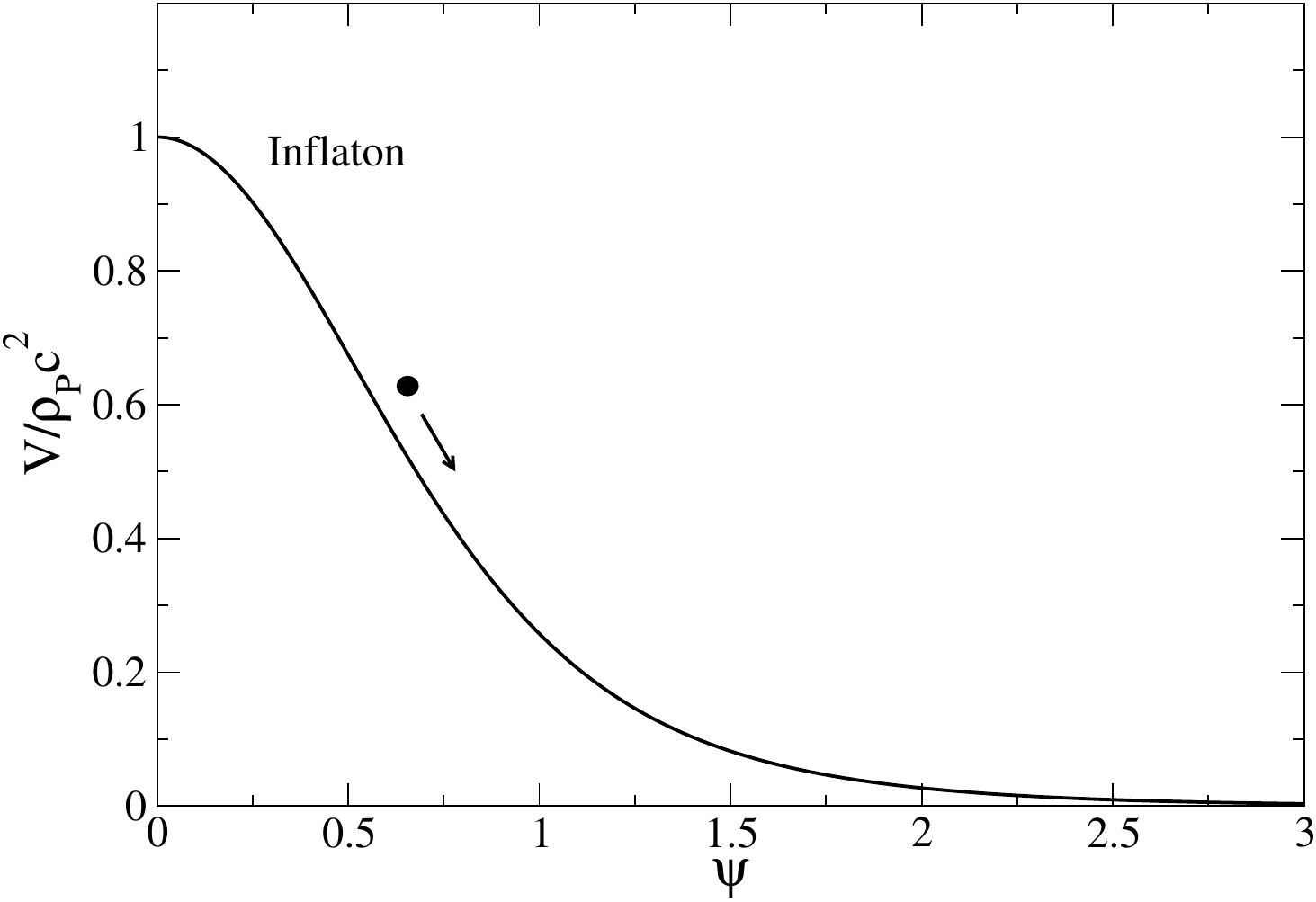}
\caption{Potential of the scalar field (inflaton) in the early universe. The
field tends to
run down the potential.}
\label{Vpsiearly}
\end{figure}

\begin{figure}[H]
\centering
\includegraphics[clip,scale=0.45]{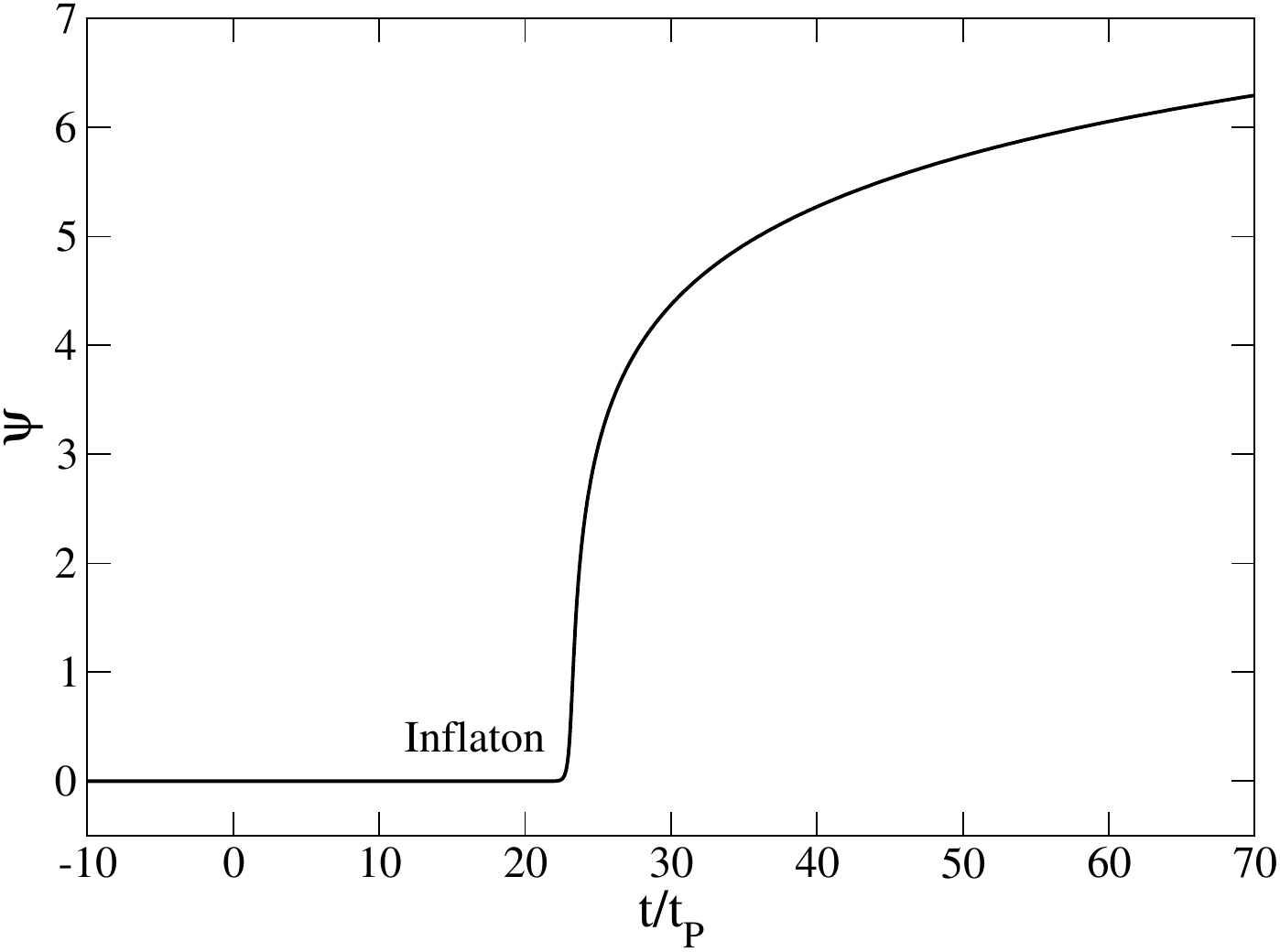}
\caption{Evolution of the scalar field (inflaton) as a function of time in the
early
universe. It displays a phase of early inflation before entering in the
$\alpha$-era.}
\label{tpsiearly}
\end{figure}

We can also show \cite{chavanis2}  that the relation between the scalar field
and the scale factor is
\begin{equation}
\label{early15}
\left (\frac{a}{a_1}\right )^{3(\alpha+1)/2}=\sinh\psi.
\end{equation}
The end of the inflation, and the beginning of the $\alpha$-era, corresponds to
$a=a_1$, hence to $\psi=\psi_1=\sinh^{-1}(1)=\ln(1+\sqrt{2})=0.881374$.
Combining Equations (\ref{early1}),
(\ref{early2}) and (\ref{early15}), we find that the energy density and the
pressure of
the universe are related to the scalar field by
\begin{eqnarray}
\rho=\frac{\rho_P}{\cosh^{2}\psi}, \qquad p=\frac{\rho_P
c^2}{\cosh^2\psi}\left\lbrack\alpha-(\alpha-1)\frac{1}{\cosh^2\psi}
\right\rbrack.
\label{inflation16}
\end{eqnarray}
Using Equation~(\ref{early15}), and the results of the previous
sections, we can
obtain the temporal evolution of the scalar field.
In the vacuum energy era ($t\rightarrow -\infty$), using Equation~(\ref{early6}), we
get
\begin{equation}
\label{early16a}
\psi\sim \left (\frac{l_P}{a_1}\right
)^{3(\alpha+1)/2}e^{\frac{3}{2}(\alpha+1)\left (\frac{8\pi}{3}\right
)^{1/2}t/t_{P}}\rightarrow 0.
\end{equation}
In the $\alpha$-era ($t\rightarrow +\infty$), using Equation~(\ref{isa2}), we
get
\begin{equation}
\label{early16b}
\psi\simeq \ln\left (\frac{t}{t_{P}}\right )+\frac{1}{2}\ln\left
(\frac{8\pi}{3}\right )+\ln\left\lbrack 3(\alpha+1)\right\rbrack\rightarrow
+\infty.
\end{equation}
More generally, using Equation~(\ref{early8}), the evolution of the scalar field
$\psi(t)$ in the early universe is given by
\begin{eqnarray}
\label{early17}
\cosh\psi-\ln \left (\frac{1+\cosh\psi}{\sinh\psi}\right
)=\frac{3}{2}(\alpha+1)\left (\frac{8\pi}{3}\right )^{1/2}
\frac{t}{t_P}+C.
\end{eqnarray}
These results are illustrated in Figures \ref{Vpsiearly} and \ref{tpsiearly} for
$\alpha=1/3$ (radiation).

{\it Remark:} The canonical  scalar field potential of Equation
(\ref{early15b}) describing the
transition between the inflation (vacuum energy) era and the $\alpha$-era
involves hyperbolic functions. The inflation due to another scalar field
with a hyperbolic potential was investigated recently in \cite{basilakos} and
confronted to the Planck 2015 data set, giving promising results. We are
carrying out a similar study with the scalar field potential of Equation
(\ref{early15b}) to
test its performance against Planck inflationary parameters.

\vspace{-24pt}

\section{The Late Universe}
\label{sec_late}

In Reference \cite{chavanis2}, we have proposed to describe the  transition between
the $\alpha$-era and the dark energy era in the late universe by a single
equation of state of the form of Equation (\ref{intro1}) with $k<0$ and $n<0$.
It can be
written as
\begin{equation}
\label{newearly1bb}
p/c^2=\alpha\rho-|k|{\rho}^{1-1/|n_l|}.
\end{equation}
Assuming $w>-1$, the equation of continuity (\ref{b7}) can be
integrated into
\begin{equation}
\label{newearly2}
\rho=\rho_\Lambda \lbrack (a_2/a)^{3(1+\alpha)/|n_l|}+1\rbrack^{|n_l|},
\end{equation}
where $\rho_{\Lambda}=\lbrack |k|/(\alpha+1)\rbrack^{|n_l|}$ and $a_2$ is a
constant of integration. We see that $\rho_{\Lambda}$ corresponds to a lower
bound (minimum value) for the density reached for $a\rightarrow +\infty$. Since
this solution describes the late universe, it is natural to identify
$\rho_{\Lambda}$ with the cosmological density
$\rho_{\Lambda}={\Lambda}/{8\pi
G}=7.02\times 10^{-24}\, {\rm g}/{\rm
m}^3$. As a result, the equation of state (\ref{newearly1bb}) can be
rewritten
as
 \begin{equation}
\label{newearly1bis}
p/c^2=\alpha\rho-(\alpha+1)\rho \left
(\frac{\rho_{\Lambda}}{\rho}\right
)^{1/|n_l|}.
\end{equation}
For the sake of simplicity, and for definiteness, we shall select the index
$n_l=-1$. The general case $n_l<0$ has been treated in \cite{chavanis2} and
leads
to qualitatively similar results. Therefore, we propose to describe the
transition between the $\alpha$-era and the dark energy era in the late universe
by a single equation of state of the form
\begin{equation}
\label{late1}
p=\alpha\rho c^2-(\alpha+1)\rho_{\Lambda} c^2,
\end{equation}
where $\rho_{\Lambda}$ is the cosmological density. This equation of state
corresponds to a generalized polytropic  equation of state (\ref{intro1})  with
$n=-1$ and $k=-(\alpha+1)\rho_{\Lambda}$. It can be viewed as the
``symmetric''
version of the equation of state (\ref{early1}) in the early universe.
For
$\rho\gg\rho_{\Lambda}$, we recover the linear equation of state $p\sim
\alpha\rho c^2$. For $\rho\rightarrow \rho_{\Lambda}$, we get $p\simeq
-\rho_{\Lambda} c^2=-6.31\times 10^{-7}{\rm g/m s^2}$ corresponding to the
dark
energy. The relation in Equation~(\ref{newearly2}) between the density and the
scale factor
becomes
\begin{equation}
\label{late2}
\rho=\rho_{\Lambda}\left\lbrack \left (\frac{a_2}{a}\right )^{3(1+\alpha)}+
1\right\rbrack.
\end{equation}
The characteristic scale $a_2$ marks the transition between the $\alpha$-era and
the dark energy era. The~equation of
state (\ref{late1}) interpolates smoothly between the $\alpha$-era
($p=\alpha\rho c^2$ and
\mbox{$\rho_{\alpha}=\rho_{\Lambda}(a_2/a)^{3(\alpha+1)}$})
and the dark energy era ($p=-\rho c^2$ and $\rho=\rho_{\Lambda}$). It
provides
therefore a ``unified'' description of the $\alpha$-era and  dark energy
era (de Sitter) in the late universe. This amounts to summing the density of
these two
phases. Indeed, Equation~(\ref{late2}) can be rewritten as
\begin{equation}
\label{late16de}
{\rho}={\rho_{\alpha}}+{\rho_{\Lambda}}.
\end{equation}
At $a=a_2$ we have $\rho_{\alpha}=\rho_{\Lambda}$ so that
$\rho=2\rho_{\Lambda}$. Writing
$\rho_{\alpha}=\rho_{\alpha,0}/(a/a_0)^{3(1+\alpha)}$, where $a_0$ is
the
present
value of the scale factor and $\rho_{\alpha,0}$ is the present density of the
$\alpha$-fluid,
and using the asymptotic
expression $\rho_{\alpha}\sim\rho_{\Lambda}(a_2/a)^{3(\alpha+1)}$, we find
that the
transition scale factor
$a_2$ is
determined by the relation $\rho_{\Lambda}
a_2^{3(1+\alpha)}=\rho_{\alpha,0}a_0^{3(1+\alpha)}$. On the other hand,
comparing Equations~(\ref{early2}) and (\ref{late2}) in
the \mbox{$\alpha$-era $a_1\ll a\ll a_2$} (\emph{i.e.},
\mbox{$\rho_{\Lambda}\ll\rho\ll\rho_P$}) where
the two equations of state overlap, we get
$\rho_{\Lambda}a_2^{3(\alpha+1)}=\rho_{P}a_1^{3(\alpha+1)}$.

The equation of state parameter $w=p/\rho c^2$ and the deceleration parameter
$q$ are given by
\begin{equation}
\label{early3}
w=\alpha-(\alpha+1)\frac{\rho_{\Lambda}}{\rho}, \qquad
q=\frac{1+3\alpha}{2}-\frac{3}{2}(\alpha+1)\frac{\rho_{\Lambda}}{\rho}.
\end{equation}
The velocity of sound $c_s^2=p'(\rho)$ is given by
\begin{equation}
\label{early3qw}
\frac{c_s^2}{c^2}=\alpha.
\end{equation}
As the universe expands from $a=0$ to $a=+\infty$, the density decreases from
$+\infty$ to $\rho_{\Lambda}$, the equation of state parameter $w$ decreases
from $\alpha$ to $-1$, the deceleration parameter $q$ decreases from
$(1+3\alpha)/2$ to $-1$, and the ratio $(c_s/c)^2$ remains constant with the
value $\alpha$ (see Figures \ref{taLOGLOG} and \ref{tpressLINLIN} of
\cite{chavanis2}).

\subsection{The Dark Energy Era: Late Inflation}

When $a\gg a_2$, the density tends to a minimum value
\begin{equation}
\label{late3}
\rho=\rho_{min}=\rho_{\Lambda}
\end{equation}
and the pressure tends to $p=-\rho_{\Lambda}c^2$. The cosmological density
$\rho_{\Lambda}={\Lambda}/{8\pi G}=7.02\times 10^{-24}\, {\rm g}/{\rm
m}^3$ (dark energy) represents a
fundamental lower bound for the density. A constant value of the
density  $\rho\simeq \rho_{\Lambda}$ gives rise to a phase of late inflation
(accelerating expansion). It is convenient to define
a~cosmological time
$t_{\Lambda}=1/(G\rho_{\Lambda})^{1/2}=(8\pi/\Lambda)^{1/2}=1.46\times
10^{18}\, {\rm s}$ and a cosmological length $l_{\Lambda}=c
t_{\Lambda}=(8\pi c^2/\Lambda)^{1/2}=4.38\times 10^{26}\, {\rm m}$. These
are
the
counterparts of the Planck scales for the late universe (see
Appendix \ref{sec_further} of \cite{chavanis2}). From
the Friedmann
equation (\ref{b9}), we find that the Hubble parameter is constant
$H=(8\pi/3)^{1/2}t_{\Lambda}^{-1}$. Numerically, $H=1.98\times 10^{-18}\,
{\rm
s}^{-1}$. Therefore, the scale factor increases exponentially rapidly with time
as
\begin{equation}
\label{late4}
a(t)\propto e^{(8\pi/3)^{1/2}t/t_{\Lambda}}.
\end{equation}
This exponential growth corresponds to the de Sitter solution \cite{bt}. The
timescale of the exponential growth is the  cosmological time
$t_{\Lambda}=1.46\times 10^{18}~{\rm s}$. This solution exists at any time in
the
future ($a\rightarrow +\infty$ and $\rho\rightarrow \rho_{\Lambda}$ for
$t\rightarrow +\infty$), so there is no future
singularity [Note 7: This is not the case
of all cosmological models. In a  ``phantom
universe''
\cite{ghosts1,ghosts2,ghosts3,ghosts4,ghosts5,ghosts6,ghosts7,ghosts8,ghosts9,
ghosts10,ghosts11,ghosts12,ghosts13,ghosts14,ghosts15,ghosts16,ghosts17,
ghosts18}, violating
the null dominant energy condition ($w<-1$),  the density increases as the
universe expands. The models based on
phantom dark energy usually predict a future singularity in which the
scale factor, the energy density, and the pressure of the universe
become infinite in a finite time. This would lead to the death of the
universe in a singularity called Big Smash \cite{ghosts3}, Big
Rip, or Cosmic Doomsday \cite{ghosts2}. Contrary to the Big
Crunch, the universe is destroyed not by excessive contraction but
rather by excessive expansion. Every
gravitationally bound system ({e.g.}, the solar system, the Milky
Way, the local group, galaxy clusters) is dissociated before the
singularity \cite{ghosts2,ghosts8}, and the black holes gradually lose
their mass and finally vanish at the Big Rip time~
\cite{ghosts9,ghosts10,ghosts11}. This
scenario allows the explicit calculation of the rest of the lifetime
of the universe. Actually, as we approach the singularity, the energy
scale may grow up to the Planck scale, giving rise to a second quantum
gravity era. Eventually, quantum effects may moderate or even prevent
the singularity~\cite{ghosts12}. It has also been proposed that, before
reaching
the Big Rip singularity, the universe could be swallowed by a
wormhole~\cite{wormhole1}. This is due to the accretion of the
ever-increasing phantom energy by the wormhole. This
accretion induces an increase of the wormhole throat so rapid that the
size of the wormhole throat becomes infinite in a finite time preceding the Big
Rip. As a result, the wormhole can engulf the entire universe before it
reaches the Big Rip singularity. The wormhole would then act as a spacetime
tunnel (Einstein-Rosen bridge) allowing the universe to pass to another, more
gentle,
universe without future singularity. This could be a way to escape the
programmed death of the universe. This has been called the Big Trip.
 The possibility that we live in a
phantom universe is not excluded by the observations. Indeed,
observational data indicate that the equation of
state parameter $w$ lies in a narrow strip around $w=-1$, possibly
being below this value.  It is
important to stress, however, that the phantom models with $w<-1$ do not
necessarily
lead to finite-time future singularities, as discussed in \cite{astashenok}.
For~example, the scale
factor and the density may increase indefinitely and become infinite in
infinite time. This~has been called Little
Rip \cite{ghosts18}. In these models, the black
holes disappear in infinite time while the wormhole's throat still becomes
infinite
in a finite time. The phantom models without future singularity are
attractive from the physical viewpoint because the occurrence of a finite time
singularity may lead to some inconsistencies.].

\subsection{The $\alpha$-era}

When $a\ll a_2$, we recover the equation
\begin{equation}
\label{ph1}
\rho\sim \frac{\rho_{\Lambda}a_2^{3(1+\alpha)}}{a^{3(1+\alpha)}}
\end{equation}
corresponding to the pure linear equation of state $p=\alpha\rho c^2$. When
$a\ll a_2$, the Friedmann \mbox{equation (\ref{b9}) yields}
\begin{equation}
\label{ph2}
\frac{a}{a_2}\sim \left\lbrack \frac{3}{2}(\alpha+1)\left
(\frac{8\pi}{3}\right
)^{1/2}\frac{t}{t_{\Lambda}}\right\rbrack^{2/[3(\alpha+1)]}.
\end{equation}
We then have
\begin{equation}
\label{ph3}
\frac{\rho}{\rho_{\Lambda}}\sim \frac{1}{\left\lbrack
\frac{3}{2}(\alpha+1)\left
(\frac{8\pi}{3}\right )^{1/2}\frac{t}{t_{\Lambda}}\right\rbrack^{2}}.
\end{equation}
During the $\alpha$-era, the scale factor increases algebraically as
$t^{2/[3(\alpha+1)]}$ and the density decreases algebraically as $t^{-2}$.

\subsection{The General Solution}

For the equation of state (\ref{late1}), the density is related
to the scale factor by Equation~(\ref{late2}). It is possible to solve the Friedmann
equation (\ref{b9}) with the density-radius relation of Equation~{(\ref{late2})
analytically. Introducing $R=a/a_2$, we obtain
\begin{eqnarray}
\label{new1qq}
\int \frac{dR}{R\sqrt{\frac{1}{R^{3(1+\alpha)}}+1}}=\left
(\frac{8\pi}{3}\right
)^{1/2}t/t_{\Lambda}
\end{eqnarray}
which can be integrated into \cite{chavanis2}:
\begin{equation}
\label{hel1}
\frac{a}{a_2}=\sinh^{2/[3(1+\alpha)]}\left\lbrack \frac{3}{2}(1+\alpha)\left
(\frac{8\pi}{3}\right )^{1/2}\frac{t}{t_{\Lambda}}\right\rbrack.
\end{equation}

The density is then given by
\begin{equation}
\label{hel1b}
\frac{\rho}{\rho_{\Lambda}}=\frac{1}{\tanh^2\left\lbrack
\frac{3}{2}(1+\alpha)\left (\frac{8\pi}{3}\right
)^{1/2}\frac{t}{t_{\Lambda}}\right\rbrack}.
\end{equation}
For $t\rightarrow 0$, Equation~(\ref{hel1}) reduces to Equation
(\ref{ph2}) and for
$t\rightarrow +\infty$ we obtain Equation~(\ref{late4}) with a~prefactor
$a_2/2^{2/[3(1+\alpha)]}$.

The time $t_2$ corresponding to the transition between the $\alpha$-era and the
dark energy era is
obtained by substituting $a=a_2$
in Equation~(\ref{hel1}). On the other hand, according to Equation~(\ref{early3}), we find
that
the universe is decelerating when $a<a'_c$ (\emph{i.e.},  $\rho>\rho'_c$) and
accelerating when $a>a'_c$ (\emph{i.e.},  $\rho<\rho'_c$)
where $a'_c/a_2=[(1+3\alpha)/2]^{1/[3(\alpha+1)]}$ and
$\rho'_c/\rho_{\Lambda}=3(\alpha+1)/(1+3\alpha)$. The time $t'_c$ at which
the
universe starts
accelerating is obtained by substituting $a=a'_c$ in Equation~(\ref{hel1}).
This corresponds to the time
at which the curve $a(t)$ presents an inflexion point.
For $\alpha=1/3$ (radiation), this inflexion point $a'_c$ coincides
with $a_2$ ($t'_c=t_2$). For $\alpha\neq 1/3$ the two points differ.

\subsection{The Pressure}

The pressure is given by Equation~(\ref{late1}). Using Equation~(\ref{late2}), we get
\begin{equation}
\label{press2}
p=\left\lbrack \alpha\left (\frac{a_2}{a}\right
)^{3(\alpha+1)}-1\right\rbrack\rho_{\Lambda}c^2.
\end{equation}
For $\alpha>0$, the pressure decreases algebraically during the $\alpha$-era
and tends to a negative constant value  $p=-\rho_{\Lambda} c^2$ for
$t\rightarrow +\infty$.   The point at which the pressure vanishes ($w=0$)
corresponds to
$\rho'_w/\rho_{\Lambda}=(\alpha+1)/\alpha$ and
$a'_w/a_2=\alpha^{1/[3(\alpha+1)]}$.
For $\alpha=0$, the pressure is a constant $p=-\rho_{\Lambda}c^2$. At~the
transition point $t=t_2$,
we have $p_2/(\rho_{\Lambda} c^2)=-(1-\alpha)$. At the acceleration point
$t=t'_c$, we have
$p'_c/(\rho_{\Lambda} c^2)=-(1+\alpha)/(1+3\alpha)$.

\subsection{The Evolution of the Late Universe}
\label{sec_evolution}

When $a\ll a_2$, the universe is in the $\alpha$-era. Its
radius increases algebraically as $t^{2/[3(\alpha+1)]}$ while its density
decreases algebraically as $\rho\propto t^{-2}$. For $\alpha>-1/3$, the
expansion of the universe is decelerating. When $a\gg a_2$, the universe is in
the dark energy era. It undergoes a phase of accelerating expansion (late
inflation) during which its radius increases exponentially rapidly with
time while its
density remains constant and equal to the cosmological density
$\rho_{\Lambda}=7.02\times 10^{-24}\, {\rm g}/{\rm
m}^3$. The Hubble constant, the Hubble time, the Hubble  radius,
and the density of the present universe are $H_0=2.27\times 10^{-18}~{\rm
s}^{-1}$, $H_0^{-1}=0.302 t_{\Lambda}=4.41\times 10^{17}~{\rm
s}$, $a_0=c/H_0=0.302 \, l_{\Lambda}=1.32\times 10^{26}~{\rm
m}$, and $\rho_0=3H_0^2/8\pi G=1.31\rho_{\Lambda}=9.20\times
10^{-24}~{\rm
g}/{\rm m}^3$.
For $\alpha=0$ (pressureless matter), the transition between the matter
era and the dark energy era takes place at $a_2=0.677 a_0=0.204
l_{\Lambda}=8.95\times 10^{25}\, {\rm m}$, $\rho_2=1.52\,
\rho_0=2\rho_{\Lambda}=1.40\times 10^{-23}~{\rm g}/{\rm m}^3$,  and
$t_2=0.674
H_0^{-1}=0.203 t_{\Lambda}=2.97\times 10^{17}~{\rm s}$ \cite{chavanis2}. The
moment at which the
universe starts accelerating corresponds to $t'_{c}=0.504 H_0^{-1}=0.152
t_{\Lambda}=2.22\times 10^{17}\, {\rm s}$,  $a'_c=0.538 a_0=0.162
l_{\Lambda}=0.794
a_2=7.11\times 10^{25}~{\rm m}$, and  $\rho'_c=2.29\,
\rho_0=3\rho_{\Lambda}=2.11\times
10^{-23}~{\rm g}/{\rm m}^3$.   The age of the universe
is $t_0=1.03\,
H_0^{-1}=0.310t_{\Lambda}=4.54\times 10^{17}{\rm s}\sim 14\, {\rm
Gyrs}$ ($t_0=13.7\, {\rm
Gyrs}$ if we use a more precise value of $H_0$ \cite{chavanis2}). The present
values of the deceleration parameter and of the equation of state parameter are
 $q_0=-0.645$ and $w_0=-0.763$. The evolution of the scale factor and density as
a function of time are represented in Figures~\ref{tasansradiationLOGLOGprime}--\ref{trhosansradiationLINLINprime} in
logarithmic and linear scales (the figures correspond to pressureless matter
$\alpha=0$).

\begin{figure}[H]
\begin{center}
\includegraphics[clip,scale=0.45]{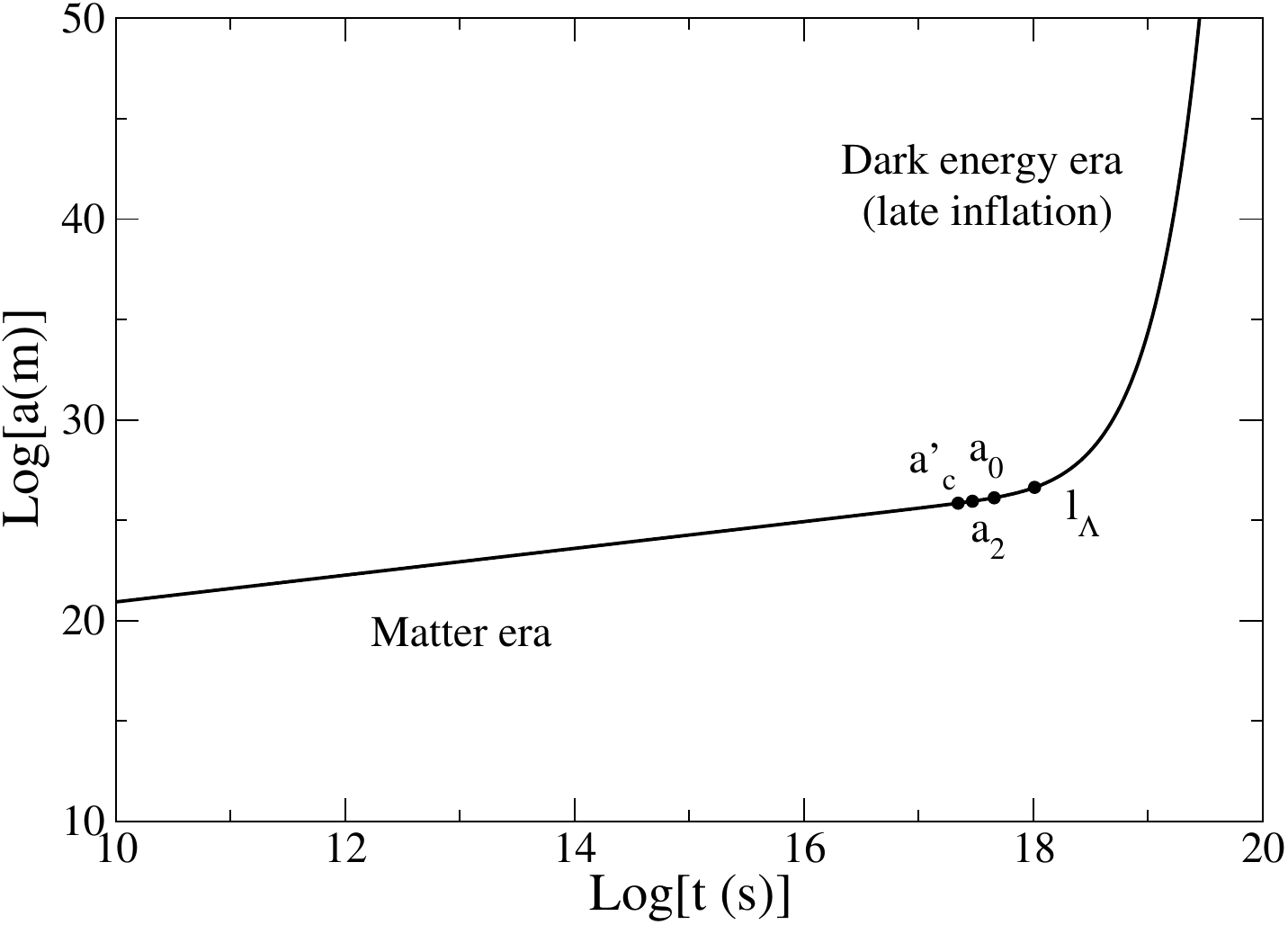}
\caption{Evolution of the scale factor $a$ with the time $t$ in logarithmic
scales. This~figure clearly shows the transition between the $\alpha$-era and
the dark energy era (de Sitter). In~the $\alpha$-era, the radius increases as
$t^{2/[3(\alpha+1)]}$. In the dark energy era, the radius increases
exponentially rapidly on a timescale of the order of the cosmological time
$t_{\Lambda}=1.46\times 10^{18}~{\rm s}$. This corresponds to a phase of late
inflation. The universe is decelerating for $a<a'_c$ and accelerating for
$a>a'_c$. The transition between the $\alpha$-era and the dark energy era
\mbox{takes
place at $a_2$}.}
\label{tasansradiationLOGLOGprime}
\end{center}
\end{figure}

\vspace{-24pt}

\begin{figure}[H]
\begin{center}
\includegraphics[clip,scale=0.45]{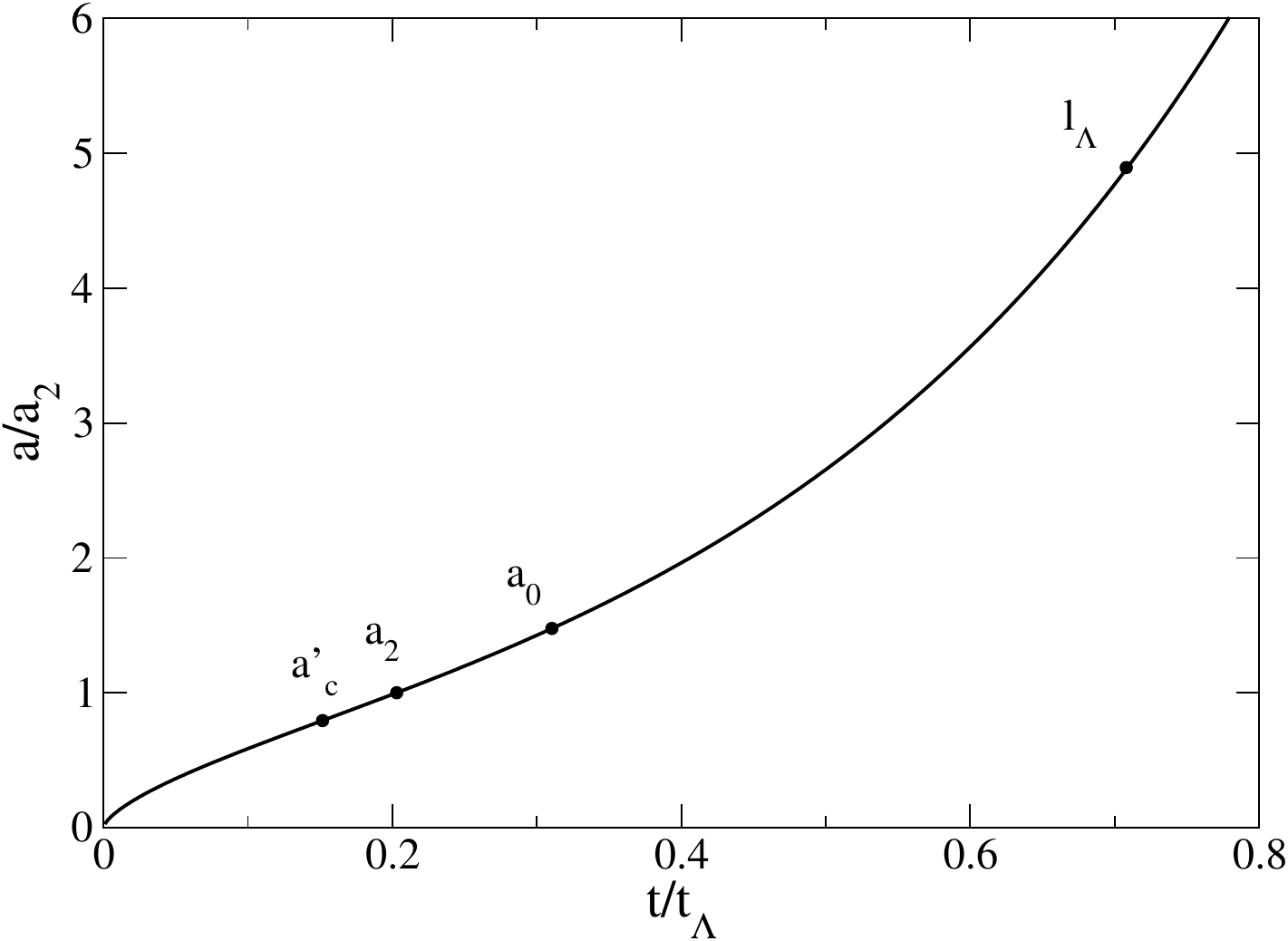}
\caption{Evolution of the scale factor $a$ with the time $t$ in linear scales.}
\label{tasansradiationLINLINprime}
\end{center}
\end{figure}

\begin{figure}[H]
\begin{center}
\includegraphics[clip,scale=0.43]{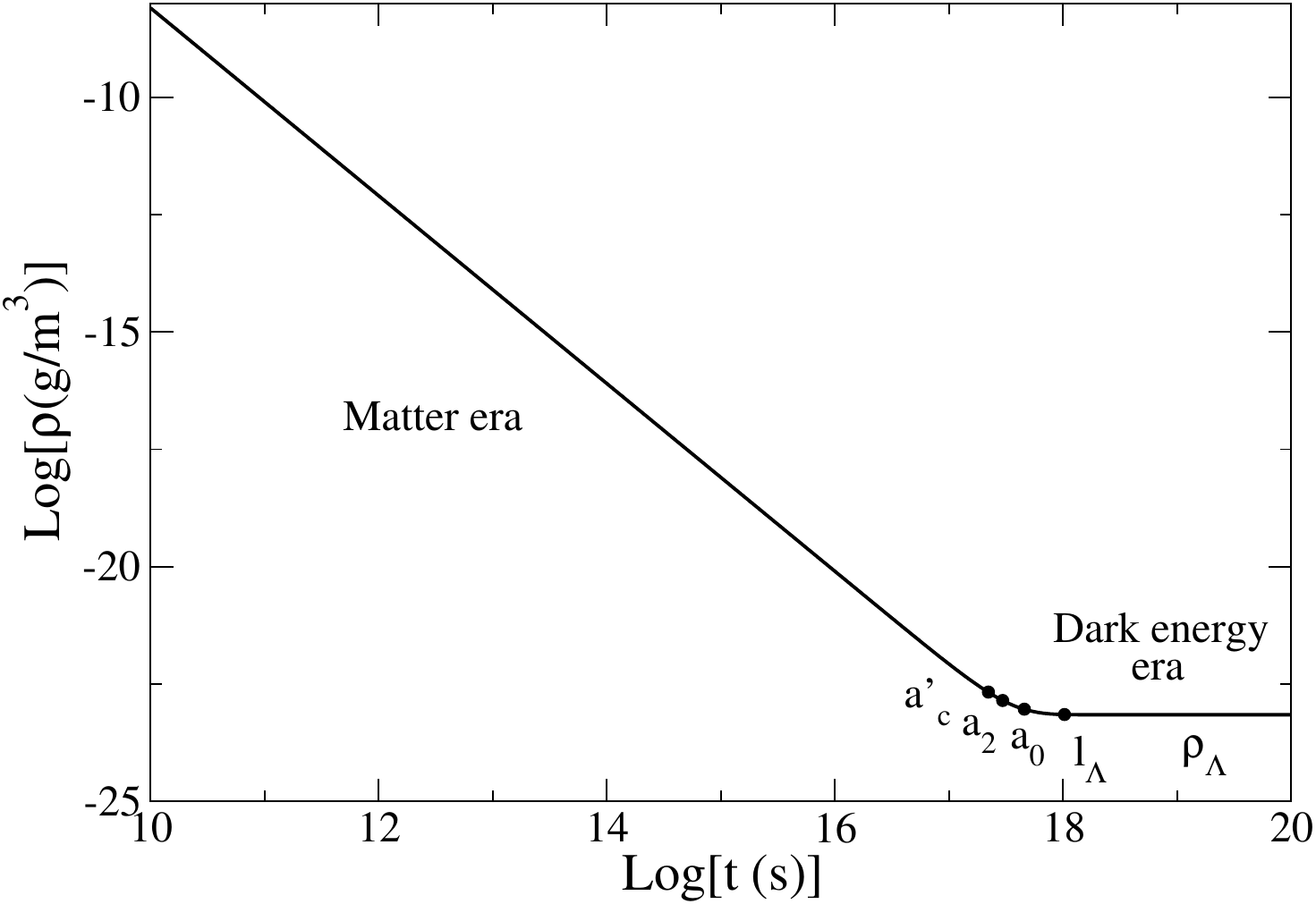}
\caption{Evolution of the density $\rho$ with the time $t$ in logarithmic
scales. In the $\alpha$-era, the density decreases as $t^{-2}$. During the late
inflation (accelerating expansion), the density remains approximately constant
with the cosmological
value $\rho_{min}=\rho_{\Lambda}$ representing a~lower~bound.}
\label{trhosansradiationLOGLOGprime}
\end{center}
\end{figure}
\vspace{-24pt}
\begin{figure}[H]
\begin{center}
\includegraphics[clip,scale=0.43]{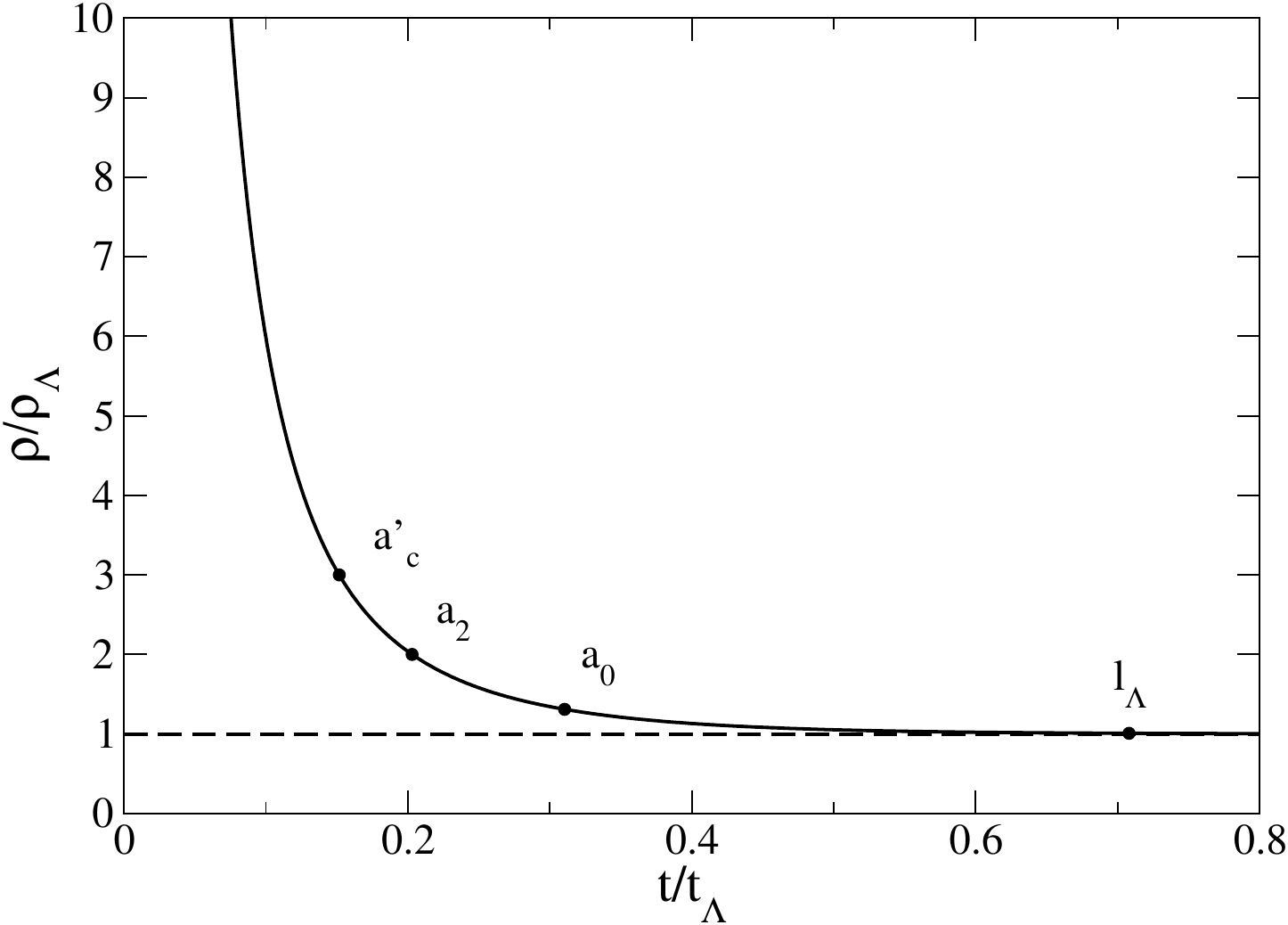}
\caption{Evolution of the density $\rho$ with the time $t$ in linear scales.
General~relativity (cosmological constant) limits the decay of the density to
the cosmological value \mbox{$\rho_{\Lambda}=7.02\times 10^{-24}\, {\rm
g}/{\rm
m}^3$.}}
\label{trhosansradiationLINLINprime}
\end{center}
\end{figure}

\subsection{Scalar Field Theory}
\label{sec_sftl}

In  alternative theories to the cosmological constant, the dark energy is
usually described by a scalar field, defined by Equations (\ref{early13b}) and
(\ref{early14}), called quintessence
\cite{quintessence1,quintessence2,quintessence3,quintessence4,
quintessence5,quintessence6,quintessence7,quintessence8,quintessence9,
quintessence10,quintessence11,quintessence12,quintessence13}.
 Using
standard
techniques  \cite{cst,bamba}, we find that the quintessence potential
corresponding to
the equation of state (\ref{late1}) is (see {Section~8.1} of
\cite{chavanis2}):

\vspace{-12pt}
\begin{equation}
\label{hel2}
V(\psi)=\frac{1}{2}\rho_{\Lambda}c^2 \left\lbrack
(1-\alpha)\cosh^2\psi+\alpha+1\right \rbrack \qquad (\psi\le 0),
\end{equation}
where we have defined
\begin{equation}
\label{hel3b}
\psi=\left (\frac{8\pi G}{3c^2}\right
)^{1/2}\frac{3\sqrt{\alpha+1}}{2}\phi.
\end{equation}
For $\psi\rightarrow -\infty$,
\begin{equation}
\frac{V(\psi)}{\rho_{\Lambda} c^2}\simeq \frac{1}{8}(1-\alpha)e^{-2\psi}
\qquad (\alpha\neq 1).
\label{sun1}
\end{equation}
For $\psi\rightarrow 0$,
\begin{equation}
\frac{V(\psi)}{\rho_{\Lambda} c^2}\simeq
1+\frac{1-\alpha}{2}\psi^2+\frac{1-\alpha}{6}\psi^4+...
\label{sun3}
\end{equation}
For $\alpha=1$, the scalar field potential  $V(\psi)=\rho_{\Lambda}c^2$ is
constant. We can also show \cite{chavanis2} that the relation between the scalar
field
and the scale factor is
\begin{equation}
\label{hel3}
\left (\frac{a_2}{a}\right )^{3(\alpha+1)/2}=-\sinh\psi.
\end{equation}
The end of the $\alpha$-era, and the beginning of the dark energy era,
corresponds to
$a=a_2$, hence to $\psi=\psi_2=\sinh^{-1}(-1)=-\ln(1+\sqrt{2})=-0.881374$.
Combining Equations (\ref{late1}),
(\ref{late2}) and (\ref{hel3}), we find that the energy density and the
pressure of the universe are related to the scalar field by
\begin{eqnarray}
\rho=\rho_{\Lambda}\cosh^{2}\psi,\qquad
p=\rho_{\Lambda}c^2(\alpha\cosh^2\psi-\alpha-1).
\label{sun2}
\end{eqnarray}
Using Equation~(\ref{hel3}), and the results of the previous
sections, we can
obtain the temporal evolution of the scalar field.
For $t\rightarrow 0$, using Equation~(\ref{ph2}), we get
\begin{equation}
\label{hel4}
\psi\simeq \ln\left (\frac{t}{t_{\Lambda}}\right )+\frac{1}{2}\ln\left
(\frac{3\pi}{2}\right )+\ln(1+\alpha)\rightarrow -\infty.
\end{equation}
For $t\rightarrow +\infty$, using Equation~(\ref{late4}) with the prefactor
$a_2/2^{2/[3(1+\alpha)]}$, we get
\begin{equation}
\label{hel5}
\psi\sim -2e^{-\frac{3}{2}(1+\alpha)\left (\frac{8\pi}{3}\right
)^{1/2}t/t_{\Lambda}}\rightarrow 0.
\end{equation}
More generally, using Equation~(\ref{hel1}), the evolution of the scalar field
$\psi(t)$ in the late universe \mbox{is
given by}
\begin{equation}
\label{hel6}
\psi=-\sinh^{-1}\left\lbrace 1/\sinh\left \lbrack\frac{3}{2}(1+\alpha)\left
(\frac{8\pi}{3}\right )^{1/2}\frac{t}{t_{\Lambda}}\right \rbrack\right \rbrace.
\end{equation}
These results are illustrated in Figures \ref{Vpsilate} and \ref{tpsilate} for
$\alpha=0$ (pressureless matter).

Some authors have proposed to represent the dark energy by a rolling tachyon
condensate appearing in a class of string theories \cite{sen}. A tachyonic
scalar field
\cite{tachyon1,tachyon2,tachyon3} has an equation of state $p=w(t)\rho c^2$
with
$-1\le w(t)\le 0$. This scalar field evolves according to the equation
\begin{equation}
\label{tachyon1}
\frac{\ddot \phi}{1-\dot\phi^2}+3H\dot\phi+\frac{1}{V}\frac{dV}{d\phi}=0.
\end{equation}
The tachyonic field tends to run down the potential towards lower energies.
The density and the pressure of the universe are related to the tachyonic field
by
\begin{equation}
\label{tachyon2}
\rho c^2=\frac{V(\phi)}{\sqrt{1-\dot\phi^2}},\qquad
p=-V(\phi)\sqrt{1-\dot\phi^2}.
\end{equation}

\begin{figure}[H]
\begin{center}
\includegraphics[clip,scale=0.42]{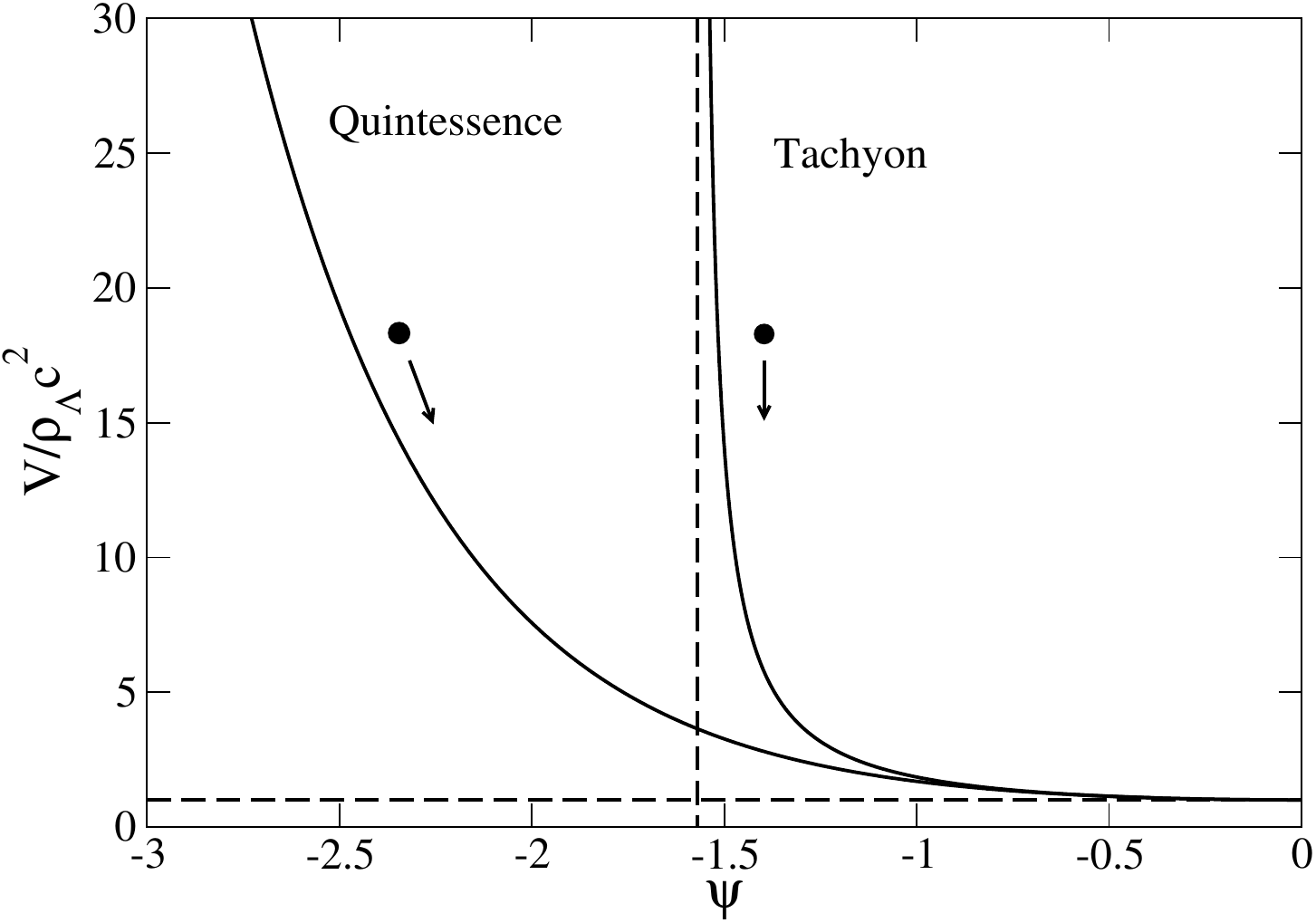}
\caption{Potential of the quintessence field and tachyonic field in the late
universe. The~field tends to run down the potential.}
\label{Vpsilate}
\end{center}
\end{figure}

\begin{figure}[H]
\begin{center}
\includegraphics[clip,scale=0.42]{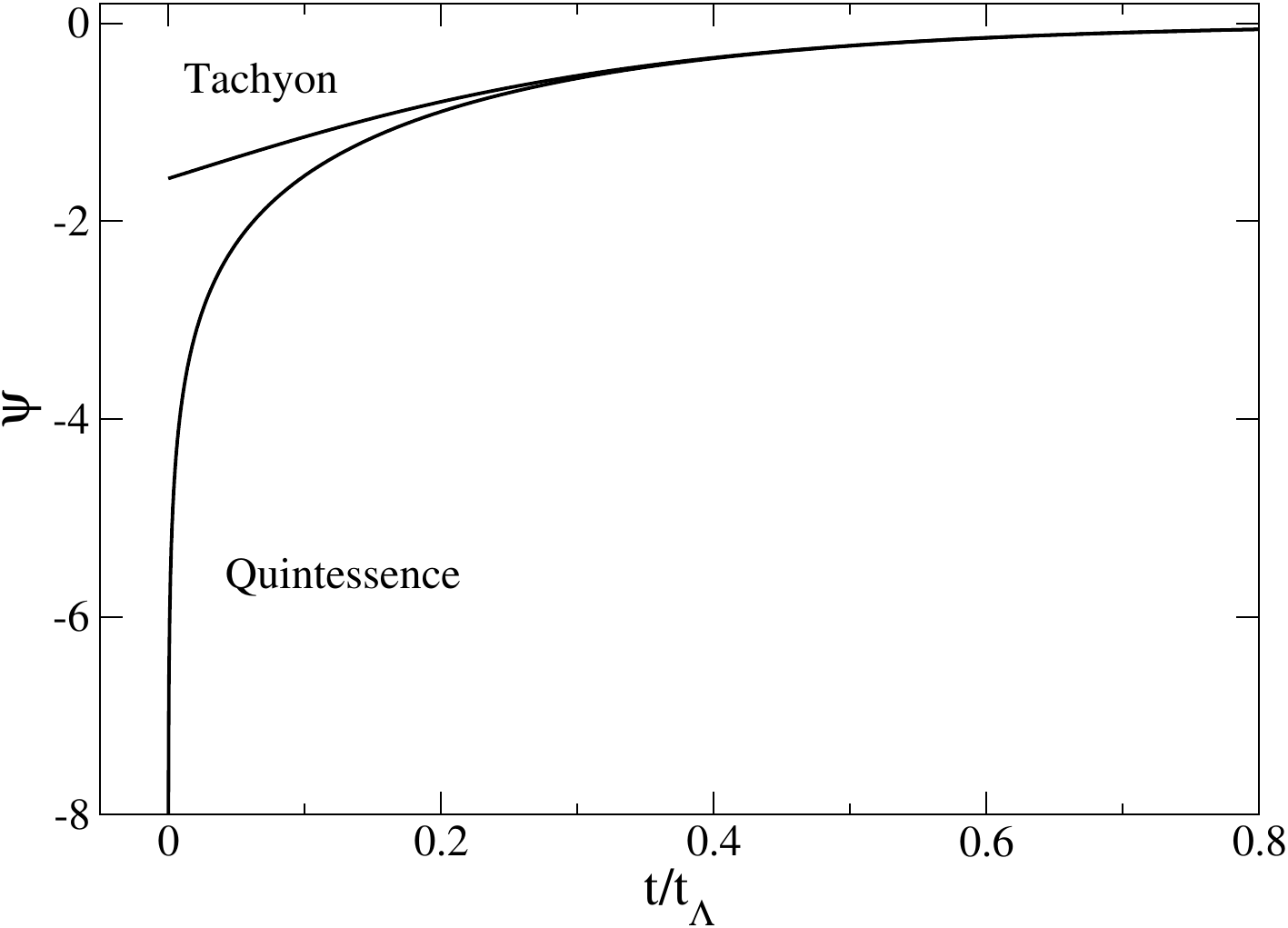}
\caption{Evolution of the quintessence field and tachyonic field as a function
of
time in the late universe.}
\label{tpsilate}
\end{center}
\end{figure}

Let us consider the equation of state (\ref{late1}) with the specific
index
$\alpha=0$ (pressureless matter). Since $w=p/\rho
c^2=-\rho_{\Lambda}/\rho$ is
between $-1$ and $0$, the equation of state  (\ref{late1}) can be
associated with
a tachyonic field.
Using
standard
techniques \cite{cst,bamba}, we find that its potential is (see {Section
8.2}~\cite{chavanis2}):
\begin{equation}
\label{a22}
V(\psi)=\frac{\rho_{\Lambda} c^2}{\cos\psi} \qquad (-\pi/2\le \psi\le
0),
\end{equation}
where we have defined
\begin{equation}
\label{nom9wa}
\psi=\sqrt{\rho_{\Lambda}
c^2}\left (\frac{6\pi G}{c^2}\right )^{1/2}\phi.
\end{equation}
For $\psi\rightarrow -\frac{\pi}{2}^+$,
\begin{equation}
\label{ab1}
\frac{V(\psi)}{\rho_{\Lambda}c^2}\simeq \frac{1}{\psi+\frac{\pi}{2}}.
\end{equation}
For $\psi\rightarrow 0$,
\begin{equation}
\label{ab2}
\frac{V(\psi)}{\rho_{\Lambda}c^2}\simeq
1+\frac{\psi^2}{2}+\frac{5\psi^4}{24}+...
\end{equation}
We can also show \cite{chavanis2} that the relation between
the scalar
field
and the scale factor is
\begin{equation}
\label{nom9w}
\left (\frac{a_2}{a}\right )^{3/2}=-\tan\psi.
\end{equation}
The end of the matter-era ($\alpha=0$), and the beginning of the dark energy
era,
corresponds to
$a=a_2$, hence to $\psi=\psi_2=\tan^{-1}(-1)=-\pi/4=-0.785398$. Combining
Equations~(\ref{late1}), (\ref{late2}) and (\ref{nom9w}), we find that the energy
density and the pressure of
the universe are related to the scalar field by
\begin{eqnarray}
\rho=\frac{\rho_{\Lambda}}{\cos^{2}\psi},\qquad p=-\rho_{\Lambda}c^2.
\label{sun5}
\end{eqnarray}
Using Equation~(\ref{nom9w}), and the results of the previous
sections, we can
obtain the temporal evolution of the scalar field.
In the matter era ($t\rightarrow 0$), using Equation~(\ref{ph2}), we get
\begin{equation}
\label{nom9aa}
\psi\simeq -\frac{\pi}{2}+\sqrt{6\pi}\frac{t}{t_{\Lambda}}\rightarrow
-\frac{\pi}{2}.
\end{equation}
In the dark energy era ($t\rightarrow +\infty$), using Equation~(\ref{late4})
with a
prefactor $a_2/2^{2/3}$, we get
\begin{equation}
\label{nom9bb}
\psi\sim -2e^{-\sqrt{6\pi}t/t_{\Lambda}}\rightarrow 0.
\end{equation}
More generally, using Equation~(\ref{hel1}), the evolution of the tachyonic
field
$\psi(t)$ in the late universe is given by
\begin{equation}
\label{nom10c}
\psi=-{\rm Arctan}\left\lbrack {1}/{\sinh\left
(\sqrt{6\pi}\frac{t}{t_{\Lambda}}\right )}\right\rbrack.
\end{equation}
These results are illustrated in Figures \ref{Vpsilate} and \ref{tpsilate} for
$\alpha=0$ (pressureless matter).

{\it Remark:} The canonical  scalar field potential of Equation (\ref{hel2})
describing the transition between the $\alpha$-era and the dark energy era
involves hyperbolic functions. It corresponds to the scalar field potential
associated  with the generalized Chaplygin gas (see Note 1)
\cite{chaplygin1,chaplygin2,chaplygin3,chaplygin4,chaplygin5,chaplygin6,
chaplygin7,chaplygin8,chaplygin9}. Other types of hyperbolic scalar field
potentials have been
introduced in the past. The
first potential with hyperbolic functions driving the observed acceleration of
the universe was proposed by Sahni and Wang \cite{sahni} and has
been confronted to the observations in \cite{shahalam}. Related potentials
were considered in \cite{chimento,ulm,rubano}. Models with hyperbolic
potentials exploiting dynamical Noether symmetries of the field equations and
admiting analytical solutions were constructed  and
compared with cosmological data in \cite{paliathanasis}, finding a good
agreement. On the other hand, the tachyonic scalar field potential of Equation
(\ref{a22})
describing
the
transition between the $\alpha$-era and the dark energy era involves
trigonometric functions. A~tachyonic field with a~trigonometric potential
was proposed in \cite{gorini} to describe the transition between dark
matter and dark energy. Interestingly, this model admits different solutions
depending on the initial condition.  It can either describe an eternally
expanding universe or lead to a future finite time singularity,
called Big Brake, which is characterized by an infinite deceleration.
This model has been recently subjected  to several cosmological tests in
\cite{keresztes}.

\section{The General Model}
\label{sec_g}
\vspace{-12pt}
\subsection{The Quadratic Equation of State}
\label{sec_quadra}

We propose to describe the vacuum energy, the $\alpha$-fluid, and the dark
energy by a unique \mbox{equation of state}
\begin{equation}
\label{newg1}
p/c^2=-(\alpha+1)\rho\left (\frac{\rho}{\rho_P}\right
)^{1/|n_e|}+\alpha\rho-(\alpha+1)\rho \left
(\frac{\rho_{\Lambda}}{\rho}\right
)^{1/|n_l|},
\end{equation}
where $n_e>0$ and $n_l<0$ are the polytropic indices of the early and late
universe, respectively. For the sake of simplicity, and for definiteness, we
shall select the indices $n_e=+1$ and $n_l=-1$. Therefore, we consider the
quadratic equation of state
\begin{equation}
\label{g1}
p=-(\alpha+1)\frac{\rho^2}{\rho_P} c^2+\alpha\rho
c^2-(\alpha+1)\rho_{\Lambda}
c^2.
\end{equation}
For $\rho\rightarrow \rho_P$, $p\rightarrow -\rho_P c^2$ (vacuum energy).
For
$\rho_{\Lambda}\ll\rho\ll\rho_P$, $p\sim \alpha\rho c^2$
($\alpha$-fluid). For
$\rho\rightarrow \rho_{\Lambda}$, $p\rightarrow -\rho_{\Lambda} c^2$ (dark
energy). The pressure vanishes at
$\rho_{w}=[\alpha/(\alpha+1)]\rho_P$ and at
$\rho'_w=[(\alpha+1)/\alpha]\rho_{\Lambda}$. It~has a maximum
$p_e=\alpha^2\rho_P c^2/[4(\alpha+1)]$ at
$\rho_e=[\alpha/2(\alpha+1)]\rho_P$.
The pressure is negative for $\rho>\rho_w$, positive for
$\rho'_w<\rho<\rho_w$ and negative for $\rho<\rho'_w$ (to
obtain these results, we have used the fact that $\rho_P\gg\rho_{\Lambda}$).
The quadratic equation of state (\ref{g1}) combines the properties of
the
equation of state (\ref{early1}) valid in the early universe, and of the
equation of state  (\ref{late1}) valid in the late universe. A nice
feature of
this equation of state
is that both the Planck density (vacuum energy) and the cosmological
density (dark energy) explicitly appear. Therefore, this equation of
state reproduces both the early inflation and the late inflation, described by
an equation of state $p=-\rho c^2$, in which the scale factor increases
exponentially rapidly. They are connected by the $\alpha$-era, with a linear
equation of state $p=\alpha\rho c^2$, in which the scale factor increases
algebraically. For $\alpha>-1/3$, the universe is decelerating during this
period.

The equation of state parameter $w=p/\rho c^2$ and the deceleration parameter
$q$ are given by
\begin{equation}
\label{fearly3a}
w=-(\alpha+1)\frac{\rho}{\rho_P}+\alpha-(\alpha+1)\frac{\rho_{\Lambda}
}{\rho},
\qquad
q=\frac{1+3\alpha}{2}-\frac{3}{2}(\alpha+1)\frac{\rho}{\rho_P}-\frac{3}{2}
(\alpha+1)\frac{\rho_{\Lambda}}{\rho}.
\end{equation}
The velocity of sound $c_s^2=p'(\rho)$ is given by
\begin{equation}
\label{fvs1}
\frac{c_s^2}{c^2}=-2(\alpha+1)\frac{\rho}{\rho_P}+\alpha.
\end{equation}

Using the equation of continuity (\ref{b7}), we get
\begin{equation}
\label{g2}
\rho=\frac{\rho_P}{(a/a_1)^{3(\alpha+1)}+1}+\rho_{\Lambda}.
\end{equation}
To obtain this
expression, we have used the fact that $\rho_P\gg \rho_{\Lambda}$ so
that $p/c^2+\rho\simeq
[(\alpha+1)/\rho_P](\rho-\rho_{\Lambda})(\rho_P-\rho)$ (see Appendix
\ref{sec_exactQ}). When
$a\rightarrow 0$, $\rho\rightarrow
\rho_{P}$ (vacuum energy). When
$\rho_{\Lambda}\ll\rho\ll\rho_{P}$, $\rho\sim
\rho_P (a_1/a)^{3(1+\alpha)}$ ($\alpha$-era). When $a\rightarrow +\infty$,
$\rho\rightarrow
\rho_{\Lambda}$ (dark energy). In the early universe, the contribution of dark
energy is negligible
and we recover Equation~(\ref{early2}). In the late universe, the contribution of
vacuum energy is negligible and we recover Equation~(\ref{late2}) with
$\rho_Pa_1^{3(1+\alpha)}=\rho_{\Lambda}a_2^{3(1+\alpha)}$.

Using $\rho_P a_1^{3(1+\alpha)}=\rho_{\alpha,0}a_0^{3(1+\alpha)}$, where
$\rho_{\alpha,0}$ is the present density of the $\alpha$-fluid and
$a_0=c/H_0=1.32\times 10^{26}~{\rm m}$ is the present distance of cosmological
horizon determined by the Hubble constant $H_0=2.27\times 10^{-18}~{\rm s}^{-1}$
(the Hubble time is $H_0^{-1}=4.41\times 10^{17}~{\rm s}$), we can rewrite
Equation
(\ref{g2}) in the form
\begin{equation}
\label{g3}
\rho=\frac{\rho_{\alpha,0}}{(a/a_0)^{3(1+\alpha)}+
(a_1/a_0)^{3(1+\alpha)}}+\rho_{\Lambda}.
\end{equation}
Substituting this relation in the Friedmann equation (\ref{b9}), and writing
$\rho_{\alpha,0}=\Omega_{\alpha,0}\rho_0$ and
$\rho_{\Lambda}=\Omega_{\Lambda,0}\rho_0$, where $\rho_0=3H_0^2/8\pi
G=9.20\times
10^{-24}~{\rm g/m^3}$ is the present density of the universe, we obtain
\begin{equation}
\label{g6}
\frac{H}{H_0}=\sqrt{\frac{\Omega_{\alpha,0}}{(a/a_0)^{3(1+\alpha)}+(a_1/a_0)^{
3(1+\alpha)}}
+\Omega_{\Lambda,0}}.
\end{equation}
Using Equations (\ref{g1}) and (\ref{g2}), the pressure can be written in
very good approximation as
\begin{equation}
\label{press3}
p=\frac{\alpha(a/a_1)^{3(\alpha+1)}-1}{\left\lbrack
(a/a_1)^{3(\alpha+1)}+1\right\rbrack^2}\rho_P c^2-\rho_{\Lambda}c^2.
\end{equation}
Finally, if we assume that the $\alpha$-fluid represents the radiation
($\alpha=1/3$),
we find that the temperature associated with the quadratic equation of state
(\ref{g1}) is given by a generalized Stefan-Boltzmann law (see {Equation~(C.4)}
in
\cite{chavanis2}).

It may be useful to re-express the results of Sections \ref{sec_early} and
\ref{sec_late} in terms of the present-day
variables. To~that purpose, we use the following
relations
$a_1/a_0=(\rho_{\alpha,0}/\rho_P)^{1/[3(1+\alpha)]}$,
$a_2/a_0=(\Omega_{\alpha,0}/\Omega_{\Lambda,0})^{1/[3(1+\alpha)]}$,
$H_0t_P=(8\pi/3)^{1/2}(\rho_0/\rho_P)^{1/2}$, and
$H_0t_{\Lambda}=(8\pi/3)^{1/2}(\Omega_{\Lambda,0})^{-1/2}$.

\subsection{The Early Universe}
\label{sec_gearly}

In the early universe, we can neglect the contribution of the dark energy in the
density. We only consider the contribution of the vacuum energy and of the
$\alpha$-fluid. Equation~(\ref{g3}) then reduces to
\begin{equation}
\label{g12}
\rho=\frac{\rho_{\alpha,0}}{(a/a_0)^{3(1+\alpha)}+
(a_1/a_0)^{3(1+\alpha)}}.
\end{equation}
The Friedmann equation (\ref{g6}) becomes
\begin{equation}
\label{g14}
\frac{H}{H_0}=\sqrt{\frac{\Omega_{\alpha,0}}{(a/a_0)^{3(1+\alpha)}+(a_1/a_0)^{
3(1+\alpha)}}}.
\end{equation}
It has the analytical solution given by Equation~(\ref{early8}).

In the vacuum energy era ($a\ll a_1$), the density is constant:
\begin{equation}
\label{ph4}
\rho\simeq \rho_P.
\end{equation}
Therefore, the Hubble parameter is also a constant $H=(8\pi/3)^{1/2}t_P^{-1}$.
In that case, the scale factor increases exponentially rapidly with time
according to Equation
(\ref{early6}).

In the $\alpha$-era ($a\gg a_1$), we have
\begin{equation}
\label{ph5}
\rho\simeq \frac{\rho_{\alpha,0}}{(a/a_0)^{3(1+\alpha)}}.
\end{equation}
The Friedmann equation (\ref{g14}) reduces to
\begin{equation}
\label{g14b}
\frac{H}{H_0}=\sqrt{\frac{\Omega_{\alpha,0}}{(a/a_0)^{3(1+\alpha)}}}.
\end{equation}
The scale factor increases algebraically rapidly as
\begin{eqnarray}
\label{g15}
\frac{a}{a_0}\sim \left\lbrack \frac{3}{2}(\alpha+1)
\sqrt{\Omega_{\alpha,0}}H_0
t\right\rbrack^{2/[3(1+\alpha)]},
\end{eqnarray}
and the density decreases as
\begin{eqnarray}
\label{g15b}
\frac{\rho}{\rho_0}\sim \frac{1}{\left\lbrack \frac{3}{2}(\alpha+1)H_0
t\right\rbrack^2}.
\end{eqnarray}

The transition
between the vacuum energy era and the $\alpha$-era corresponds to
$\rho_{\alpha}=\rho_P$. This yields
$a_1/a_0=(\rho_{\alpha,0}/\rho_{P})^{1/[3(1+\alpha)]}$ and
$\rho_1/\rho_P=1/2$.
The
inflation ends at the time $t_1$. The universe starts decelerating when
$a_c/a_0=\lbrace
2\rho_{\alpha,0}/[(1+3\alpha)\rho_P]\rbrace^{1/[3(1+\alpha)]}$ and
$\rho_c/\rho_P=(1+3\alpha)/[3(\alpha+1)]$. This~corresponds to the first
inflexion point of the curve $a(t)$. The universe is accelerating for
$t<t_c$ and decelerating for $t>t_c$. For $\alpha=1/3$, $t_c=t_1$.

\subsection{The Late Universe}
\label{sec_glate}

In the late universe, we can neglect the contribution of the vacuum energy in
the
density. We only consider the contribution of the $\alpha$-fluid and of the dark
energy. Equation~(\ref{g3}) then reduces to
\begin{equation}
\label{g18}
\rho=\frac{\rho_{\alpha,0}}{(a/a_0)^{3(1+\alpha)}}+\rho_{\Lambda}.
\end{equation}
The Friedmann equation (\ref{g6}) becomes
\begin{equation}
\label{g20}
\frac{H}{H_0}=\sqrt{\frac{\Omega_{\alpha,0}}{(a/a_0)^{3(1+\alpha)}}
+\Omega_{\Lambda,0}}.
\end{equation}
It has the analytical solution
\begin{equation}
\label{g21}
\frac{a}{a_0}=\left (\frac{\Omega_{\alpha,0}}{\Omega_{\Lambda,0}}\right
)^{1/[3(1+\alpha)]}\sinh^{2/[3(1+\alpha)]}\left
\lbrack\frac{3}{2}(1+\alpha)\sqrt{\Omega_{\Lambda,0}}H_0 t\right \rbrack.
\end{equation}
This corresponds to the $\Lambda$CDM model. The density evolves as
\begin{equation}
\label{g22}
\frac{\rho}{\rho_0}=\frac{\Omega_{\Lambda,0}}{\tanh^2\left
\lbrack\frac{3}{2}(1+\alpha)\sqrt{\Omega_{\Lambda,0}}H_0 t\right \rbrack}.
\end{equation}
Setting $a=a_0$ in Equation~(\ref{g21}), we find the age of the universe [Note
8: Of course, for the determination of the age of the universe, we can neglect
the contribution of the vacuum energy in the early universe and take $a_1=0$
(strictly speaking, the age of the universe is infinite since it has no origin;
however, we define the age of the universe from the time $t=0$ at which
$a=l_P$).]:
\begin{equation}
\label{g30}
t_{0}=\frac{1}{H_0}\frac{2}{3(1+\alpha)}\frac{1}{\sqrt{\Omega_{\Lambda,0}}}
\sinh^{-1}\left\lbrack \left (\frac{\Omega_{\Lambda,0}}{\Omega_{\alpha,0}}\right
)^{1/2}\right\rbrack.
\end{equation}
The present values of the equation of state parameter and
deceleration parameter are
\begin{equation}
\label{early3now}
w_0=\alpha-(\alpha+1)\Omega_{\Lambda,0}, \qquad
q_0=\frac{1+3\alpha}{2}-\frac{3}{2}(\alpha+1)\Omega_{\Lambda,0}.
\end{equation}

In the dark energy era ($a\gg a_2$), the density is constant:
\begin{equation}
\label{ph6}
\rho\simeq \rho_{\Lambda}.
\end{equation}
Therefore, the Hubble parameter is also a constant
$H=H_0\sqrt{\Omega_{\Lambda,0}}$. Using Equation~(\ref{g21}), we find that  the scale
factor  increases exponentially rapidly as
\begin{equation}
\label{g23}
\frac{a}{a_0}\sim \left (\frac{\Omega_{\alpha,0}}{4\Omega_{\Lambda,0}}\right
)^{1/[3(1+\alpha)]}e^{\sqrt{\Omega_{\Lambda,0}}H_0 t}.
\end{equation}
This corresponds to de Sitter's solution.

In the $\alpha$-era ($a\ll a_2$), we recover Equations
(\ref{ph5})--(\ref{g15b}).

The transition between the $\alpha$-era and the dark energy era corresponds to
$\rho_{\alpha}=\rho_{\Lambda}$. This yields
${a_2}/{a_0}=({\Omega_{\alpha,0}}/{\Omega_{\Lambda,0}})^{1/[3(1+\alpha)]}$ and
$\rho_2/\rho_{\Lambda}=2$. The universe starts  accelerating when
${a'_c}/{a_0}=[(1+3\alpha)\Omega_{\alpha,0}/2\Omega_{\Lambda,0}]^{1/[
3(1+\alpha)
]}$ and $\rho'_c/\rho_0=3(\alpha+1)\Omega_{\Lambda,0}/(1+3\alpha)$. This
corresponds to the second inflexion point of the curve $a(t)$.  The universe is
decelerating for $t<t'_c$ and accelerating for $t>t'_c$. \mbox{For $\alpha=1/3$,
$t'_c=t_2$}.

\subsection{The General Solution}

For the quadratic equation of state (\ref{g1}), the density is related
to the scale factor by Equation~(\ref{g2}).  It is possible to solve the Friedmann
equation (\ref{b9})
with the density-radius relation of Equation~(\ref{g2})
analytically. Introducing $R=a/a_1$ and
$\lambda=\rho_{\Lambda}/\rho_{P}=1.36\times 10^{-123}\ll 1$, we obtain
\begin{eqnarray}
\label{hel7}
\int \frac{\sqrt{1+R^{3(\alpha+1)}}}{R\sqrt{1+\lambda R^{3(\alpha+1)}}}\,
dR=\left (\frac{8\pi}{3}\right )^{1/2}t/t_P
\end{eqnarray}
which can be integrated into
\begin{eqnarray}
\label{hel8}
\frac{1}{\sqrt{\lambda}}\ln\left \lbrack 1+2\lambda
R^{\kappa}+2\sqrt{\lambda(1+R^{\kappa}+\lambda
R^{2\kappa})}\right\rbrack\qquad\nonumber\\
-\ln\left\lbrack \frac{2+R^{\kappa}+2\sqrt{1+R^{\kappa}+\lambda
R^{2\kappa}}}{R^{\kappa}}\right\rbrack
=\kappa\left (\frac{8\pi}{3}\right )^{1/2}\frac{t}{t_P}+C,
\end{eqnarray}
where $\kappa=3(\alpha+1)$ and $C$ is a constant determined such that $a=l_P$
at
$t=0$. This solution is interesting because it describes
analytically a phase of early inflation and a phase of late inflation (or late
accelerating expansion) connected by a power-law expansion ($\alpha$-era). The
corresponding scalar field theory is developed in Appendix
\ref{sec_sft}. We obtain a scalar field potential (see Equations
(\ref{sft6})---(\ref{unif1})) that unifies the inflaton potential
of Section
\ref{sec_sfe} and the quintessence potential of Section \ref{sec_sftl}.

\subsection{The Whole Evolution of the Universe}
\label{sec_connection}

In our model, the evolution of the scale factor is given by Equation~(\ref{g6}). For
$a_1=0$, we obtain the same equation as in the $\Lambda$CDM model (slightly
generalized to account for a possibly non-zero value of $\alpha$) where the
contributions
of matter and dark energy
are added individually \cite{bt}. However, the $\Lambda$CDM model does not
describe
the phase of early inflation and presents a singularity at $t=0$ (Big
Bang). For $a_1\neq 0$, we obtain a generalized model
which does not present a primordial singularity and which displays a
phase of early inflation. In this model, the universe always existed
in the past but, for $t<0$, it has a very small radius, smaller than
the Planck length [Note 9: Our semi-classical
model certainly breaks down in this period since it does not take
quantum fluctuations into account. The Planck era may not be described
in terms of an equation of state $p(\rho)$, or even in terms of the
Einstein equations, as we have assumed. It probably requires the
development of a theory of quantum gravity that does not exist for the
moment. An interesting description of the early inflation has been
given by Monerat  \emph{et al}. \cite{monerat} in terms of a quantized
model based on a simplified Wheeler-DeWitt (WdW) equation. In that model, a
quantum tunneling process explains the birth of the universe with a
well defined size after tunneling.]. At~$t=0$, it undergoes an inflationary
expansion in a very short lapse of time of the order of the Planck
time. For~$t\gg t_P$, this model gives the same results as the
$\Lambda$CDM model (see Equation~(\ref{g20})): the universe first undergoes an
algebraic
expansion in the $\alpha$-era, then an exponential
expansion (late inflation) in the dark energy era. A nice feature of
this model is its simplicity since it incorporates a phase of early
inflation in a very simple manner. We just have to add a term
$+(a_1/a_0)^{3(1+\alpha)}$ in the standard Equation (\ref{g20})
of the $\Lambda$CDM model. This term has the effect of a small
scale cut-off. Therefore, the
modification implied by Equation~(\ref{g6}) to describe the early
inflation is very natural. On the other hand, this model gives
the same results  in the late universe as the
standard $\Lambda$CDM model, so this does not bring any modification to the
usual
Equation (\ref{g20}). Therefore, our model completes the
standard $\Lambda$CDM model by incorporating the phase of early
inflation in a natural manner.  This is an interest of this
description since the standard $\Lambda$CDM model gives results that agree with
the observations after the Planck era.

In Figures \ref{taLOGLOGcomplet} and \ref{trhoLOGLOGcomplet}, we have represented
the evolution of the scale factor and density of the universe as a function of
time. The universe exhibits two types of inflations: an early inflation due to
the Planck density $\rho_P=5.16\times 10^{99}~{\rm g}/{\rm m}^3$ (vacuum
energy)
and a late inflation due to the cosmological density $\rho_{\Lambda}=7.02\times
10^{-24}~{\rm g}/{\rm m}^3$ (dark energy). They are connected by an
$\alpha$-era
during which the scale factor and the density evolve algebraically. In this
period, the
universe is decelerating (for $\alpha>-1/3$). There~exists a striking
``symmetry'' between the early and the late evolution of the universe. The early
universe is described by a~polytropic equation of state of index $n=+1$ and the
late universe is described by a~polytropic equation of state of index $n=-1$. On
the other hand, the cosmological constant $\Lambda$ in the late universe plays
the same role as the Planck constant $\hbar$ in the early universe. In
particular, Figure~\ref{trhoLOGLOGcomplet} shows that the density varies between
two bounds $\rho_{max}=\rho_P$ and $\rho_{min}=\rho_{\Lambda}$ that are
fixed by
fundamental constants [Note 10: these bounds come from the
properties of the polytropic equation of state  (\ref{intro1}). This
is a bit
similar to the bound on the mass of white dwarf stars
(Chandrasekhar's mass) which also comes from the properties of a polytropic
equation of state with index $n=3$ describing an~ultra-relativistic degenerate
fermionic gas \cite{chandramass}.]. These values differ by a factor of the order
of $10^{122}$. The early universe is governed by quantum mechanics ($\hbar$) and
the late universe by general relativity ($\Lambda$).

\begin{figure}[H]
\begin{center}
\includegraphics[clip,scale=0.45]{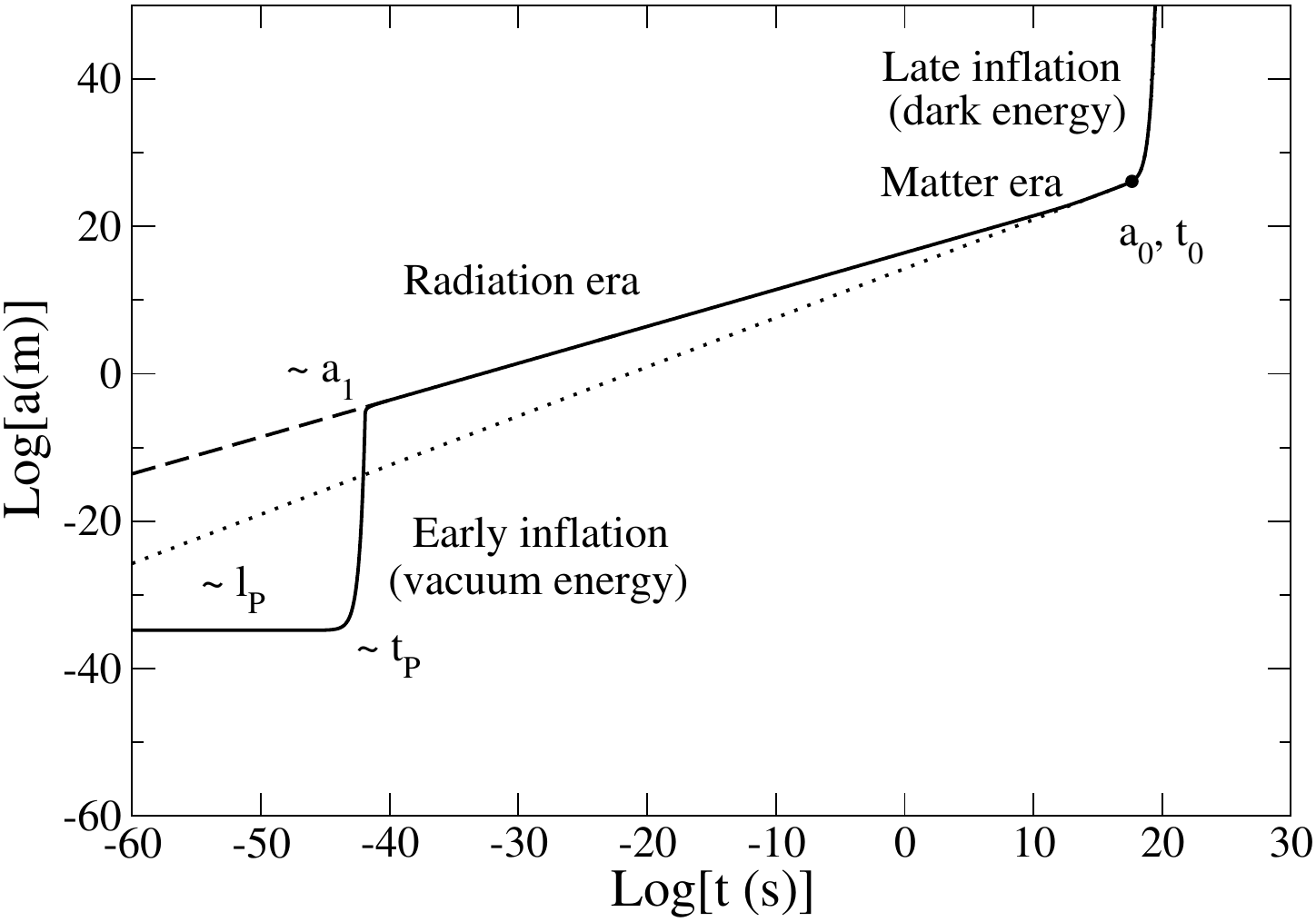}
\caption{Evolution of the scale factor $a$ as a function of time in logarithmic
scales. The universe exists at all times in the past and in
the future and there is no singularity (aioniotic universe). The universe first
undergoes a phase of early inflation (Planck era)
due to the vacuum energy during which the scale factor increases exponentially
rapidly on a timescale of the order of the Planck time $t_P$ (for $\alpha=1/3$
it increases by $29$ orders of magnitude in less than $10^{-42}$ s).  This is
followed by the $\alpha$-era during which the scale factor increases
algebraically ($a\sim t^{2/[3(\alpha+1)]}$). The dashed line corresponds to
$\alpha=1/3$ (radiation) and the dotted line corresponds to $\alpha=0$
(matter).
Without the early inflation, the universe would exhibit a primordial singularity
(Big Bang). Finally, the universe undergoes a phase of late inflation (de Sitter
era) due to the dark energy during which the scale factor increases
exponentially rapidly on a timescale of the order of the cosmological  time
$t_{\Lambda}$. The universe exhibits two types of inflation: an early inflation
corresponding to the Planck density $\rho_P$ (vacuum energy) due to quantum
mechanics (Planck constant) and a~late inflation corresponding to the
cosmological density $\rho_{\Lambda}$ (dark energy) due to general relativity
(cosmological constant). The evolution of the early and late universe is
remarkably symmetric. In our model, it is described by two
polytropic equations of state with index $n=+1$ and $n=-1$ that can be combined
into a single quadratic equation of state (\ref{g1}). The expansion of
the
universe is accelerating during the phases of inflation and decelerating during
the $\alpha$-era (if $\alpha>-1/3$).  We have also represented the location of
the present universe
(bullet). It happens to be just at the transition between the matter era and the
dark energy era (see Section \ref{sec_ccp}).}
\label{taLOGLOGcomplet}
\end{center}
\end{figure}

\begin{figure}[H]
\begin{center}
\includegraphics[clip,scale=0.45]{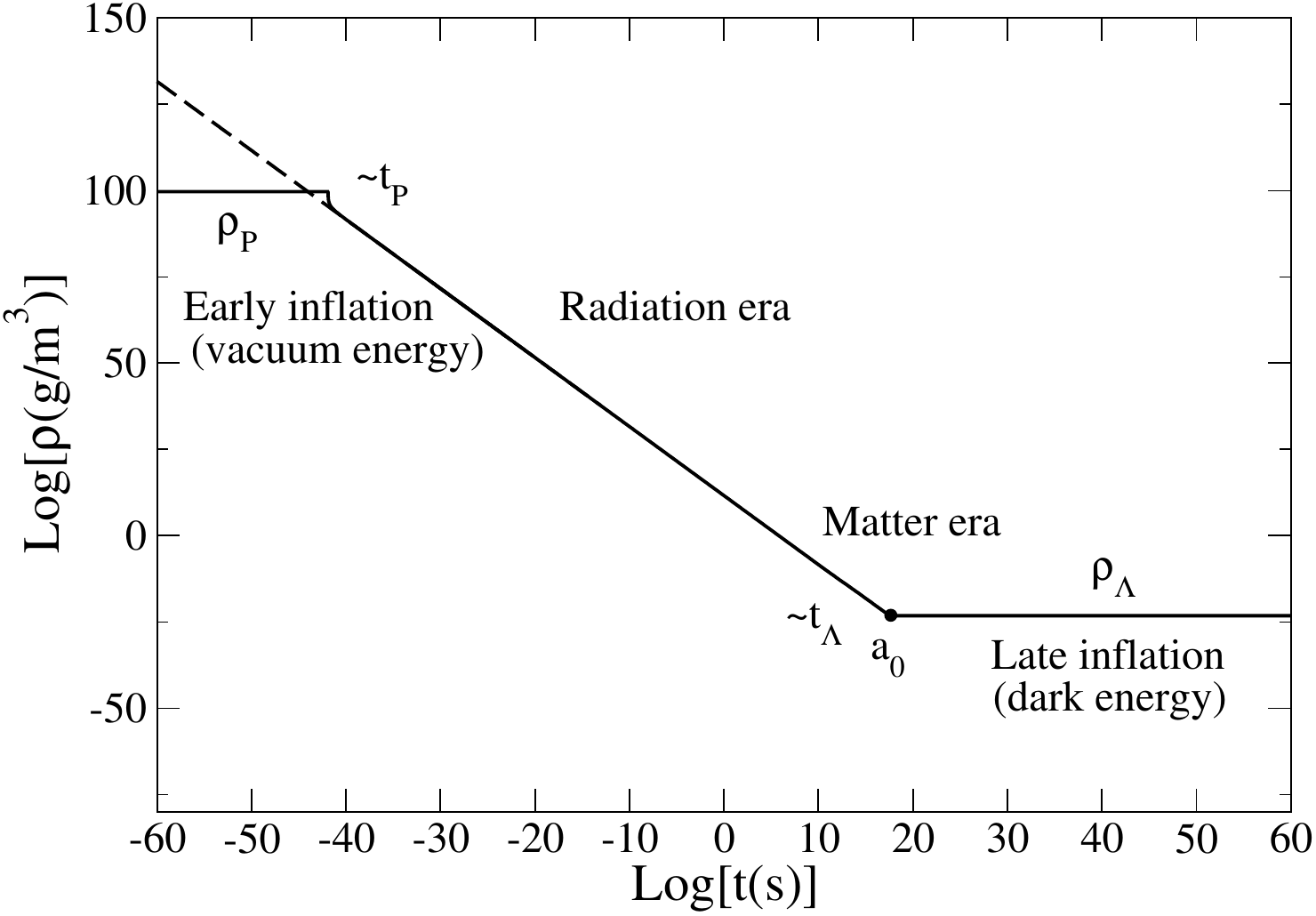}
\caption{Evolution of the density $\rho$ as a function of time in logarithmic
scales. The~density goes from a maximum value $\rho_{max}=\rho_P$ determined
by
the Planck constant (quantum mechanics) to a minimum value
$\rho_{min}=\rho_{\Lambda}$ determined by the cosmological constant (general
relativity). These two bounds, which are fixed by fundamental constants of
physics, are responsible for the early and late inflation of the universe. In
between, the density decreases as $t^{-2}$.}
\label{trhoLOGLOGcomplet}
\end{center}
\end{figure}

\subsection{The Cosmic Coincidence Problem}
\label{sec_ccp}

Let us consider the transition between the matter era ($\alpha=0$) and the dark
energy era described by Equation~(\ref{g20}). It is striking to note that the present
size of the universe
$a_0=0.302 l_{\Lambda}=1.32\times 10^{26}~{\rm m}$ is precisely of the order of
the transition
scale $a_2=0.204 l_{\Lambda}=8.95\times 10^{25}\, {\rm m}$ ($a_0=1.48 a_2$).  We
have $a_0\sim a_2\sim l_{\Lambda}$ and
$t_0\sim t_2\sim t_{\Lambda}$. Therefore, we live just at the
transition between the matter era and the dark energy era (see the bullets
in Figures
\ref{tasansradiationLOGLOGprime}--\ref{trhosansradiationLINLINprime}, \ref{taLOGLOGcomplet} and \ref{trhoLOGLOGcomplet}). In the
context
of the standard $\Lambda$CDM model, the way to state this observation is to say
that the present ratio $\Omega_{\Lambda,0}/\Omega_{m,0}=3.22$ between dark
energy and
matter is of order unity. Since the matter density changes as
$a^{-3}$ the ratio between matter and dark energy is of order unity
only during a ``brief'' period of time in the evolution of the universe. It
turns
out that we live precisely in this period.  This coincidence [Note 11: We should
be careful about this statement. It~is true that the order of magnitude of
$t_0=4.54\times 10^{17}~{\rm s}$ and $t_2=2.97\times 10^{17}~{\rm s}$ is the
same.
As a
result, these two times appear very close to each other on a logarithmic scale
(see Figures \ref{tasansradiationLOGLOGprime}
and~\ref{trhosansradiationLOGLOGprime}). However, on a linear scale, these
two times
differ by $t_2-t_0\sim 5\, {\rm Gyrs}$ (see Figures
\ref{tasansradiationLINLINprime} and \ref{trhosansradiationLINLINprime}).
In~this sense, there is no real coincidence.] is intriguing and often
referred to as the ``cosmic coincidence problem''~\cite{ccp}. Several
theories have been proposed to explain why
$\Omega_{\Lambda,0}/\Omega_{m,0}\sim 1$ \cite{ccpg1,ccpg2,ccpg3,ccpg4}. However,
this may be just a pure coincidence without deeper reason.
Life (and researchers inquiring about cosmology) may have
emerged at this special moment of the history of the universe, $\sim 14$ Gyrs
after the Big Bang, precisely at the epoch where
$\Omega_{\Lambda}/\Omega_m\sim 1$. Life cannot have emerged much
earlier because galaxies, stars and planets are not created, and it
may not persist much later because planets, stars, and galaxies will
die. Therefore, this coincidence is not really a
``problem''. Furthermore, we note that, in our model, matter and dark
energy are two manifestations of the {\it same} fluid described by a single
equation of state (\ref{late1}).  Since we are
considering a single ``dark'' fluid, the cosmic coincidence problem discussed
above \mbox{may be alleviated}.

\subsection{The Cosmological Constant Problem}

The cosmological constant $\Lambda$ is equivalent to a constant density
$\rho_{\Lambda}=\Lambda/(8\pi G)$ called dark energy. Its value resulting from
observations is $\rho_{\Lambda}=7.02\times 10^{-24}~{\rm g}/{\rm m}^3$.
It is
oftentimes argued that the cosmological density $\rho_{\Lambda}$  should
correspond to the  vacuum energy density due to quantum
fluctuations~\cite{vacuum1,vacuum2,vacuum3}. However, according to particle
physics and quantum field theory,
the vacuum energy density is of the order of the Planck density
$\rho_P=5.16\times
10^{99}~{\rm g}/{\rm m}^3$ which is $10^{122}$ times larger than the
cosmological
density. This leads to the so-called cosmological constant problem
\cite{weinbergcosmo,paddycosmo}.

Actually, as illustrated in Figure \ref{trhoLOGLOGcomplet}, the Planck density and
the cosmological density represent fundamental upper and lower density bounds
acting in the early and late universe, respectively. It is not surprising
therefore that they are so different: $\rho_{\Lambda}\ll \rho_P$. Because of
these bounds, the universe undergoes two phases of inflation. The inflation in
the early universe is due to quantum mechanics (Planck constant) and is related
to the Planck density $\rho_P$ (vacuum energy). The inflation in the late
universe is due to general relativity (cosmological constant) and is related to
the cosmological density  $\rho_{\Lambda}$ (dark energy). Quantum mechanics is
negligible in the late universe.  Therefore, we should not identify the
dark energy (or the cosmological constant) with the vacuum
energy. The cosmological constant should be interpreted as a fundamental
constant of physics. It applies to the very large universe (cosmophysics)
exactly like the Planck constant applies to the very small universe
(microphysics). Actually, there is a complete symmetry between the small and
large universe where $\hbar$ and $\Lambda$ play symmetric roles. Therefore, we
propose to interpret  the cosmological constant as a fundamental constant of
physics describing the cosmophysics (late universe) in the same sense that the
Planck constant describes the microphysics (early universe).

If this interpretation is correct, the origin of the  dark energy density
$\rho_{\Lambda}$ should not be sought in quantum mechanics, but in pure general
relativity. In this sense, the cosmological constant ``problem'' may be a false
problem. If $\Lambda$ is a fundamental constant of physics, independent from the
others, its value should not cause problem. It is fixed by nature, just like the
values of $G$, $c$, and $\hbar$. In other words, it has the value that it has!
One can just expect that $\rho_P$ is ``very large'' and $\rho_{\Lambda}$ is
``very small''. Of course, the origin of the cosmological constant still needs
to be understood by developing a theory of cosmophysics. In addition, it would
be important to understand why $\rho_P$ and $\rho_{\Lambda}$ represent upper
and
lower bounds, and if these bounds are as fundamental as, for example, the bound
on the velocity fixed by the speed of light.

\subsection{A Generalization of the $\Lambda$CDM Model Incorporating a Phase of
Early Inflation}
\label{sec_genx}

In the standard $\Lambda$CDM model, radiation, baryonic matter,
dark matter and dark energy are treated as four independent species
characterized by different equations of state \cite{bt}. The equation of state
is $p=\rho c^2/3$ for radiation, $p=0$ for baryonic matter and dark
matter, and $p=-\rho c^2$ for dark energy. We~then solve the equation of
continuity (\ref{b7}) for these four species in order to obtain the
relation
between their density and the scale factor,  \emph{i.e.},
$\rho_{rad}=\rho_{rad,0}/(a/a_0)^4$, $\rho_{B}=\rho_{B,0}/(a/a_0)^3$,
$\rho_{DM}=\rho_{DM,0}/(a/a_0)^3$, and $\rho_{\Lambda}=\rho_{\Lambda,0}$.
Finally, we add their individual densities
$\rho=\rho_{rad}+\rho_B+\rho_{DM}+\rho_{\Lambda}$ and substitute in the
Friedmann
equation (\ref{b9}). This leads to the standard equation
\cite{bt}:
\begin{equation}
\label{smx1}
\frac{H}{H_0}=\sqrt{\frac{\Omega_{rad,0}}{(a/a_0)^4}+\frac{\Omega_{B,0}}{
(a/a_0)^3}
+\frac{\Omega_{DM,0}}{(a/a_0)^3}+\Omega_{\Lambda,0}}
\end{equation}
with $\Omega_{rad}+\Omega_{B}+\Omega_{DM}+\Omega_{\Lambda}=1$. From observations
$\Omega_{rad,0}=8.48\times 10^{-5}$, $\Omega_{B,0}=0.0455$,
$\Omega_{DM,0}=0.1915$,
$\Omega_{m,0}=0.237$, and $\Omega_{\Lambda,0}=0.763$ \cite{bt}. The
$\Lambda$CDM model provides a good description of the universe after the Planck
era. It describes the radiation era, the matter era, and the dark energy era
(late accelerating
expansion). However, it exhibits a primordial singularity at
$t=0$ (Big Bang). Furthermore, it does not describe the phase of early
inflation. This phase is usually described by another theory, e.g., a
scalar field theory \cite{guth1,guth2,guth3,linde}, that is then connected to
the $\Lambda$CDM
model.

Our approach is physically different from the $\Lambda$CDM model
and generalizes it. In our model, based on
the quadratic equation of state (\ref{g1}), the evolution of the scale
factor is
given by Equation~(\ref{g6}). For $a_1=0$, Equation~(\ref{g6}) takes a form similar to
Equation~(\ref{smx1}) of the $\Lambda$CDM model. For $\alpha=1/3$, it
describes the transition between the radiation era and the dark energy era. For
$\alpha\simeq 0$, it describes the transition between the matter era (having
possibly a small temperature) and the dark energy era. However, this approach is
incomplete. Indeed, the equation of state (\ref{g1}) is not able to
describe the
transition between the radiation era and the matter era. It can only describe
one of these two phases depending on the chosen value of $\alpha$. This is not a
drawback of the model. Indeed, there is no reason to ``unify'' radiation and
matter in a single equation of state. Radiation and matter are different
entities that should be treated as two different species as in the standard
$\Lambda$CDM model. By contrast, our approach suggests that vacuum energy,
radiation, and dark energy may be the manifestation of a unique form of
``generalized radiation''. Therefore, by selecting $\alpha=1/3$
in Equation~(\ref{g1}) we obtain an equation of state
\begin{equation}
\label{g1select}
p=-\frac{4\rho^2}{3\rho_P} c^2+\frac{1}{3}\rho
c^2-\frac{4}{3}\rho_{\Lambda} c^2
\end{equation}
that unifies vacuum energy, radiation, and dark energy. This leads to the
density (see Equation~(\ref{g3})):
\begin{equation}
\label{g3select}
\rho=\frac{\rho_{rad,0}}{(a/a_0)^{4}+
(a_1/a_0)^{4}}+\rho_{\Lambda}.
\end{equation}
This ``generalized radiation'' is one component
of the Friedmann equation (\ref{b9}). Then, the contribution of baryonic matter
and dark matter is introduced in the Friedmann equation (\ref{b9}) as other
species described by the equation of state $p=0$  leading to a density
$\rho_{m}=\rho_{m,0}/(a/a_0)^3$ (or by the equation of
state $p=\alpha \rho c^2$ with
$\alpha\ll 1$ leading to a density
$\rho_{m}=\rho_{m,0}/(a/a_0)^{3(1+\alpha)}$
if we take into account the effect of a small temperature). As a result, we
obtain the complete equation [Note 12: By selecting
$\alpha=0$ in Equation~(\ref{g1}) we could unify vacuum energy, matter and
dark
energy, then introduce
radiation as an independent species with an equation of state $p=\rho c^2/3$
leading to a density $\rho_{rad}=\rho_{rad,0}/(a/a_0)^4$. However, we believe
that it is more relevant to unify vacuum energy, radiation and dark energy, and
treat baryonic matter and dark matter as independent species. We note that
baryonic matter and dark matter must be introduced at sufficiently late time
in Equation~(\ref{smx2}) in
order to avoid a spurious divergence when $a\rightarrow 0$ due to the fact that
$a_1\neq 0$ in our model. For exemple, we can replace $\Omega_{B,0}$ and
$\Omega_{DM,0}$ by $\Omega_{B,0}H(a-a_1)$ and  $\Omega_{DM,0}H(a-a_1)$, where
$H$ is the Heaviside function.]:
\begin{equation}
\label{smx2}
\frac{H}{H_0}=\sqrt{\frac{\Omega_{rad,0}}{(a/a_0)^4+(a_1/a_0)^4}+\frac{\Omega_{B
,0}}{(a/a_0)^3}
+\frac{\Omega_{DM,0}}{(a/a_0)^3}+\Omega_{\Lambda,0}}
\end{equation}
with $a_1/a_0=(\rho_{rad,0}/\rho_P)^{1/4}$. For $a_1=0$, our
model gives the same result as the
$\Lambda$CDM model of Equation~(\ref{smx1}), although the justification is
different. This
is a good thing because it ensures that our model produces results that agree
with the observations after the Planck era (\emph{i.e.}, for $a\gg a_1$). For~$a_1\neq
0$, our model generalizes the $\Lambda$CDM model of Equation~(\ref{smx1}) by
naturally incorporating
 a phase of early inflation and avoiding the primordial singularity, as
discussed in Section \ref{sec_connection}. Equation~(\ref{smx2}) is
interesting because it describes the early inflation (Planck era), the radiation
era, the matter era, and the dark energy era (de Sitter era) in a
self-consistent manner. The evolution of the scale factor with time is obtained
by solving the first order differential Equation (\ref{smx2}). This yields
\begin{equation}
\label{pr8}
\int_{a_i/a_0}^{a/a_0} \frac{dx}{x\sqrt{\frac{\Omega_{rad,0}}{x^4+
(a_1/a_0)^4}+\frac{(\Omega_{B,0}+\Omega_{DM,0})}{x^3}+\Omega_{\Lambda,0}}}=H_0
t,
\end{equation}
where $a_i=a_1=0$ in the standard $\Lambda$CDM model and $a_i=l_P$ in our model.
This is how
we have obtained the solid curves of Figures \ref{taLOGLOGcomplet} and
\ref{trhoLOGLOGcomplet}. They provide the complete evolution of the universe
from the early inflation to the late acceleration. The age of the universe is
\begin{equation}
\label{pr9}
t_{0}=\frac{1}{H_0}\int_{0}^{1}
\frac{dx}{x\sqrt{\frac{\Omega_{rad,0}}{x^4}+\frac{\Omega_{B,0}}{x^3}+\frac{
\Omega_{DM,0}}{x^3}
+\Omega_{\Lambda,0}}}.
\end{equation}
Of course, for the determination of the age of the universe we can neglect the
vacuum energy era and take $a_1=0$. We obtain the standard result $t_{0}=1.03\,
H_0^{-1}=4.53\times 10^{17}\, {\rm s}=14.4 \, {\rm
Gyr}$ [Note 13: The Hubble constant is usually written as
$H_0=2.268\, h_7\times 10^{-18}~{\rm s}^{-1}$ where the dimensionless parameter
$h_7$
is about $10\%$ of unity \cite{bt}. For simplicity, we have taken $h_7=1$ in
the
numerical applications. The current value is $h_7=1.05\pm 0.05$. If we take
$h_7=1.05$, the age of the universe is $t_0=13.7 \, {\rm Gyr}$.].  Actually,
we
find the same result if we neglect radiation and use the analytical expression
of 
Equation~(\ref{g30}) with $\alpha=0$ instead of Equation~(\ref{pr9}).

\subsection{Discussion}
\label{sec_discu}

The main novelty of our paper is the quadratic equation of state (\ref{intro2}).
To the best of
our knowledge, this equation of state has
never been introduced before. A nice feature of this equation of state is that
it
incorporates both the Planck density $\rho_P$ (vacuum energy) and the
cosmological density $\rho_{\Lambda}$
(dark energy). We are not aware of any equation of state of that kind. For
example, the
usual equation of state $p=-\rho c^2$ used to model the primordial inflation or
the late acceleration of the universe leads to a constant density but the value
of this density
cannot be read from the equation of state itself and  it has to be adapted,
rather
arbitrarily, to the situation (e.g., one takes $\rho_P$ to describe the early
inflation
and $\rho_{\Lambda}$ to describe the late acceleration).  In
Equation~(\ref{intro2})
there is no
arbitrariness since this equation of state explicitly contains $\rho_P$ and
$\rho_{\Lambda}$
as parameters. This equation of state naturally imposes $\rho=\rho_P$ in the
early universe
and $\rho=\rho_{\Lambda}$ in the late universe.  These two densities appear as
fundamental
upper and lower bounds in the model. In  this sense, the equation of state
(\ref{intro2}) with $\alpha=1/3$ suggests
a sort of ``unification'' of vacuum  energy, radiation, and dark energy. This
may be the
same entity, a ``generalized radiation'', described by a unique equation  of
state (\ref{intro2}),
but its density is very different in the early  and in the late universe causing
an early
inflation and a late acceleration.

Our aim was to show that this equation of state is ``viable'' and
that it can reproduce the basic features of the universe. To that purpose, we
considered its limiting behavior in the early and in the late universe:

(i) In the late universe, the equation of state (\ref{intro2})
reduces to Equation (\ref{late1}) which can be viewed as a generalized
Chaplygin/polytropic gas model with $n_{l}=-1$. This model is physically
different from the usual $\Lambda$CDM model but it gives the same
results. Therefore, it is able to reproduce the observations. This shows that
the equation of state (\ref{intro2}) is viable in the late universe.

(ii) In the early universe, the equation of state (\ref{intro2})
reduces to Equation~(\ref{early1}) which can be viewed as a generalized
Chaplygin/polytropic gas model with $n_{e}=+1$. This model is different from the
usual models of inflation \cite{guth1,guth2,guth3,linde} but it gives similar
results. In
particular, it can account for an~increase of the scale factor by about $29$
orders of magnitude in less than $10^{-42}\, {\rm s}$. It would
be important to
determine whether this equation of state can reproduce the observed primordial
fluctuations and power spectrum. This problem has been considered recently in
\cite{setare1,setare2,fg}. The conclusion of these works is that the polytropic
equation of state represents a robust and interesting scenario to study the
evolution of the universe.
Therefore, the equation of state (\ref{intro2}) may be viable in the
early
universe.

(iii) In between, the equation of state (\ref{intro2}) reduces to
a linear equation of state $p=\alpha\rho c^2$
describing a phase of decelerating expansion (when $\alpha>-1/3$).

As a whole, the equation of state (\ref{intro2}) describes an
early and late acceleration connected by a~deceleration phase.
Therefore, it accounts for the basic features of the universe and it may be
viable. The~drawback of the $\Lambda$CDM
model is that it does not describe a phase of early inflation and 
displays instead a primordial singularity (Big Bang).
The equation of state (\ref{intro2}) returns the result of the
$\Lambda$CDM
model for sufficiently late times but it also naturally
incorporates  a phase of early inflation. \mbox{As a result}, it describes the
early inflation and the late
acceleration simultaneously. This is the main interest of this model.

Interestingly, the phase of early inflation and the phase of
late acceleration are described by two~symmetric (or dual) Chaplygin/polytropic
gas models with index $+1$ and $-1$, respectively. The~quadratic equation of
state (\ref{intro2}) combines these two symmetric models into a single
equation
of state. As a result, this equation of state displays both a phase of early
inflation driven by the Planck density $\rho_P$ and a phase of late acceleration
driven by the cosmological density $\rho_{\Lambda}$. These two phases are
bridged by a phase of deceleration. The
pressure is successively negative
(vacuum energy), positive ($\alpha$-era),
and negative again (dark
energy).

There has been many works attempting to unify two phases
(deceleration $+$ acceleration) in the evolution of the universe with a unique
equation of state of the Chaplygin form
\cite{chaplygin1,chaplygin2,chaplygin3,chaplygin4,chaplygin5,
chaplygin6,chaplygin7,chaplygin8,chaplygin9}. We have  proposed in
this paper to unify {\it three} phases (inflation $+$ deceleration $+$
acceleration) with the
quadratic equation of state (\ref{intro2}), or more generally with
Equation
(\ref{newg1}), which can be viewed as a
generalized
Chaplygin gas model.

\section{Conclusions}

We have constructed a cosmological model based on the quadratic
equation of state (\ref{intro2}). This~equation of state describes the
evolution
of a universe presenting a phase of early inflation (Planck era), a phase of
decelerating expansion  ($\alpha$-era), and a phase of late accelerating
expansion (de Sitter era). An interest of this model is its simplicity (while
being already quite rich) and the fact that it admits a fully analytical
solution (see Equation~(\ref{hel8})). It provides a particular solution of the
Friedmann equations. In~addition, it is in qualitative agreement with the evolution of our own universe.
Finally, it admits a~scalar field interpretation based on an
inflaton, quintessence, or tachyonic field.

This model does not
present any singularity and exists ``eternally'' in the past and in the future,
so that it has no origin nor end (we call it aioniotic universe). In
particular, the phase of
early inflation avoids the Big Bang singularity and replaces it by a sort of
second order phase transition where the non-zero value of the Planck constant
$\hbar$ plays the same role as finite size effects in statistical mechanics. Of
course, it is probably
incorrect to extrapolate the results in the infinite past since our model is
purely classical, or semi-classical, and does not take quantum gravity into
account. In our model, the universe starts from $t=-\infty$ but, for $t<0$, its
size  is less than the Planck length $l_P=1.62\times 10^{-35}\, {\rm m}$. In the
Planck era, the classical Einstein equations may be incorrect and should be
replaced by a theory of quantum gravity that still has to be constructed.

Another interest of this model is that it describes
the early universe and the late universe in a~symmetric manner. The early
universe is described by a polytrope $n=+1$ and the late universe by a~polytrope
$n=-1$. The mathematical formulae are then strikingly symmetric (we sum the
inverse of the densities in the early universe and we sum the densities in the
late universe). Furthermore, the Planck density $\rho_P$ in the early universe
plays the
same role as the cosmological density $\rho_{\Lambda}$ in the late universe.
They~represent
fundamental upper and lower density bounds differing by $122$~orders of
magnitude. They~lead to phases of early and late inflation with very different
timescales $t_P=1/(G\rho_P)^{1/2}=5.39\times 10^{-44}~{\rm s}$ and
$t_{\Lambda}=1/(G\rho_{\Lambda})^{1/2}=1.46\times 10^{18}~{\rm s}$.
The densities $\rho_P$ and $\rho_{\Lambda}$ (together with $\alpha$) appear
as
the coefficients of the  equation of state (\ref{intro2}). Therefore,
this
equation of state
provides a ``unification'' of vacuum energy and dark energy.

A limitation of this equation of state is that it cannot describe
the transition between the radiation era and the  matter era because they are
both
described by a linear equation of state $p=\alpha\rho c^2$ with $\alpha=1/3$
and
$\alpha=0$ respectively. Therefore, the equation of state
(\ref{intro2}) with
$\alpha=1/3$ describes a~universe without matter while the equation of state
(\ref{intro2}) with $\alpha=0$ describes a~universe without
radiation. A better
model can be achieved by letting $\alpha(t)$ vary with time and decrease
smoothly from $1/3$ to $0$. Alternatively, we can ``unify'' vacuum energy $+$
radiation $+$ dark energy into a~``generalized radiation'' described by the
equation of
state (\ref{intro2}) with $\alpha=1/3$ and treat baryonic matter and
dark matter
as independent species described by the equations of state $p_{B}=0$ and
$p_{DM}=0$. This general approach (see Section \ref{sec_genx}) allows us to
describe the complete evolution of the universe (inflation $+$ radiation $+$
matter $+$ acceleration). It leads to an interesting  generalization
of the $\Lambda$CDM model where Equation~(\ref{smx1}) is replaced by  Equation
(\ref{smx2}). As a result, our model generalizes the standard $\Lambda$CDM model
by incorporating naturally a phase of early inflation that avoids the primordial
singularity. Furthermore, it describes the early inflation and the late
acceleration simultaneously with a single equation of state that unifies vacuum
energy and dark energy, and a single
scalar field potential that unifies the inflaton potential and the quintessence
potential (see Appendix \ref{sec_sft}).

Dark matter is usually described by a pressureless equation of state $p=0$. This
leads to the famous Einstein-de Sitter (EdS) model. However, there are
indications that dark matter may be
described by an isothermal equation of state $p=\alpha\rho c^2$ with
$\alpha\ll
1$
\cite{muller}. Therefore, it is useful to provide general results valid for
arbitrary values of $\alpha$ as we have done here. This allows us to describe
either radiation ($\alpha=1/3$), pressureless matter ($\alpha=0$), or
isothermal
matter ($\alpha\neq 0$).

In this paper, for definiteness, we have selected the indices
$n_e=+1$ and $n_l=-1$  in Equation~(\ref{newg1}) to describe the early and
the late
universe respectively. As we have seen, the equation of state (\ref{newg1}) with
$n_l=-1$ leads to
results that coincide with the $\Lambda$CDM model in the late universe. This
equivalence is not trivial since our approach is fundamentally different from
the $\Lambda$CDM model (in particular this coincidence is not true anymore for
$n_l\neq -1$). Since the $\Lambda$CDM model provides a good description of the
universe after the Planck era, this implies that observations tend to favor the
value
$n_l=-1$ of the polytropic index. This is a reason why we have selected this
index. However, we can obtain more general models by taking $n_l$ different from
$-1$. In particular, a value of $n_l$  close to $-1$ may be consistent with the
observations and may improve upon the $\Lambda$CDM model.   In this respect,
Asadzadeli {\it et al}. \cite{karami} have shown that the observations constrain
the
index $n_l$ to be in the
range $-1.05^{+0.08}_{-0.08}$
or $-1.05^{+0.15}_{-0.16}$ at $68.3\%$ and $95.4\%$ confidence levels,
respectively. On the other hand, we have selected the index
$n_e=+1$  in Equation~(\ref{newg1}) to describe the inflation in the early universe.
More general models of
inflation can be obtained
by selecting other values of $n_e$. The value of $n_e$ could be
determined from the measurements of the cosmic microwave background (CMB)
radiation. Therefore, possible
extensions of our study would be to consider the generalized equation of
state (\ref{newg1}) with arbitrary values of $n_e$ and $n_l$. General~results
valid
for the asymptotic expressions (Equations (\ref{newearly1}) and
(\ref{newearly1bis})) of the
equation of state (\ref{newg1}) in the early and late universe are given
in
\cite{chavanis1,chavanis2,chavanis3}.

A weakness of our work is that we have not given a justification
of the quadratic equation of state (\ref{intro2}) [Note 14: It would
be of considerable interest to justify this equation of state
from a microscopic model of radiation since we
argued that this equation of state provides an effective model of ``generalized
radiation''. The quadratic equation of state (\ref{early1}) and the
analytical
solution of Equation~(\ref{early8}) describing the phase of inflation were
first given
in \cite{aabec} in relation to Bose-Einstein condensates (BECs) with
an attractive self-interaction (negative scattering length $a_s<0$). Although
this BEC model may be too simplistic (see the discussion in Appendix B of
\cite{chavanis1}), it
shows how
a quadratic equation of state with a negative pressure may arise in practice.
We may also draw an analogy between the BEC model and the primeval
atom of Lema\^itre \cite{primeval}.].
However, since this equation
of state is connected to the  Chaplygin gas model, there is some hope that this
equation of state could arise from fundamental theories since the original
Chaplygin gas model
\cite{chaplygin1,chaplygin2,chaplygin3,chaplygin4,chaplygin5,
chaplygin6,chaplygin7,chaplygin8,chaplygin9} is motivated by some works on
string
theory. It can be obtained from the Nambu-Goto (or Born-Infeld) action for
$d$-branes moving in a
$(d+2)$-dimensional spacetime in the light-cone parametrization [Note 15: We
also note that a theoretical justification
of a
polytropic equation of state of the form $p=k\rho^{1+1/n}c^2$  with $n\in
\mathbb{N}$ for a perfect
relativistic fluid in cosmology has been given
recently by Kazinski  \cite{kazinski} in relation to the quantum
gravitational
anomaly.]. We may
also argue that the quadratic
equation of state (\ref{intro2}) (and its generalization
in Equation~(\ref{newg1})) is the ``simplest''
equation of state that we can imagine in order to obtain a non singular model of
universe (see the discussion in {Section~7.4} of~\cite{chavanis2}). This ``simplest''
model is in
agreement with the known properties of our universe. It is interesting that
these properties can be reproduced by an equation of state of the form
$ax^2+bx+c$ where the three~coefficients are related to $\hbar$, $\alpha$, and
$\Lambda$.

In this paper, the quadratic equation of state (\ref{intro2})
has been
introduced heuristically in order to obtain a cosmological model that describes
simultaneously a phase of inflation, a phase of decelerating expansion, and a
phase of acceleration. The Planck constant $\hbar$ and the cosmological
constant $\Lambda$ are interpreted as two fundamental constants fixed by nature
that appear in the coefficients of this equation of state. In a sense, this
equation of state can be viewed as a simple interpolation formula between
different, well-motivated, equations of state. Interpolation formulae are often
useful in physics and the equation of state (\ref{intro2}) is also
interesting in
this respect.

The next step would be to study the growth of
perturbations in our model. In particular, it would be
important to confront the model of inflation based on the polytropic equation
of state (\ref{newearly1a}) to the Planck 2015 data set. In this
respect, it may be necessary to use the
scalar field interpretation instead of the hydrodynamic interpretation because
it is known that, although the hydrodynamic and scalar field equations
coincide at the background level,
they may show differences at the level of the
perturbations. A first study has
been made in this direction by Freitas and Gon\c calves \cite{fg} using our
model
of inflation.
They studied the  primordial quantum
fluctuations in the slow-roll approximation and calculated the scalar
perturbations, the scalar power spectrum, and the spectral index. They
obtained encouraging results, but some of their results need to be
further developed. This will be considered in future works.

{\it Remark:} In this paper, we have proposed to describe the
complete evolution of the universe (early and late inflation) by the quadratic
equation of state (\ref{intro2}). A quadratic equation of state aiming
at
describing the late acceleration of the universe has been considered in
\cite{ghosts12,ghosts13,ananda,linder} and compared to the observations by
Sharov \cite{sharov}. Sharov mentions that a quadratic equation of state with a
positive coefficient in front of $\rho^2$ can lead to past singularities. In the
present paper, we have considered a~quadratic equation of
state with a small negative coefficient in front of $\rho^2$ (of the order
of $\rho_P^{-1}$), so that the quadratic term accounts for the early inflation,
not for the late acceleration. This is why we do not have past singularities
in our model (the universe exits at any time in the past). The possibility of
past singularities when the
coefficient in front of $\rho^2$ is positive is discussed in \cite{chavanis1}.

\conflictofinterests{Conflicts of Interest}

The author declares no conflict of interest.

\appendix{\section*{\noindent Appendix}}

\setcounter{figure}{0}
\renewcommand\thefigure{A\arabic{figure}}

\section{Scalar Field Theory}
\label{sec_sft}

In this Appendix, we determine the canonical scalar field
potential and the tachyonic
potential corresponding to the quadratic equation of state (\ref{g1}).
Interestingly, this scalar field theory describes, with a single potential, the
whole evolution of the
universe, from the early inflation to the late acceleration, passing through a
phase of  algebraically decelerating expansion.

\subsection{Canonical Scalar Field}

We consider a canonical scalar field defined by Equations
(\ref{early13b}) and (\ref{early14}).
From Equation~(\ref{early14}), we get
\begin{eqnarray}
\label{newsft1}
\dot\phi^2=(w+1)\rho c^2,
\end{eqnarray}
where we recall that $w=p/\rho c^2$. Using $\dot\phi=(d\phi/da)Ha$ and the
Friedmann equation (\ref{b9}) valid for a~flat universe, we find that the
relation between the scalar field and the
scale factor is given by
\begin{eqnarray}
\label{sft1}
\frac{d\phi}{da}=\left (\frac{3c^2}{8\pi G}\right )^{1/2}\frac{\sqrt{1+w}}{a}.
\end{eqnarray}
On the other hand, according to Equation~(\ref{early14}), the potential of the
scalar
field is given by
\begin{eqnarray}
\label{sft2}
V=\frac{1}{2}(1-w)\rho c^2.
\end{eqnarray}
For the quadratic equation of state (\ref{g1}), we have
\begin{eqnarray}
\label{sft3}
w=-(\alpha+1)\frac{\rho}{\rho_P}+\alpha-(\alpha+1)\lambda\frac{
\rho_P}{\rho}
\end{eqnarray}
and
\begin{eqnarray}
\label{sft4}
\frac{\rho}{\rho_P}=\frac{1}{R^{3(\alpha+1)}+1}+\lambda,
\end{eqnarray}
where $R=a/a_1$ and $\lambda=\rho_{\Lambda}/\rho_P=1.36\times 10^{-123}$.
With
the
change of variables
\begin{equation}
\label{sft5}
x=R^{3(\alpha+1)/2},\qquad \psi=\left (\frac{8\pi G}{3c^2}\right
)^{1/2}\frac{3\sqrt{\alpha+1}}{2}\phi,
\end{equation}
we obtain after simple calculations (using the fact that  $\lambda\ll 1$ to
simplify some terms):
\begin{equation}
\label{sft6}
\psi=\int_0^x\frac{ds}{(\lambda
s^2+1)^{1/2}(s^2+1)^{1/2}}={\rm sc}^{-1}(x,1-\lambda),
\end{equation}
\begin{equation}
\label{sft7}
V=\frac{1}{2}\rho_P c^2\frac{2+(1-\alpha)x^2+2\lambda x^4}{(x^2+1)^2},
\end{equation}
where ${\rm sc}$ is the Jacobian Elliptic function. We
have ${\rm
sc}^{-1}(x,1-\lambda)=-iF(i\sinh^{-1}(x),\lambda)$
where $F(\phi,\lambda)=\int_0^{\phi}(1-\lambda\sin^2\theta)^{ -1/2}\,
d\theta=\int_0^{\sin(\phi)}(1-t^2)^{-1/2}(1-\lambda t^2)^{-1/2}\, dt$ is the
Elliptic integral of the first kind. We have taken $\psi=0$ when $x=0$.
Equations
(\ref{sft6}) and (\ref{sft7}) define the potential $V(\psi)$ in parametric form
with parameter $x$ going from $0$ to $+\infty$. The scalar
field goes from $\psi=0$ to
$\psi_{max}=\int_0^{+\infty}{ds}/[(\lambda
s^2+1)^{1/2}(s^2+1)^{1/2}]=K(1-\lambda)$
where $K$ is the
complete Elliptic integral of the first kind [Note 16: For $\lambda\ll 1$, we
have $\psi_{max}=K(1-\lambda)\simeq (1/2)\ln(16/\lambda)$ as will be found
below
by a different method.]. Eliminating $x$ between Equations (\ref{sft6}) and
(\ref{sft7}), we get
\begin{equation}
\label{unif1}
V(\psi)=\frac{1}{2}\rho_P c^2\frac{2+(1-\alpha){\rm
sc}(\psi,1-\lambda)^2+2\lambda {\rm
sc}(\psi,1-\lambda)^4}{\left\lbrack {\rm
sc}(\psi,1-\lambda)^2+1\right\rbrack^2}\qquad
(0\le \psi\le\psi_{max}).
\end{equation}
In the early universe, we can
neglect the dark energy ($\lambda=0$) so that Equations (\ref{sft6}) and (\ref{sft7})
reduce~to
\begin{equation}
\label{sft8}
\psi=\sinh^{-1}(x),
\end{equation}
\begin{equation}
\label{sft9}
V=\frac{1}{2}\rho_P c^2\frac{2+(1-\alpha)x^2}{(x^2+1)^2}.
\end{equation}
Equations (\ref{sft8}) and (\ref{sft9}) are equivalent to Equations
(\ref{early15b}) and (\ref{early15}) defining
the inflaton potential
\begin{eqnarray}
\label{a}
V(\psi)=\frac{1}{2}\rho_P c^2
\frac{(1-\alpha)\cosh^2\psi+\alpha+1}{\cosh^4\psi} \qquad (\psi\ge 0).
\end{eqnarray}
In the late universe, we
can neglect the vacuum energy (which amounts to taking $x\gg 1$) so that Equations (\ref{sft6}) and (\ref{sft7}) reduce to
\begin{equation}
\label{sft10}
\psi-\psi_{max}=-\sinh^{-1}\left (\frac{1}{\sqrt{\lambda}x}\right ),
\end{equation}
\begin{equation}
\label{sft11}
V=\frac{1}{2}\rho_P c^2 \left (\frac{1-\alpha}{x^2}+2\lambda\right ),
\end{equation}
where $\psi_{max}$ is a constant of integration. Equations
(\ref{sft10}) and (\ref{sft11}) are equivalent to
Equations (\ref{hel2}) and~(\ref{hel3}) defining the quintessence potential
\begin{eqnarray}
\label{b}
V(\psi)=\frac{1}{2}\rho_{
\Lambda}c^2 \left\lbrack
(1-\alpha)\cosh^2(\psi_{max}-\psi)+\alpha+1\right\rbrack\qquad
(\psi\le\psi_{max}),
\end{eqnarray}
where
we have arbitrarily taken $\psi_{max}=0$ in Equations (\ref{hel2}) and
(\ref{hel3}).
Therefore, the
scalar field potential defined by Equation
(\ref{unif1}) provides a unification of the inflaton and
quintessence potentials defined by Equations (\ref{a}) and (\ref{b})
respectively.

The constant $\psi_{max}$ can be obtained by matching the solutions
from Equations
(\ref{sft8}), (\ref{sft9}), (\ref{sft10}) and~(\ref{sft11}) in the intermediate
region of algebraic expansion ($\alpha$-era). First, we note that $V=\rho_P
c^2$
for $\psi=0$
($x=0$) and $V=\rho_{\Lambda} c^2$ for $\psi=\psi_{max}$ ($x\rightarrow
+\infty$). For $x\gg 1$, Equations  (\ref{sft8}) and~(\ref{sft9}) reduce
to
$\psi\sim\ln(2x)$ and $V\sim (1/2)\rho_P c^2(1-\alpha)x^{-2}$ yielding
\begin{equation}
\label{sft9b}
V(\psi)\simeq 2\rho_P c^2(1-\alpha)e^{-2\psi}\qquad (\psi\gg 1).
\end{equation}
On the other hand, for $x\ll 1$, Equations  (\ref{sft10}) and (\ref{sft11})
reduce to
$\psi_{max}-\psi\sim\ln(2/\sqrt{\lambda}x)$ and $V\sim (1/2)\rho_P
c^2(1-\alpha)x^{-2}$ yielding
\begin{equation}
\label{sft9c}
V(\psi)\simeq \frac{1}{8}\rho_P c^2(1-\alpha)\lambda
e^{2(\psi_{max}-\psi)}\qquad (\psi\ll \psi_{max}).
\end{equation}
This exponential potential corresponds to the potential of a scalar
field with a linear equation of state $p=\alpha\rho c^2$ (see, e.g., {Section
8.1} of
\cite{chavanis2}). Comparing Equations (\ref{sft9b}) and (\ref{sft9c}), we find that
\begin{equation}
\psi_{max}=\frac{1}{2}\ln\left (\frac{16}{\lambda}\right )\simeq 142.84.
\end{equation}
We note that $\psi_{max}$ is of the order of the famous number
$\log(\rho_P/\rho_{\Lambda})\simeq 123$ appearing in relation to the
cosmological constant problem \cite{weinbergcosmo,paddycosmo}. Using Equations
(\ref{hel8}) and
(\ref{sft6}), we can obtain the evolution of the scalar field with time. Using
the asymptotic results of Sections \ref{sec_early} and \ref{sec_late}, we find that
the
scalar  field evolves as $\psi\simeq (l_P/a_1)^{3(\alpha+1)/2}{\rm
exp}[(3/2)(\alpha+1)(8\pi/3)^{1/2}t/t_P]$ in the vacuum energy era, as
$\psi\simeq \ln(t/t_P)+(1/2)\ln(8\pi/3)+\ln[3(\alpha+1)]$ in the
$\alpha$-era,
and as $\psi\simeq \psi_{max}-2{\rm
exp}[-(3/2)(\alpha+1)(8\pi\lambda/3)^{1/2}t/t_P]$ in the dark energy era.
Some representative curves are shown in Figures \ref{xV}--\ref{psit} for
$\alpha=0$.

\begin{figure}[H]
\begin{center}
\includegraphics[clip,scale=0.5]{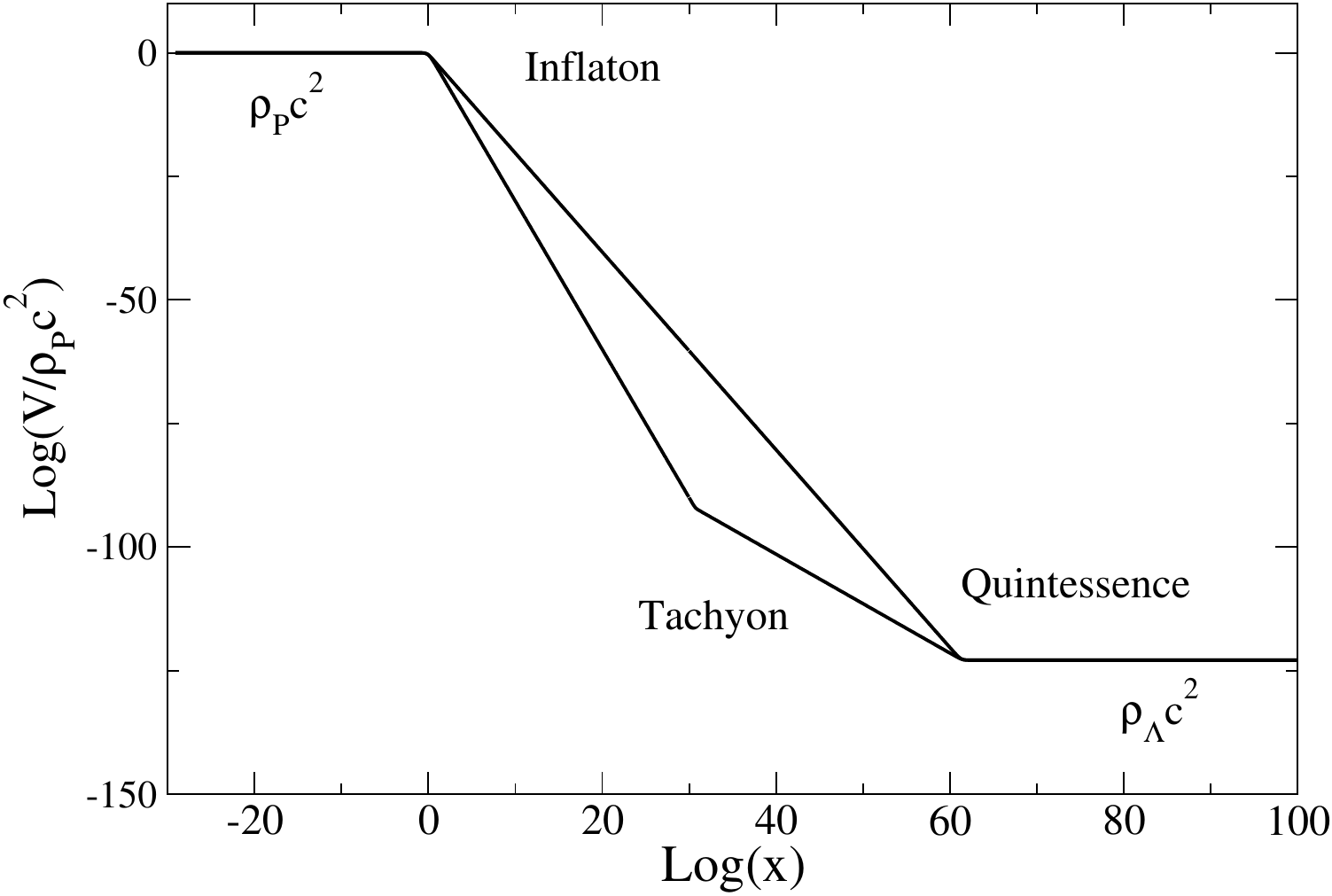}
\caption{Potential $V(x)$ in logarithmic coordinates.}
\label{xV}
\end{center}
\end{figure}

\begin{figure}[H]
\begin{center}
\includegraphics[clip,scale=0.5]{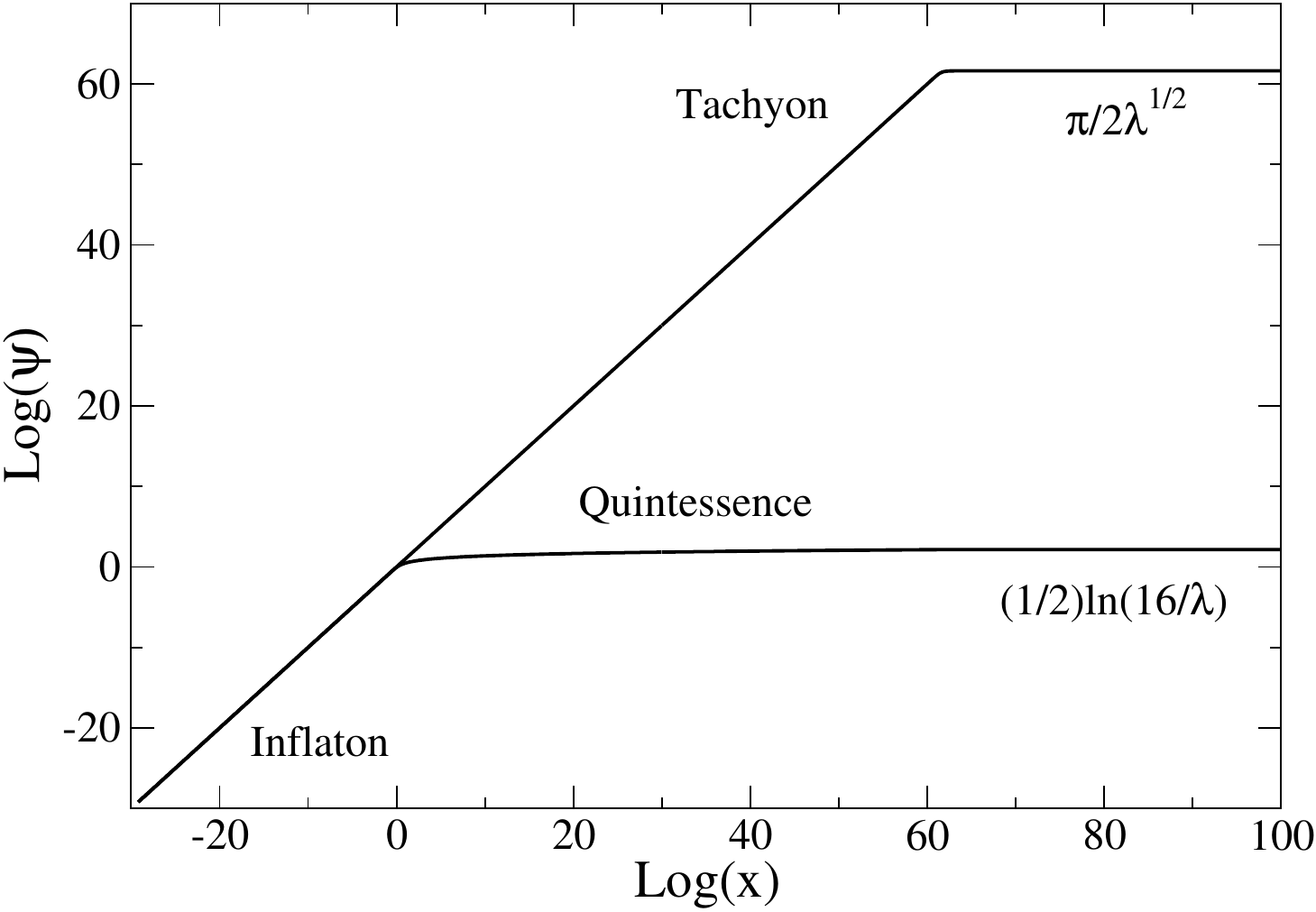}
\caption{Evolution of the scalar field $\psi$ with the normalized scale factor
$x$ in logarithmic~coordinates.}
\label{psix}
\end{center}
\end{figure}
\vspace{-12pt}

\begin{figure}[H]
\begin{center}
\includegraphics[clip,scale=0.5]{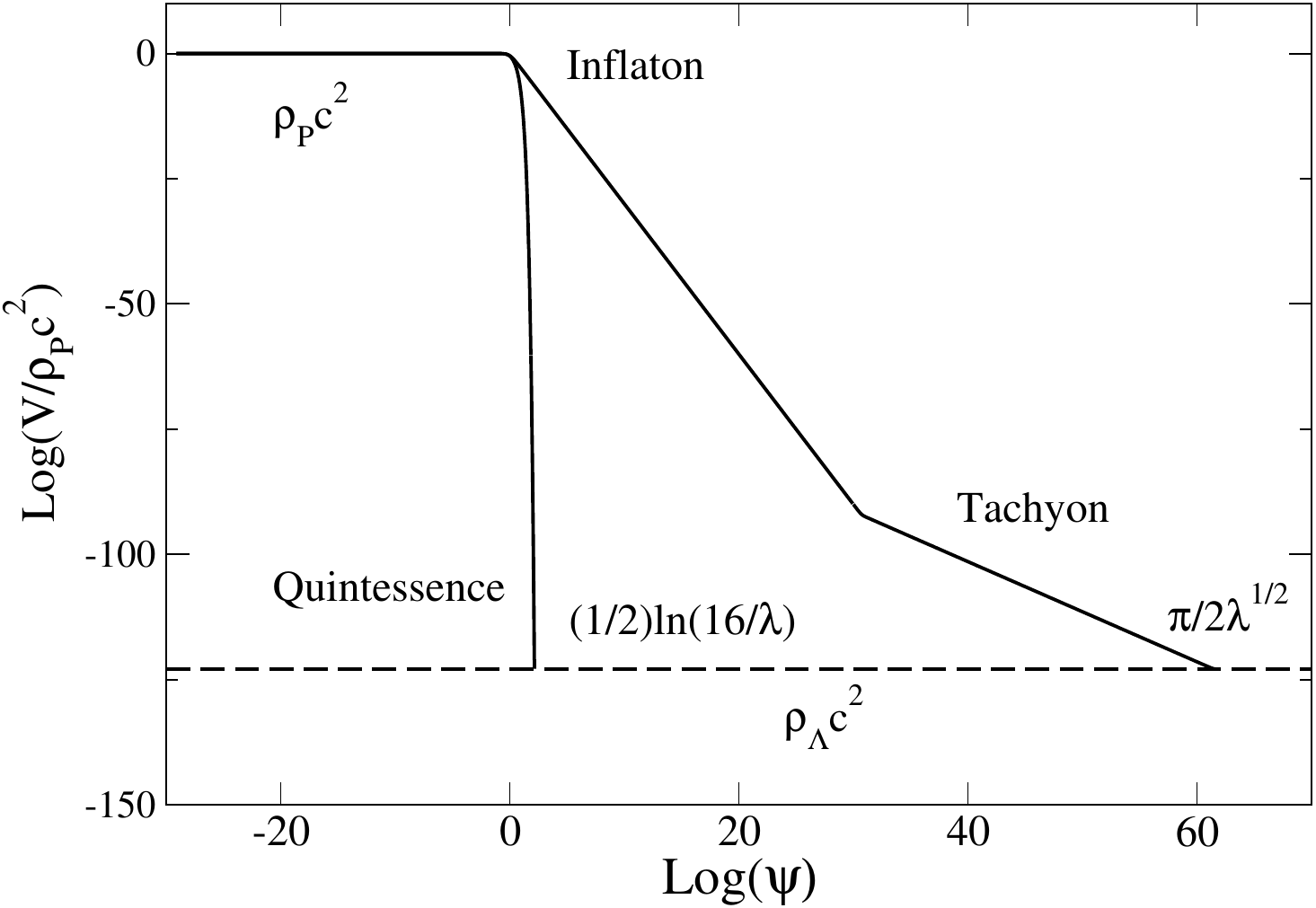}
\caption{Potential $V(\psi)$ in logarithmic coordinates.}
\label{psiV}
\end{center}
\end{figure}
\vspace{-12pt}

\begin{figure}[H]
\begin{center}
\includegraphics[clip,scale=0.5]{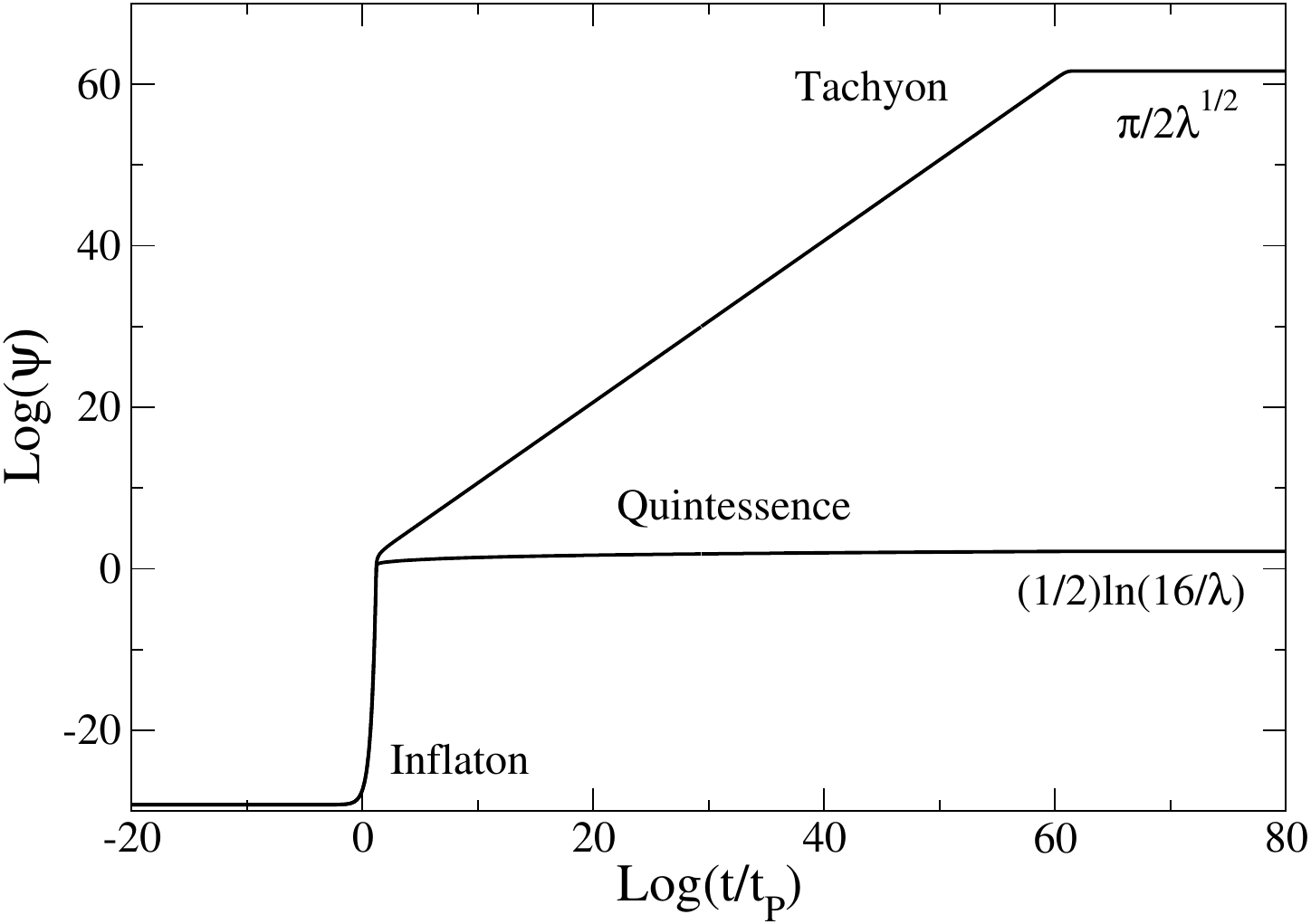}
\caption{Evolution of the scalar field $\psi$ with the normalized time $t/t_P$
in logarithmic~coordinates.}
\label{psit}
\end{center}
\end{figure}

{\it Remark:} Using matched asymptotics, we can propose a simple analytical
expression of the scalar field potential that unifies the inflaton and
quintessence potentials. Indeed, the following potential
\begin{eqnarray}
V(\psi)=\frac{1}{2}\rho_P c^2
\frac{(1-\alpha)\cosh^2\psi+\alpha+1}{\cosh^4\psi}+\frac{1}{2}\rho_{
\Lambda}c^2 \left\lbrack
(1-\alpha)\cosh^2(\psi_{max}-\psi)+\alpha+1\right \rbrack \nonumber\\
-2\rho_P
c^2(1-\alpha)e^{-2\psi}\qquad (0\le\psi\le\psi_{max})
\end{eqnarray}
provides  a good approximation of the exact potential given by Equation
(\ref{unif1}).

\subsection{Tachyonic field}

We consider a tachyonic
field defined by Equations (\ref{tachyon1}) and (\ref{tachyon2}).
From Equation~(\ref{tachyon2}), we get
\begin{eqnarray}
\label{newsft12}
\dot\phi^2=1+w.
\end{eqnarray}
Using $\dot\phi=(d\phi/da)Ha$ and the Friedmann equation (\ref{b9}) valid for
a
flat universe, we find that the relation between the tachyonic field and the
scale factor is given by
\begin{eqnarray}
\label{sft12}
\frac{d\phi}{da}=\left (\frac{3c^2}{8\pi G}\right
)^{1/2}\frac{\sqrt{1+w}}{\sqrt{\rho c^2}a}.
\end{eqnarray}
On the other hand, according to Equation~(\ref{tachyon2}), the potential of the
scalar field is given by
\begin{eqnarray}
\label{sft13}
V^2=-w\rho^2 c^4.
\end{eqnarray}
The tachyonic field is defined provided that $-1\le w\le 0$. This is the case
for all
times
with the quadratic equation of state (\ref{g1}) when $\alpha=0$. Using Equations
(\ref{sft3}) and (\ref{sft4}) and making the change of variables
\begin{equation}
\label{sft14}
x=R^{3(\alpha+1)/2},\qquad \psi=\sqrt{\rho_P c^2}\left (\frac{8\pi
G}{3c^2}\right )^{1/2}\frac{3\sqrt{\alpha+1}}{2}\phi,
\end{equation}
we obtain after simple calculations
(using the fact that $\lambda\ll 1$ to
simplify some terms):
\begin{equation}
\label{sft15}
\psi=\int_0^x\frac{ds}{1+\lambda s^2}=\frac{1}{\sqrt{\lambda}}\tan^{-1}\left
(\sqrt{\lambda}x\right ),
\end{equation}
\begin{equation}
\label{sft16}
V=\rho_P c^2\frac{\sqrt{1-\alpha x^2-(\alpha-1)\lambda
x^4+\lambda^2x^6}}{(x^2+1)^{3/2}}.
\end{equation}
We have taken $\psi=0$ when $x=0$. Equations (\ref{sft15}) and (\ref{sft16})
define the
potential $V(\psi)$ in parametric form with parameter $x$ going from $0$ to
$+\infty$. The scalar field goes from $\psi=0$ to
$\psi_{max}=\pi/(2\sqrt{\lambda})$. Eliminating $x$ between Equations
(\ref{sft15}) and (\ref{sft16}), we obtain
\begin{equation}
\frac{V(\psi)}{\rho_P
c^2}=\frac{\sqrt{1-\alpha\frac{1}{\lambda}\tan^2(\sqrt{\lambda}
\psi)-(\alpha-1)\frac { 1 } { \lambda }
\tan^4(\sqrt{\lambda}\psi)+\frac{1}{
\lambda }
\tan^6(\sqrt{\lambda}\psi)}}{\left\lbrack
\frac{1}{\lambda}\tan^2(\sqrt{\lambda}\psi)+1\right\rbrack^{3/2}}.
\end{equation}
For $\alpha=0$, we get
\begin{equation}
\frac{V(\psi)}{\rho_P
c^2}=\frac{\sqrt{1+\frac{1}{\lambda}\tan^4(\sqrt{\lambda}\psi)+\frac{1}{
\lambda}
\tan^6(\sqrt{\lambda}\psi)}}{\left\lbrack
\frac{1}{\lambda}\tan^2(\sqrt{\lambda}\psi)+1\right\rbrack^{3/2}}.
\end{equation}
In the early universe, we can neglect the
dark energy ($\lambda=0$) so that Equations (\ref{sft15}) and (\ref{sft16})
\mbox{reduce to}
\begin{equation}
\label{sft17}
\psi=x,\qquad V=\rho_P c^2\frac{\sqrt{1-\alpha x^2}}{(x^2+1)^{3/2}},
\end{equation}
giving
\begin{equation}
V(\psi)=\rho_P c^2\frac{\sqrt{1-\alpha\psi^2}}{(\psi^2+1)^{3/2}}.
\end{equation}
For $\alpha=0$, we get
\begin{equation}
\label{sft18}
V(\psi)=\frac{\rho_P c^2}{(\psi^2+1)^{3/2}}.
\end{equation}
For $\psi=x=0$, we have $V=\rho_P c^2$. For $\psi=x\gg 1$ we
obtain  $V\sim \rho_P c^2\sqrt{-\alpha}/\psi^2$ if $\alpha\neq 0$ and
$V\sim
\rho_P c^2/\psi^3$ if $\alpha=0$.
In the late universe, we can neglect the vacuum energy (which amounts to taking
$x\gg 1$) so that Equation  (\ref{sft15}) and (\ref{sft16}) reduce to
\begin{equation}
\label{sft15b}
\psi=\frac{1}{\sqrt{\lambda}}\tan^{-1}\left (\sqrt{\lambda}x\right
),\qquad V=\rho_P c^2\frac{\sqrt{-\alpha-(\alpha-1)\lambda
x^2+\lambda^2x^4}}{x^2},
\end{equation}
giving
\begin{equation}
\frac{V(\psi)}{\rho_P
c^2}=\frac{\sqrt{-\alpha-(\alpha-1) \tan^2(\sqrt{\lambda}\psi)+
\tan^4(\sqrt{\lambda}\psi)}}{\frac{1}{\lambda}\tan^2(\sqrt{\lambda}\psi)
}.
\end{equation}
For $\alpha=0$, we get
\begin{equation}
\label{sft19b}
V(\psi)=\frac{\rho_{\Lambda}c^2}{\sin(\sqrt{\lambda}\psi)}.
\end{equation}
Equations (\ref{sft15b}) and (\ref{sft19b}) are equivalent to Equations
(\ref{a22}) and (\ref{nom9w}) where we have arbitrarily
shifted $\sqrt{\lambda}\psi$ by $-\pi/2$. For
$\psi_{max}=\pi/(2\sqrt{\lambda})=4.26\times 10^{61}$ ($x\rightarrow
+\infty$),
we
have $V=\rho_{\Lambda} c^2$. For $\psi\ll\psi_{max}$ ($x\ll
1$), we have
$\psi\sim x$ and we obtain $V\sim \rho_P c^2\sqrt{-\alpha}/\psi^2$ if
$\alpha\neq 0$ and $V\sim \rho_P c^2\sqrt{\lambda}/\psi$ if
$\alpha=0$. The algebraic potential  $V\sim \rho_P
c^2\sqrt{-\alpha}/\psi^2$
corresponds to the potential of a tachyonic field with a~linear equation of
state $p=\alpha\rho c^2$ (see, e.g., Section 8.2 of
\cite{chavanis2}) [Note 17: For $\alpha=0$,
the tachyonic scalar field potential behaves as $\psi^{-3}$ at the end of the
early
universe and as $\psi^{-1}$ at the beginning of the late universe (this is
because the potential of a tachyonic field with an equation of state
$p=\alpha\rho c^2$ degenerates to $V(\phi)=0$ when $\alpha=0$; see, e.g.,
Section
8.2 of \cite{chavanis2}). This implies
a discontinuity of $d\ln V/d\ln\psi$ during the matter era as is apparent on
Figures \ref{xV} and \ref{psiV}.]. Using Equations~(\ref{hel8})
and (\ref{sft15}), we can obtain the evolution of the scalar field with time.
Using the asymptotic results of Sections \ref{sec_early} and \ref{sec_late}, we
find that the tachyonic field evolves as $\psi\simeq (l_P/a_1)^{3/2}{\rm
exp}[(6\pi)^{1/2}t/t_P]$ in the vacuum energy era, as $\psi\simeq
(6\pi)^{1/2}t/t_P$ in the $\alpha$-era, and as $\psi\simeq
\psi_{max}-(2/\sqrt{\lambda}){\rm exp}[-(6\pi\lambda)^{1/2}t/t_P]$ in
the dark
energy era. Some representative curves are shown in Figures \ref{xV}--\ref{psit}
for
$\alpha=0$.

\section{Further Generalizations of the Cosmological Model}
\label{sec_further}

In the standard model of cosmology, the universe undergoes a
radiation era, a  (baryonic $+$ dark) matter era, and a dark energy era.
These different phases are described by a linear equation of state
$p=\alpha\rho
c^2$ with $\alpha=1/3$ for the radiation, $\alpha=0$ for the
pressureless matter (including baryonic matter and dark matter), and
$\alpha=-1$ for the dark energy. The corresponding evolution of the universe
is described by Equation~(\ref{smx1}). It has been proposed in certain
cosmological models \cite{zeldovich,zeldocosmo,shapiro,mlbec,stiff} that the
universe also experiences a primordial stiff matter era during which
the speed of sound is equal to the speed of light ($c_s=c$). This phase is
described by a linear equation of state $p=\alpha\rho c^2$
with $\alpha=1$. If we
take into account a stiff matter era,
Equation~(\ref{smx1}) is generalized into
\begin{eqnarray}
\frac{H}{H_0}=\sqrt{\frac{\Omega_{s,0}}{(a/a_0)^6}
+\frac{\Omega_{rad,0}}{ (a/a_0)^4}
+\frac{\Omega_{m,0}}{(a/a_0)^3}+\Omega_{\Lambda,0}}.
\label{further1}
\end{eqnarray}
For convenience, we have written $\Omega_{m,0}=\Omega_{B,0}+\Omega_{DM,0}$.
Even more generally, we can
introduce an~$\alpha$-era with an equation of state $p=\alpha\rho c^2$ so
that
Equation~(\ref{further1}) becomes
\begin{eqnarray}
\frac{H}{H_0}=\sqrt{\frac{\Omega_{\alpha,0}}{(a/a_0)^{3(1+\alpha)}}
+\frac{\Omega_{s,0}}{(a/a_0)^6}
+\frac{\Omega_{rad,0}}{ (a/a_0)^4}
+\frac{\Omega_{m,0}}{(a/a_0)^3}+\Omega_{\Lambda,0}}.
\label{further2}
\end{eqnarray}
The $\alpha$-fluid may describe a new hypothetical fluid that dominates in the
very early universe (if $\alpha\ge 0$). Alternatively, the exponent $\alpha$
may
be a convenient
notation standing for stiff matter ($\alpha=1$), radiation  ($\alpha=1/3$) or
pressureless matter ($\alpha=0$) depending on the
situation contemplated.

The
cosmological model defined by Equation~(\ref{further2}) generates a primordial
Big Bang singularity in which the energy density diverges as the scale
factor goes to zero. Using the procedure introduced in this paper, it is easy to
modify Equation~(\ref{further2}) so as to incorporate an inflation era that replaces
the initial Big Bang singularity. Indeed, we can describe the transition between
an inflation era and
a generic $\alpha$-era by an equation of the form
\begin{eqnarray}
\frac{H}{H_0}=\sqrt{\frac{\Omega_{\alpha,0}}{
(a/a_0)^{3(1+\alpha)}+(a_1/a_0)^{3(1+\alpha)}}
+\frac{\Omega_{s,0}}{(a/a_0)^6}+\frac{\Omega_{rad,0}}{(a/a_0)^4}
+\frac{\Omega_{m,0}}{(a/a_0)^3}+\Omega_{\Lambda,0}}
\label{further3}
\end{eqnarray}
with $a_1/a_0=(\rho_{\alpha,0}/\rho_P)^{1/[3(1+\alpha)]}$.
This corresponds to the model of Section \ref{sec_early}  based on the quadratic
equation of state (\ref{early1}). We can generalize this model further
by
considering a
polytropic equation of state of the form of Equation~(\ref{newearly1}) with an
arbitrary index $n>0$.
In that case, we obtain
\begin{eqnarray}
\frac{H}{H_0}=\sqrt{\frac{\Omega_{\alpha,0}}{\left\lbrack
(a/a_0)^{3(1+\alpha)/n}+(a_1/a_0)^{3(1+\alpha)/n}\right\rbrack^n}
+\frac{\Omega_{s,0}}{(a/a_0)^6}+\frac{\Omega_{rad,0}}{(a/a_0)^4}
+\frac{\Omega_{m,0}}{(a/a_0)^3}+\Omega_{\Lambda,0}}.
\label{further4}
\end{eqnarray}
The cosmological model defined by Equation~(\ref{further4}) exhibits a phase of
early
inflation followed by an $\alpha$-era (which may represent a new exotic era, a
stiff matter era, or a radiation era), a radiation era (if different from the
$\alpha$ era), a matter era, and a dark energy era.

{\it Remark:} An additional generalization may be introduced if dark matter is
made of Bose-Einstein condensates (BECs) instead of being pressureless. In that
case, the term $\Omega_{DM,0}/(a/a_0)^3$ should be replaced by
$\Omega_{DM,0}/[(a/a_0)^3\mp(a_{BEC}/a_0)^3]$ as proposed in References
\cite{harkocosmo,aabec}. It is shown in these papers that the presence
of BECs can substantially accelerate the formation of large-scale structures in
the universe.

\section{Exact Solution of the Quadratic Equation of State}
\label{sec_exactQ}

In this Appendix, we present the exact solution of the quadratic equation of
state (\ref{g1}) and show the extreme relevance of the approximations
made in the core of the paper.

Substituting the equation of state (\ref{g1}) in the equation of
continuity 
(\ref{b7}) and solving the resulting differential equation by retaining all
the terms, we obtain after elementary calculations
\begin{eqnarray}
\rho=\frac{\rho_P\sqrt{1-4\lambda}}{1+({a}/{a_1})^{3(\alpha+1)\sqrt{1-4\lambda}}
}+\frac{1}{2}\rho_P\left
(1-\sqrt{1-4\lambda}\right ),
\label{exactQ1}
\end{eqnarray}
where we recall that $\lambda=\rho_{\Lambda}/\rho_P=1.36\times 10^{-123}$.
Since $\lambda\ll 1$ by $123$ orders of magnitude (!), we see that the
approximations leading to Equation (\ref{g2}) are excellent.

Substituting  Equation~(\ref{exactQ1}) into the Friedmann
equation (\ref{b9}) and  introducing the notation $R=a/a_1$, we obtain
\begin{eqnarray}
\label{exactQ2}
\int \frac{\sqrt{1+R^{\kappa_*}}}{R\sqrt{1+\lambda_* R^{\kappa_*}}}\,
dR=\left (\frac{8\pi}{3}\right )^{1/2}t/t_P^*,
\end{eqnarray}
with
\begin{eqnarray}
\label{exactQ3}
\kappa_*=3(\alpha+1)\sqrt{1-4\lambda},\qquad
\lambda_*=\frac{1-\sqrt{1-4\lambda}}{1+\sqrt{1-4\lambda}},
\end{eqnarray}
\begin{eqnarray}
\label{exactQ4}
t_P^*=\frac{t_P}{\sqrt{\frac{1}{2}+\frac{1}{2}\sqrt{1-4\lambda}}},
\qquad
t_{\Lambda}^*=\frac{t_P^*}{\sqrt{\lambda_*}}=t_{\Lambda}\sqrt{\frac{2\lambda}{
1-\sqrt { 1-4\lambda } } } .
\end{eqnarray}
Since $\lambda\ll 1$, we have  $\kappa_*\simeq\kappa$,
$\lambda_*\simeq \lambda$, $t_P^*\simeq t_P$, and
$t_{\Lambda}^*\simeq t_{\Lambda}$ in excellent approximation. We note
that
Equation~(\ref{exactQ2}) has the same form as Equation~(\ref{hel7}) with the star
variables. Therefore, its solution is given by Equation~(\ref{hel8}) where
$\kappa$,
$\lambda$ and $t_P$ are replaced by $\kappa_*$, $\lambda_*$ and $t_P^*$. For
$t\rightarrow -\infty$ (early inflation), we get
\begin{eqnarray}
\label{exactQ5}
a\propto e^{(8\pi/3)^{1/2}t/t_{P}^*}.
\end{eqnarray}
For
$t\rightarrow +\infty$ (late inflation), we get
\begin{eqnarray}
\label{exactQ6}
a\propto e^{(8\pi/3)^{1/2}t/t_{\Lambda}^*}.
\end{eqnarray}
Again, since  $\lambda=1.36\times 10^{-123}$, we commit almost no mistake by
replacing the star variables by \mbox{the naked ones.}

\section{Scalar Field Theory Taking into Account the Presence of Radiation and
Matter}
\label{sec_sfg}

We assume that the universe contains radiation
\begin{eqnarray}
\label{sfg1}
P_{rad}=\frac{1}{3}\rho_{rad}c^2 \qquad \Rightarrow \qquad
\frac{\rho_{rad}}{\rho_0}=\frac{\Omega_{rad,0}}{(a/a_0)^4}
\end{eqnarray}
and (baryonic $+$ dark) matter
\begin{eqnarray}
\label{sfg2}
P_{m}=0 \qquad \Rightarrow \qquad
\frac{\rho_{m}}{\rho_0}=\frac{\Omega_{m,0}}{(a/a_0)^3}.
\end{eqnarray}
The relations on the right between the density and the scale factor have been
obtained from the equation of continuity
(\ref{b7}) applied to each species considered as a perfect fluid.

In order to unify vacuum energy ($\rho_P$) and dark energy
($\rho_{\Lambda}$), we assume that the universe contains a~scalar field with
a quadratic equation of state
\begin{equation}
\label{sfg3}
p_{\phi}=-(\alpha+1)\frac{\rho_{\phi}^2}{\rho_P}
c^2+\alpha\rho_{\phi}
c^2-(\alpha+1)\rho_{\Lambda}
c^2.
\end{equation}
The $\alpha$-component may represent radiation ($\alpha=1/3$) or dark matter
($\alpha=0$), or be a new species. The scalar field evolves according to the
Equations (\ref{early13b}) or (\ref{tachyon1}) that describe a canonical
or a tachyonic scalar field, respectively. We can check that they imply
the equation of continuity for the scalar field
\begin{equation}
\label{sfg4}
\frac{d\rho_{\phi}}{dt}+3\frac{\dot a}{a}\left
(\rho_{\phi}+\frac{p_{\phi}}{c^2}\right )=0.
\end{equation}
Therefore, the energy density of the scalar field is given by (see Section
\ref{sec_quadra}):
\begin{equation}
\label{sfg5}
\frac{\rho_{\phi}}{\rho_0}=\frac{\Omega_{\alpha,0}}{(a/a_0)^{3(\alpha+1)}
+\frac{
\Omega_{\alpha,0}}{\Omega_{P,0}}} +\Omega_{\Lambda,0}.
\end{equation}
The Friedmann equation taking into account
the contribution of all the
species writes
\begin{equation}
\label{sfg6}
H^2=\left (\frac{\dot a}{a}\right )^2=\frac{8\pi
G}{3}(\rho_{rad}+\rho_m+\rho_{\phi}).
\end{equation}
If radiation or dark matter is included in the scalar field, it should not be
counted as a new species, \emph{i.e.}, one should set $\rho_{rad}=0$ or
$\rho_{DM}=0$
in Equation~(\ref{sfg6}). Following the suggestion of Section \ref{sec_genx}, we
should include radiation in the scalar field so that $\alpha=1/3$ [and
$\rho_{rad}=0$ in Equation~(\ref{sfg6})] but, for the sake of generality, we
leave
$\alpha$ unspecified.

Generalizing the procedure of Appendix \ref{sec_sft} by taking into account
the presence of radiation and matter, we find that the potential of the
canonical scalar field is determined by the equations
\begin{equation}
\label{sfg7}
\phi(R)=\left (\frac{3c^2}{8\pi G}\right )^{1/2}\int_1^R
\frac{\sqrt{1+w_{\phi}(x)}}{x}\sqrt{\frac{\rho_{\phi}(x)}{\rho_{rad}
(x)+\rho_{m
}(x)+\rho_{\phi}(x)}}\, dx,
\end{equation}
\begin{equation}
\label{sfg8}
\frac{V(R)}{\rho_0 c^2}=\frac{1}{2}\left\lbrack 1-w_{\phi}(R)\right\rbrack
\frac{\rho_{\phi}(R)}{\rho_0}.
\end{equation}
We have defined $w_{\phi}=p_{\phi}/\rho_{\phi}c^2$, $R=a/a_0$, and
$x=a'/a_0$. We have assumed that $\phi=0$ at the present epoch ($R=1$). If we
take $\rho_{rad}=\rho_m=0$, considering a single scalar field, we recover the
results of Appendix  \ref{sec_sft}. Therefore, the present Appendix generalizes
the results of Appendix  \ref{sec_sft} by taking into account the presence of
radiation and matter. Although it does not seem possible to obtain the scalar
field potential $V(\phi)$ analytically, it can be obtained numerically
and it provides a unification of vacuum energy (inflaton) and dark energy
(quintessence) in the presence of radiation and matter. In very good
approximation (using the fact that $\lambda=\rho_{\Lambda}/\rho_P=1.36\times
10^{-123}\ll 1$), we obtain
\begin{eqnarray}
\label{sfg9}
\phi(R)=\left (\frac{3c^2}{8\pi G}\right )^{1/2}\sqrt{1+\alpha}\int_1^R
\frac{\left
(\frac{\Omega_{P,0}}{\Omega_{\alpha,0}}\right
)^{1/2}x^{3(\alpha+1)/2}}{\sqrt{1+\frac{\Omega_{P,0}}{\Omega_{\alpha,0}}
x^{3(\alpha+1)}}\sqrt{1+\frac{\Omega_{\Lambda,0}}{\Omega_{\alpha,0}}
x^{3(\alpha+1)}}}\nonumber\\
\times \sqrt{\frac{\frac{\Omega_{\alpha,0}}{x^{3(\alpha+1)}+\frac{
\Omega_{\alpha,0}}{\Omega_{P,0}}}
+\Omega_{\Lambda,0}}{\frac{\Omega_{rad,0}}{x^4}+\frac{\Omega_{m,0}}{x^3}+
\frac{\Omega_{\alpha,0}}{x^{3(\alpha+1)}+\frac{
\Omega_{\alpha,0}}{\Omega_{P,0}}}+\Omega_{\Lambda,0}}}\, \frac{dx}{x},
\end{eqnarray}
\begin{eqnarray}
\label{sfg10}
\frac{V(R)}{\rho_0 c^2}=\frac{1}{2}\Omega_{P,0}
\frac{2+(1-\alpha)\frac{\Omega_{P,0}}{\Omega_{\alpha,0}}(a/a_0)^{3(\alpha+1)
}
+2\frac{\Omega_{\Lambda,0}\Omega_{P,0}}{(\Omega_{\alpha,0})^2}(a/a_0)^{
6(\alpha+1) } } {\left\lbrack
\frac{\Omega_{P,0}}{\Omega_{\alpha,0}}(a/a_0)^{3(\alpha+1)}+1\right\rbrack^2}.
\end{eqnarray}
In the early universe ($\Omega_{\Lambda}=0$), we can make the approximation
\begin{eqnarray}
\label{sfg11}
\phi(R)=\left (\frac{3c^2}{8\pi G}\right )^{1/2}\sqrt{1+\alpha}\int_1^R
\frac{\left
(\frac{\Omega_{P,0}}{\Omega_{\alpha,0}}\right
)^{1/2}x^{3(\alpha+1)/2}}{\sqrt{1+\frac{\Omega_{P,0}}{\Omega_{\alpha,0}}
x^{3(\alpha+1)}}}
\sqrt{\frac{\frac{\Omega_{\alpha,0}}{x^{3(\alpha+1)}+\frac{
\Omega_{\alpha,0}}{\Omega_{P,0}}}}{\frac{\Omega_{rad,0}}{x^4}+\frac{\Omega_{m,0}
}{x^3}+
\frac{\Omega_{\alpha,0}}{x^{3(\alpha+1)}+\frac{
\Omega_{\alpha,0}}{\Omega_{P,0}}}}}\, \frac{dx}{x},
\end{eqnarray}
\begin{eqnarray}
\label{sfg12}
\frac{V(R)}{\rho_0 c^2}=\frac{1}{2}\Omega_{P,0}
\frac{2+(1-\alpha)\frac{\Omega_{P,0}}{\Omega_{\alpha,0}}(a/a_0)^{3(\alpha+1)
}}
{\left\lbrack
\frac{\Omega_{P,0}}{\Omega_{\alpha,0}}(a/a_0)^{3(\alpha+1)}+1\right\rbrack^2}.
\end{eqnarray}
In the late universe ($\Omega_{P}\rightarrow +\infty$), we can make the
approximation
\begin{eqnarray}
\label{sfg13}
\phi(R)=\left (\frac{3c^2}{8\pi G}\right )^{1/2}\sqrt{1+\alpha}\int_1^R
\frac{1}{\sqrt{1+\frac{\Omega_{\Lambda,0}}{\Omega_{
\alpha,0}}
x^{3(\alpha+1)}}} \sqrt{\frac{\frac{\Omega_{\alpha,0}}{x^{3(\alpha+1)}}
+\Omega_{\Lambda,0}}{\frac{\Omega_{rad,0}}{x^4}+\frac{\Omega_{m,0}}{x^3}+
\frac{\Omega_{\alpha,0}}{x^{3(\alpha+1)}}+\Omega_{\Lambda,0}}}\,
\frac{dx}{x},
\end{eqnarray}
\begin{eqnarray}
\label{sfg14}
\frac{V(R)}{\rho_0 c^2}=
\frac{(1-\alpha)\Omega_{\alpha,0}}{2(a/a_0)^{3(\alpha+1)}}
+\Omega_{\Lambda,0}.
\end{eqnarray}

Generalizing the procedure of Appendix \ref{sec_sft} by taking into account
the presence of radiation and matter, we find that the potential of the
tachyonic scalar field is determined by the equations
\begin{equation}
\label{sfg15}
\phi(R)=\left (\frac{3c^2}{8\pi G \rho_0 c^2}\right )^{1/2}\int_1^R
\frac{\sqrt{1+w_{\phi}(x)}}{x\sqrt{\frac{\rho_{\phi}(x)}{\rho_0}}}
\sqrt{\frac{\rho_{\phi}(x)}{\rho_{rad}(x)+\rho_{m
}(x)+\rho_{\phi}(x)}}\, dx,
\end{equation}
\begin{equation}
\label{sfg16}
\frac{V(R)}{\rho_0
c^2}=\sqrt{|w_{\phi}(R)|}\frac{\rho_{\phi}(R)}{\rho_0}.
\end{equation}
If we take $\rho_{rad}=\rho_m=0$, considering a single scalar field, we
recover
the results of Appendix \ref{sec_sft}. In~very good approximation (using the
fact that
$\lambda=\rho_{\Lambda}/\rho_P=1.36\times 10^{-123}\ll 1$), we obtain
\begin{eqnarray}
\label{sfg17}
\phi(R)=\left (\frac{3c^2}{8\pi G\rho_0 c^2}\right
)^{1/2}\frac{\sqrt{1+\alpha}}{\sqrt{\Omega_{\alpha,0}}}\int_1^R
\frac{x^{3(\alpha+1)/2}}{1+\frac{\Omega_{\Lambda,0}}{\Omega_{\alpha,0}}
x^{3(\alpha+1)}}\nonumber\\
\times \sqrt{\frac{\frac{\Omega_{\alpha,0}}{x^{3(\alpha+1)}+\frac{
\Omega_{\alpha,0}}{\Omega_{P,0}}}
+\Omega_{\Lambda,0}}{\frac{\Omega_{rad,0}}{x^4}+\frac{\Omega_{m,0}}{x^3}+
\frac{\Omega_{\alpha,0}}{x^{3(\alpha+1)}+\frac{
\Omega_{\alpha,0}}{\Omega_{P,0}}}+\Omega_{\Lambda,0}}}\, \frac{dx}{x},
\end{eqnarray}
\begin{eqnarray}
\label{sfg18}
\frac{V(R)}{\rho_0
c^2}=\frac{\Omega_{P,0}\sqrt{1-\alpha\frac{\Omega_{P,0}}{\Omega_{\alpha,0}}
(a/a_0)^{3(\alpha+1)}-(\alpha-1)\frac{\Omega_{P,0}\Omega_{\Lambda,0}}{(\Omega_
{
\alpha,0})^2}(a/a_0)^{6(\alpha+1)}+\frac{\Omega_{P,0}(\Omega_{\Lambda,0})^2}{
(\Omega_{\alpha,0})^3}(a/a_0)^{9(\alpha+1)}}}{\left\lbrack
\frac{\Omega_{P,0}}{\Omega_{\alpha,0}}(a/a_0)^{3(\alpha+1)}+1\right\rbrack^{
3/2}
}.
\end{eqnarray}
In the early universe ($\Omega_{\Lambda}=0$), we can make the approximation
\begin{eqnarray}
\label{sfg19}
\phi(R)=\left (\frac{3c^2}{8\pi G\rho_0 c^2}\right
)^{1/2}\frac{\sqrt{1+\alpha}}{\sqrt{\Omega_{\alpha,0}}}\int_1^R
x^{3(\alpha+1)/2}
\sqrt{\frac{\frac{\Omega_{\alpha,0}}{x^{3(\alpha+1)}+\frac{
\Omega_{\alpha,0}}{\Omega_{P,0}}}}{\frac{\Omega_{rad,0}}{x^4}+\frac{\Omega_{m,0}
}{x^3}+
\frac{\Omega_{\alpha,0}}{x^{3(\alpha+1)}+\frac{
\Omega_{\alpha,0}}{\Omega_{P,0}}}}}\, \frac{dx}{x},
\end{eqnarray}
\begin{eqnarray}
\label{sfg20}
\frac{V(R)}{\rho_0
c^2}=\frac{\Omega_{P,0}\sqrt{1-\alpha\frac{\Omega_{P,0}}{\Omega_{\alpha,0}}
(a/a_0)^{3(\alpha+1)}}}
{ \left\lbrack
\frac{\Omega_{P,0}}{\Omega_{\alpha,0}}(a/a_0)^{3(\alpha+1)}+1\right\rbrack^{
3/2}
}.
\end{eqnarray}
In the late universe ($\Omega_{P}\rightarrow +\infty$), we can make the
approximation
\begin{eqnarray}
\label{sfg21}
\phi(R)=\left (\frac{3c^2}{8\pi G\rho_0 c^2}\right
)^{1/2}\frac{\sqrt{1+\alpha}}{\sqrt{\Omega_{\alpha,0}}}\int_1^R
\frac{x^{3(\alpha+1)/2}}{1+\frac{\Omega_{\Lambda,0}}{\Omega_{\alpha,0}}
x^{3(\alpha+1)}}\nonumber\\
\times \sqrt{\frac{\frac{\Omega_{\alpha,0}}{x^{3(\alpha+1)}}
+\Omega_{\Lambda,0}}{\frac{\Omega_{rad,0}}{x^4}+\frac{\Omega_{m,0}}{x^3}+
\frac{\Omega_{\alpha,0}}{x^{3(\alpha+1)}}+\Omega_{\Lambda,0}}}\,
\frac{dx}{x},
\end{eqnarray}
\begin{eqnarray}
\label{sfg22}
\frac{V(R)}{\rho_0
c^2}=\frac{\Omega_{\alpha,0}\sqrt{-\alpha-(\alpha-1)
\frac{\Omega_{\Lambda,0}}{\Omega_{\alpha,0}}(a/a_0)^{3(\alpha+1)}+\left
(\frac{\Omega_{\Lambda,0}}{\Omega_{\alpha,0}}\right
)^2(a/a_0)^{6(\alpha+1)}}}{(a/a_0)^{3(\alpha+1)}}.
\end{eqnarray}
In conclusion, we have given the general equations that allow us to construct
the scalar field potential $V(\phi)$ that unifies the inflaton and quintessence
potentials in the presence of radiation and matter. The~physical consequences
of this model will be developed elsewhere.





\bibliographystyle{mdpi}
\makeatletter
\renewcommand\@biblabel[1]{#1. }
\makeatother


\begin{thebibliography}{999} 
\bibitem{bt}{Binney, J.;  Tremaine, S.  \emph{Galactic Dynamics}; Princeton
University Press: Princeton, NJ, USA,~2008.}
\bibitem{guth1}{Guth,  A.H. Inflationary universe: A possible solution to the
horizon and flatness problems. \emph{Phys.~Rev. D} {\bf 1981}, \emph{23},
347--356.}
\bibitem{guth2}{Linde, A.D. A new inflationary universe scenario: A
possible solution of the horizon, flatness, homogeneity, isotropy and primordial
monopole problems. \emph{Phys. Lett. B} {\bf 1982}, \emph{108}, 389--393.}
\bibitem{guth3}{Albrecht, A.; Steinhardt,
P.J.; Turner, M.S.;
Wilczek, F. Reheating an inflationary universe. \emph{Phys. Rev.
Lett.} {\bf 1982}, \emph{48}, 1437--1440.}
\bibitem{linde}{Linde, A. \emph{Particle Physics and Inflationary Cosmology};
Harwood: Chur, Switzerland, 1990.}
\bibitem{weinberg}{Weinberg,  S. \emph{Gravitation and Cosmology}; John Wiley \&
Sons: Hoboken, NJ, USA, 1972.}
\bibitem{cst}{Copeland,  E.J.;  Sami, M.; Tsujikawa, S.  Dynamics of Dark
Energy. \emph{Int. J. Mod. Phys. D}
{\bf 2006}, \emph{15}, 1753--1935.}
\bibitem{novae1}{Riess, A.G.;  Filippenko, A.V.; Challis, P.; Clocchiatti, A.;
Diercks, A.; Garnavich, P.M.; Gilliland,~R.L.; Hogan, C.J.; Jha, S.; Kirshner,
R.P.; \emph{et al}. Observational Evidence from Supernovae for an Accelerating
Universe and a Cosmological Constant. \emph{Astron. J.} \textbf{1998},
\emph{116}, 1009--1038.}
\bibitem{novae2}{Perlmutter, S.; Aldering, G.; Goldhaber, G.; Knop, R. A.;
Nugent, P.;
Castro, P. G.; Deustua, S.; Fabbro, S.; Goobar, A.; Groom, D. E.; \emph{et al}. 
Measurements of $\Omega$ and $\Lambda$ from $42$ High-Redshift Supernovae.
\emph{ApJ}
\textbf{1999}, \emph{517}, 565--586.}
\bibitem{novae3}{de Bernardis, P.; Ade, P. A. R.; Bock, J. J.; Bond, J. R.;
Borrill, J.;
Boscaleri, A.; Coble, K.; Crill, B. P.; De Gasperis, G.; Farese, P. C.;
\emph{et al}. A flat Universe from high-resolution maps of the cosmic
microwave background radiation.
\emph{Nature} \textbf{2000},  \emph{404}, 955--959.}
\bibitem{novae4}{Hanany, S.; Ade, P.; Balbi, A.; Bock, J.; Borrill, J.;
Boscaleri, A.; de
Bernardis, P.; Ferreira, P. G.; Hristov, V. V.; Jaffe, A. H.; \emph{et
al}. MAXIMA-1: A Measurement of the Cosmic Microwave Background Anisotropy on
Angular Scales of $10'-5^{o}$. \emph{ApJ} \textbf{2000},  \emph{545},
L5--L9.}
\bibitem{quintessence1}{Ratra, B.; Peebles, J. Cosmological consequences of a
rolling homogeneous scalar field. \emph{Phys. Rev. D} \textbf{1988}, \emph{37},
3406--3427.}
\bibitem{quintessence2}{Starobinski,
A.A. How to determine an effective potential for a variable cosmological
term. \emph{JETP Lett.} \textbf{1998}, \emph{68},
757--763.}
\bibitem{quintessence3}{Caldwell, R.R.; Dave, R.;
Steinhardt, P.J. Cosmological Imprint of an Energy Component with General
Equation of State. \emph{Phys.~Rev. Lett.} \textbf{1998}, \emph{80}, 1582-1585.}
\bibitem{quintessence4}{Brax, 
P.; Martin,~J. Quintessence and supergravity.
\emph{Phys.~Lett. B} \textbf{1999}, \emph{468}, 40--45.}
\bibitem{quintessence5}{Albrecht, A.; Skordis, C. Phenomenology of a Realistic
Accelerating Universe Using Only Planck-Scale Physics. \emph{Phys. Rev.
Lett.} \textbf{2000}, \emph{84}, 2076-2079.}
\bibitem{quintessence6}{Barreiro,~T.; Copeland, E.J.; Nunes, N.J. Quintessence
arising from exponential potentials. \emph{Phys. Rev.
D} \textbf{2000}, \emph{61}, 127301.}
\bibitem{quintessence7}{Ure\~na-L\'opez, L.A.;  Matos, T. New cosmological
tracker solution for quintessence. \emph{Phys.~Rev. D} \textbf{2000}, \emph{62},
081302.}
\bibitem{quintessence8}{Brax, P.; Martin, J. Robustness of quintessence.
\emph{Phys. Rev. D} \textbf{2000}, \emph{61}, 103502.}
\bibitem{quintessence9}{Saini,~T.D.; Raychaudhury, S.; Sahni, V.; Starobinsky,
A.A. Reconstructing the Cosmic Equation of State from Supernova
Distances. \emph{Phys. Rev. Lett.} \textbf{2000}, \emph{85}, 1162--1165.}
\bibitem{quintessence10}{Sahni,~V.;  Starobinsky, A.A.  The Case for a Positive
Cosmological $\Lambda$-Term. \emph{Int. J. Mod. Phys. D} \textbf{2000},
\emph{9}, 373--443.}
\bibitem{quintessence11}{Sahni, V. The cosmological constant problem and
quintessence. \emph{Class. Quantum Grav.} \textbf{2002}, \emph{19}, 3435--3448.}
\bibitem{quintessence12}{Pavlov, M.; Rubano, C.;
Sazhin, M.; Scudellaro, P. Analysis of Two Quintessence Models with Type Ia
Supernova Data. \emph{Astrophys. J.} \textbf{2002}, \emph{566}, 619--622.}
\bibitem{quintessence13}{Sahni, V.;
Saini, T.D.; Starobinsky, A.A.;  Alam, U.  Statefinder--A new geometrical
diagnostic of dark energy. \emph{JETP Lett}. \textbf{2003}, \emph{77},
201--206.}
\bibitem{sen}{Sen,  A. Rolling Tachyon. \emph{J. High Energy Phys.}
\textbf{2002}, \emph{4}, 048.}
\bibitem{tachyon1}{Gibbons,   G.W. Cosmological evolution of the rolling
tachyon. \emph{Phys. Lett. B} \textbf{2002}, \emph{537}, 1--4.}
\bibitem{tachyon2}{Padmanabhan, T. Accelerated expansion of the universe driven
by tachyonic matter.
\emph{Phys. Rev. D} \textbf{2002}, \emph{66}, 021301}
\bibitem{tachyon3}{Frolov, A.; Kofman, L.;
Starobinsky, A. Prospects and problems of tachyon matter cosmology.
\emph{Phys. Lett. B} \textbf{2002}, \emph{545}, 8--16.}
\bibitem{chavanis1}{Chavanis, P.H. Models of universe with a polytropic equation
of state: I. The early universe. \emph{Eur. Phys. J. Plus} \textbf{2014},
\emph{129}, 38.}
\bibitem{chavanis2}{Chavanis, P.H. Models of universe with a polytropic equation
of state: II. The late universe. \emph{Eur. Phys. J. Plus} \textbf{2014},
\emph{129}, 222.}
\bibitem{chavanis3}{Chavanis, P.H. Models of universe with a polytropic equation
of state: III. The phantom universe. arXiv:1208.1185.}
\bibitem{chavanisAIP}{Chavanis, P.H. A simple model of universe describing the
early inflation and the late accelerated expansion in a symmetric manner.
\emph{AIP Conf.
Proc.} \textbf{2013}, \emph{1548}, 75--115.}
\bibitem{chandra}{Chandrasekhar,  S. \emph{An Introduction to the Study of
Stellar Structure}; Dover: New-York, USA, 1958.}
\bibitem{st}{Shapiro,  S.; Teukolsky, S.A. \emph{Black Holes, White Dwarfs, and
Neutron Stars}; Wiley: New-York, USA,~1983.}
\bibitem{tsallis}{Tsallis,  C. \emph{Introduction to Nonextensive Statistical
Mechanics}; Springer: New-York, USA, 2009.}
\bibitem{murray}{Murray,  J.D. \emph{Mathematical Biology}; Springer: Berlin, Germany,
1991.}
\bibitem{chaplygin1}{Kamenshchik,  A.; Moschella, U.; Pasquier, V. An
alternative to quintessence. \emph{Phys.
Lett. B} \textbf{2001}, \emph{511}, 265--268.}
\bibitem{chaplygin2}{Bilic, N.; Tuper,~G.B.; Viollier, R. Unification of dark
matter and dark energy: the inhomogeneous Chaplygin gas. \emph{Phys. Lett. B}
\textbf{2002}, \emph{535}, 17--21.}
\bibitem{chaplygin3}{Fabris, J.S.;
Gon\c calves, S.V.; de Souza,~P.E. Density Perturbations in a Universe
Dominated by the Chaplygin Gas. \emph{Gen. Relativ. Gravit}. \textbf{2002},
\emph{34}, 53--63.}
\bibitem{chaplygin4}{Bento, M.C.; Bertolami, O.; Sen,  A.A. Generalized
Chaplygin gas, accelerated expansion, and dark-energy-matter unification.
\emph{Phys. Rev. D} \textbf{2002}, \emph{66}, 043507.}
\bibitem{chaplygin5}{Benaoum, H.B. Accelerated Universe from Modified Chaplygin
Gas and Tachyonic Fluid. arXiv:hep-th/0205140.}
\bibitem{chaplygin6}{Gorini,
V.; Kamenshchik, A.; Moschella, U. Can the Chaplygin gas be a plausible model
for dark energy?
\emph{Phys.~Rev. D} \textbf{2003}, \emph{67}, 063509.}
\bibitem{chaplygin7}{Bento, M.C.; Bertolami,
O.; Sen, A.A. Revival of the unified dark energy - dark matter model?
\emph{Phys.~Rev. D} \textbf{2004}, \emph{70}, 083519.}
\bibitem{chaplygin8}{Debnath, U.; Banerjee, A.;
Chakraborty, S. Role of modified Chaplygin gas in accelerated
universe. \emph{Class. Quantum Grav}.  \textbf{2004}, \emph{21},  5609--5617.}
\bibitem{chaplygin9}{Karami,  K.; Ghaffari, S.; Fehri, J. Interacting polytropic
gas model of phantom dark energy in non-flat universe. \emph{Eur. Phys. J. C}
\textbf{2009}, \emph{64}, 85--88.}
\bibitem{weinbergcosmo}{Weinberg, S. The cosmological constant problem.
\emph{Rev. Mod. Phys}. \textbf{1989}, \emph{61}, 1--23.}
\bibitem{paddycosmo}{Padmanabhan, T. Cosmological constant -- the weight of the
vacuum. \emph{Phys. Rep}. \textbf{2003}, \emph{380}, 235-320.}
\bibitem{muller}{M\"uller,  C.M. Cosmological bounds on the equation of state of
dark matter. \emph{Phys. Rev. D} \textbf{2005}, \emph{71}, 047302.}
\bibitem{bamba}{Bamba, K.; Capozziello, S.; Nojiri, S.; Odintsov, S.D. Dark
energy cosmology: the equivalent description via different theoretical models
and cosmography tests. \emph{Astrophys. Space Sci}. \textbf{2012}, \emph{342},
155--228.}
\bibitem{planck2013}{Planck Collaboration. Planck 2013 results.
XVI. Cosmological parameters. \emph{Astron.
Astrophys}. \textbf{2014}, \emph{571}, 66.}
\bibitem{planck2015}{Planck Collaboration. Planck 2015 results. XIII.
Cosmological parameters. arXiv:1502.01589.}
\bibitem{ghosts1}{Caldwell, R.R. A phantom menace? Cosmological consequences
of a dark energy component with super-negative equation of state. \emph{Phys.
Lett. B} \textbf{2002}, \emph{545}, 23--29.}
\bibitem{ghosts2}{Caldwell, R.R.;
Kamionkowski, M.; Weinberg, N.N. Phantom Energy: Dark Energy with $w<-1$ Causes
a Cosmic Doomsday. \emph{Phys. Rev. Lett.} \textbf{2003}, \emph{91}, 071301.}
\bibitem{ghosts3}{McInnes, B. The dS/CFT Correspondence and the Big
Smash. \emph{J. High Energy Phys.} \textbf{2002}, \emph{8}, 029.}
\bibitem{ghosts4}{Carroll,~S.M.; Hoffman, M.;
Trodden, M. Can the dark energy equation-of-state parameter $w$ be less than
$-1$? \emph{Phys. Rev. D} \textbf{2003}, \emph{68}, 023509.}
\bibitem{ghosts5}{Singh, P.;
Sami, M.; Dadhich, N. Cosmological dynamics of a phantom field.
\emph{Phys.~Rev. D} \textbf{2003}, \emph{68}, 023522.}
\bibitem{ghosts6}{Cline, J.M.; Jeon, S.;
Moore, G.D. The phantom menaced: Constraints on low-energy effective ghosts.
\emph{Phys. Rev. D} \textbf{2004}, \emph{70}, 043543.}
\bibitem{ghosts7}{Sami, M.; Toporensky, A. Phantom Field and the Fate of the
Universe. \emph{Mod. Phys. Lett. A}
\textbf{2004}, \emph{19}, 1509--1517.}
\bibitem{ghosts8}{Nesseris, S.; Perivolaropoulos, L. Fate of bound systems in
phantom and quintessence cosmologies. \emph{Phys. Rev. D} \textbf{2004},
\emph{70},
123529.}
\bibitem{ghosts9}{Babichev, E.; Dokuchaev, V.;  Eroshenko, Y.  Black Hole Mass
Decreasing due to Phantom Energy Accretion. \emph{Phys. Rev. Lett}.
\textbf{2004}, \emph{93}, 021102.}
\bibitem{ghosts10}{Gonz\'alez-D\'ias,~P.F.; Sig\"uenza, C.L. The fate of black
holes in an accelerating universe. \emph{Phys. Lett. B} \textbf{2004},
\emph{589}, 78--82.}
\bibitem{ghosts11}{Gonz\'alez-D\'ias,~P.F.;
Sig\"uenza, C.L. Phantom thermodynamics. \emph{Nucl. Phys. B} \textbf{2004},
\emph{697}, 363--386.}
\bibitem{ghosts12}{Nojiri, S.; Odintsov, S.D. Final state and thermodynamics of
a dark energy universe. \emph{Phys. Rev. D} \textbf{2004}, \emph{70}, 103522.}
\bibitem{ghosts13}{Nojiri,
S.; Odintsov, S.D.; Tsujikawa, S. Properties of singularities in the (phantom)
dark energy universe. \emph{Phys. Rev. D} \textbf{2005}, \emph{71}, 063004.}
\bibitem{ghosts14}{Stefanci\'c,~H. Expansion around the vacuum equation of
state: Sudden future singularities and asymptotic behavior. \emph{Phys. Rev. D}
\textbf{2005}, \emph{71}, 084024.}
\bibitem{ghosts15}{Bouhmadi-L\'opez, M.;
Gonzalez-D\'iaz, P.F.; Mart\'in-Moruno, P. Worse than a big rip? \emph{Phys.
Lett. B} \textbf{2008}, \emph{659}, 1--5.}
\bibitem{ghosts16}{Garc\'ia-Compe\'an, H.; Garc\'ia-Jim\'enez, G.; Obreg\'on,
O.; Ram\'irez,~C. Crossing the phantom divide in an interacting generalized
Chaplygin gas. \emph{JCAP} \textbf{2008}, \emph{7}, 016.}
\bibitem{ghosts17}{Fern\'andez-Jambrina, L. $w$-cosmological
singularities. \emph{Phys. Rev. D} \textbf{2010}, \emph{82}, 124004.}
\bibitem{ghosts18}{Frampton, P.H.; Ludwick, K.J.;
Scherrer, R.J. The little rip. \emph{Phys. Rev. D}
\textbf{2011}, \emph{84}, 063003.}
\bibitem{jgrav}{Chavanis, P.H. A Cosmological Model Based on a Quadratic
Equation of State Unifying Vacuum Energy, Radiation, and Dark Energy.
\emph{Journal of Gravity} \textbf{2013},
\emph{2013}, 682451.}
\bibitem{basilakos}{{Basilakos, S.; Barrow, J.D. Hyperbolic inflation in the
light of Planck 2015 data. \emph{Phys. Rev.
D} \textbf{2015}, \emph{91}, 103517.}}
\bibitem{wormhole1}{{Gonz\'alez-D\'ias, P.F.;
Jim\'enez-Madrid, J.A. Phantom inflation and the ``Big Trip''. \emph{Phys. Lett.
B} \textbf{2004}, \emph{596}, 16--25.}}
\bibitem{astashenok}{{Astashenok, A.; Nojiri, S.; Odintsov, S.D.;
Yurov, A.V. Phantom cosmology without Big Rip singularity. \emph{Phys. Lett.
B} \textbf{2012}, \emph{709}, 396--403.}}
\bibitem{sahni}{{Sahni, V.; Wang, L. New cosmological model of quintessence and
dark matter. \emph{Phys.
Rev. D} \textbf{2000}, \emph{62}, 103517.}}
\bibitem{shahalam}{{Shahalam, M.; Sami, S.; Agarwal, A. $Om$ diagnostic applied
to scalar field models and slowing down of cosmic acceleration.
\emph{Mon. Not. R. Astron. Soc.} \textbf{2015}, \emph{448}, 2948--2959.}}
\bibitem{chimento}{{Chimento, L.P.; Jakubi, A.S. Scalar Field Cosmologies with
Perfect Fluid in Robertson-Walker Metric. \emph{Int.
J. Mod. Phys. D} \textbf{1996}, \emph{5}, 71--84.}}
\bibitem{ulm}{{Ure\~na-L\'opez, L.A.; Matos, T. New cosmological tracker
solution for quintessence. \emph{Phys. Rev.
D} \textbf{2000}, \emph{62}, 081302.}}
\bibitem{rubano}{{Rubano, C.; Barrow, J.D. Scaling solutions and reconstruction
of scalar field potentials. \emph{Phys.
Rev.
D} \textbf{2001}, \emph{64}, 127301.}}
\bibitem{paliathanasis}{{Paliathanasis, A.; Tsamparlis, M.;
Basilakos, S. Dynamical symmetries and observational constraints in scalar field
cosmology. \emph{Phys. Rev. D} \textbf{2014},\emph{ 90},  103524.}}
\bibitem{gorini}{{Gorini, V.; Kamenshchik, A.; Moschella, U.;
Pasquier, V. Tachyons, scalar fields, and cosmology. \emph{Phys. Rev. D}
\textbf{2004}, \emph{69},  123512.}}
\bibitem{keresztes}{{Keresztes, Z.;  Gergely, L.A. Combined cosmological tests
of a bivalent tachyonic dark energy scalar field model. \emph{JCAP}
\textbf{2014}, \emph{11}, 026.}}
\bibitem{monerat}{Monerat, G. A.; Oliveira-Neto, G.; Silva, E. V. Corr\^ea;
Filho, L. G. Ferreira; Romildo, P., Jr.; Fabris, J. C.; Fracalossi, R.;
Gon\c calves, S. V. B.; Alvarenga, F. G. Dynamics of the early universe and the
initial conditions for inflation in a model with radiation and a Chaplygin gas.
\emph{Phys. Rev. D} \textbf{2007}, \emph{76}, 024017.}
\bibitem{chandramass}{Chandrasekhar, S. The Maximum Mass of Ideal White
Dwarfs. \emph{Astrophys. J.}  \textbf{1931}, \emph{74}, 81.}
\bibitem{ccp}{Steinhardt, P. {\it Critical Problems in Physics};
Fitch, V.L., Marlow, D.R., Eds.; Princeton University Press: Princeton,
NJ, USA, 1997.}
\bibitem{ccpg1}{Zlatev, I.; Wang, L.; Steinhardt, P.J. Quintessence, Cosmic
Coincidence, and the Cosmological Constant. \emph{Phys. Rev. Lett.}
\textbf{1999}, \emph{82}, 896--899.}
\bibitem{ccpg2}{Amendola, L.; Tocchini-Valentini, D. Stationary dark energy: The
present universe as a global attractor. \emph{Phys. Rev. D} \textbf{2001},
\emph{64},
043509}
\bibitem{ccpg3}{Chimento, L.P.; Jakubi, A.S.; Pav\'on,~D. Dark energy,
dissipation, and the coincidence problem. \emph{Phys. Rev. D} \textbf{2003},
\emph{67}, 087302.}
\bibitem{ccpg4}{Garc\'ia-Compe\'an, H.; Garc\'ia-Jim\'enez,
G.; Obreg\'on, O.; Ram\'irez,~C. Crossing the phantom divide in an interacting
generalized Chaplygin gas.
 \emph{JCAP} \textbf{2008}, \emph{7}, 016.}
\bibitem{vacuum1}{Lema\^itre, G. Evolution of the Expanding Universe.
\emph{Proc. Nat. Acad. Sci.}  \textbf{1934}, \emph{20}, 12--17.}
\bibitem{vacuum2}{Sakharov, A.D. Vacuum quantum fluctuations in curved space and
the theory of gravitation. \emph{Dokl. Akad. Nauk SSSR} \textbf{1967},
\emph{177}, 70.}
\bibitem{vacuum3}{Zeldovich, Y.B. Special Issue: the Cosmological Constant and
the Theory of Elementary Particles.
\emph{Sov. Phys. Uspek.} \textbf{1968}, \emph{11}, 381--393.}
\bibitem{setare1}{Setare, M.R.; Kamali, V. Tachyon-polytropic inflation on the
brane. \emph{Cent. Eur. J.
Phys.} \textbf{2013}, \emph{11}, 545--552.}
\bibitem{setare2}{Setare, M.R.; Houndjo, M.J.S.; Kamali, V. Warm-Polytropic
Inflationary Universe Model. \emph{Int. J. Mod. Phys. D} \textbf{2013},
\emph{22}, 1350041.}
\bibitem{fg}{Freitas, R.C.; Gon\c calves, S.V.B. Polytropic equation of state
and primordial quantum fluctuations. \emph{EPJC} \textbf{2014}, \emph{74},
3217.}
\bibitem{karami}{Asadzadeli, S.; Safari, Z.; Karami, K.; Abdolmaleki,
A. Cosmological Constraints on Polytropic Gas Model.
\emph{Int. J. Theor. Phys.} \textbf{2014}, \emph{53}, 1248--1262.}
\bibitem{aabec}{{Chavanis, P.H.} Growth of perturbations in an expanding
universe with Bose-Einstein condensate dark matter. \emph{Astron. Astrophys.} 
\textbf{2012}, \emph{537}, A127.}
\bibitem{primeval}{Lema\^itre,  G. \emph{The Primeval Atom}; Munitz, M.K.,
 Ed.; The Free Press, 1957.}
\bibitem{kazinski}{Kazinski, P.O. Quantum gravitational anomaly as a dark
matter. {arXiv:1501.05777}.}
\bibitem{ananda}{{Ananda, K.N.; Bruni, M. Cosmological dynamics and dark energy
with a nonlinear equation of state: A quadratic model. \emph{Phys. Rev. D}
\textbf{2006}, \emph{74},  023523.}}
\bibitem{linder}{{Linder, E.V.; Scherrer, R.J. Aetherizing Lambda: Barotropic
fluids as dark energy. \emph{Phys. Rev. D}
\textbf{2009}, \emph{80}, 023008.}}
\bibitem{sharov}{{Sharov, G.S. Observational constraints on cosmological models
with Chaplygin gas and
quadratic equation of state. {arXiv:1506.05246.}}}
\bibitem{zeldovich}{Zel'dovich, Y.B. The equation of state at ultrahigh
densities and its relativistic limitations. \emph{Soviet Phys. JETP}
\textbf{1962}, \emph{14}, 1143.}
\bibitem{zeldocosmo}{Zel'dovich, Y.B. A hypothesis, unifying the structure and
the entropy of the Universe. \emph{Mon. Not. R. Astron. Soc.} \textbf{1972},
\emph{160}, 1.}
\bibitem{shapiro}{Li, B.; Rindler-Daller, T.; Shapiro, P.R. Cosmological
constraints on Bose-Einstein-condensed scalar field dark matter. \emph{Phys.
Rev. D} \textbf{2014}, \emph{89},  083536.}
\bibitem{mlbec}{Chavanis, P.H. Partially relativistic self-gravitating
Bose-Einstein condensates with a stiff equation of state. \emph{Eur. Phys. J.
Plus} \textbf{2015},
\emph{130}, 181.}
\bibitem{stiff}{Chavanis, P.H. Cosmology with a stiff matter
era. {arXiv:1412.0743}.}
\bibitem{harkocosmo}{Harko, T. Evolution of cosmological perturbations in
Bose-Einstein condensate dark matter. \emph{Mon. Not. R. Astron.
Soc.} \textbf{2011}, \emph{413},  3095--3104.}











\end{thebibliography}


%


%

\end{document}